\documentclass[12]{amsart}
\usepackage{colortbl}
\theoremstyle{definition}

\theoremstyle{remark}

\numberwithin{equation}{section}

\begin{document}


\title[Algebraic structure of Hamiltonians on the NC phase space]{On the algebraic structure of rotationally invariant two-dimensional Hamiltonians on the noncommutative phase space}%

\author{H.\ Falomir$^1$, P.A.G.\ Pisani$^1$, F.\ Vega$^1$}%
\address{$1$: Instituto de F\'{\i}sica de La Plata (IFLP), CONICET; Departamento de F\'{\i}sica, Facultad de Ciencias Exactas, Universidad Nacional de La Plata (UNLP), Argentina}%
\email{falomir@fisica.unlp.edu.ar; pisani@fisica.unlp.edu.ar; \hfill\break 
federicogaspar@gmail.com}%

\author{D.\ C\'{a}rcamo$^2$, F.\ M\'{e}ndez$^2$}%
\address{$2$: Departamento de F\'{\i}sica, Universidad de Santiago de Chile (USACH), Chile}%
\email{dacarcamo@uc.cl; feritox@gmail.com;}%

\author{M.\ Loewe$^3$}%
\address{$3$: Instituto de F\'{\i}sica, Pontificia Universidad Cat\'{o}lica de Chile, Casilla 306, Santiago 22, Chile;
Centro Cient\'{\i}fico-Tecnol\'{o}gico de Valpara\'{\i}so, Casilla 110-V, Valpara\'{\i}so, Chile;
Centre for Theoretical and Mathematical Physics, and Department of Physics,
University of Cape Town, Rondebosch 7700, South Africa}
\email{mloewelo@yahoo.com;}%


\begin{abstract}
We study two-dimen\-sional Hamiltonians in phase space with noncommutativity both in coordinates and momenta. We consider the generator of rotations on the noncommutative plane and the Lie algebra generated by Hermitian rotationally invariant quadratic forms of noncommutative dynamical variables. We show that two quantum phases are possible, characterized by the Lie algebras $sl(2,\mathbb{R})$ or $su(2)$ according to the relation between the noncommutativity parameters, with the rotation generator related with the Casimir operator. From this algebraic perspective, we analyze the spectrum of some simple models with nonrelativistic rotationally invariant Hamiltonians in this noncommutative phase space, as the isotropic harmonic oscillator, the Landau problem and the cylindrical well potential.

\hfill\break



\end{abstract}
\maketitle

\section{Introduction} 

Noncommuting coordinates appeared first in \cite{Peierls}, in the description of nonrelativistic electrons of mass $m$ on a plane subject to a strong perpendicular magnetic field $B$ in the lowest Landau level. Indeed, in the $m/B \rightarrow 0$ limit only the lowest Landau level is accesible and the coordinates on the plane, $x_i, i=1,2$, appear as canonically conjugate dynamical variables \cite{Dunne-Jackiw,jcm2,Horvathy}. On the other hand, if one imposes the modified commutator for momenta $[p_i, p_j] = i \epsilon_{ij} B$ with constant $B$, the Hamiltonian of the \emph{free} particle, $H=\frac{\mathbf{p}^2}{2m}$, becomes equivalent to that of the conventional Landau problem with $B$ playing the  role of the magnetic field orthogonal to the  plane.

In 1930, W.\ Heisenberg also suggested to consider a noncommutative space as a tool to regularize the ultraviolet divergencies appearing in the quantum theory of fields \cite{Heisenberg-1930}, but deformed Heisenberg algebras by constant terms typically break Lorentz invariance. The first approach to this problem is due to H.\ Snyder who, in 1947, constructed a manifestly Lorentz invariant model of noncommutative space-time \cite{snyder1,yang} in which the four-dimensional projection of the symmetry group of a five-dimensional de Sitter space reproduces the Lorentz algebra, with the coordinates realized as generators of compact subgroups, thus presenting discrete spectra. Snyder applied these ideas as a regularization schema to Electrodynamics \cite{snyder2} but his results were not further considered, in spite of the intrinsic beauty of his proposal, mainly due to the success of the contemporary renormalization program which gave solution to the ultraviolet divergencies problem of quantum field theory and proved to allow for accurate numerical predictions for some physical observables.

More recently, the idea of a noncommutative space has attracted new attention due to some results in string theory \cite{Green}, in which an analogous behavior appears in the low energy effective theory of D-branes in background magnetic fields \cite{witten,DouNek}. It has also been argued that Quantum Mechanics combined with Einstein's theory would require, at the Planck scale, a space-time with nontrivial uncertainty relations which impose an effective short distance cut-off, since the attempt to localize an event with extreme precision demands an energy which would produce a gravitational collapse \cite{DFR,Madore,Szabo}. Similar noncommutative aspects have been found in some quantum gravity theories \cite{quant-grav,quant-grav-1,quant-grav-2}.

The advent of noncommutative geometry \cite{connes} has also stimulated the interest in the study of quantum mechanical systems with deformed algebra of commutators    \cite{jackiw,mezincescu,jcm1}, since these systems present interesting  properties, could give rise to models with some possible phenomenological consequences and provide new techniques for studying some quantum mechanical problems.

In this context, one can speculate with the possibility of having a low energy relic of these characteristics in the form of a nonrelativistic model of \emph{noncommutative quantum mechanics} (NCQM), where the dynamical variables satisfy a deformed Heisenberg algebra with nonvanishing commutators also between coordinates and/or between momenta \cite{Carroll,Duval,Poly,Bellucci,Morariu,Chai1,Scha,Ho,Smailagic,Lukierski,Chakraborty,Scholtz2,Scholtz1,Mik-Hor-2,mezincescu,Mikhail1,Mik-5,Saha,Stern,Scholtz-formulation}. There is a huge amount of work dedicated to the study of different nonrelativistic problems in NCQM considered from different perspectives; a partial list includes  \cite{Acatrinei,Chai2,Falomir1,Banerjee,Jonke,Deriglazov,Subir-1,Kijanka,Kokado,Banerjee1,Zhang-PLB-2004,Zhang-PRL-2004,Hor-Mik-1,Bertolami2,Bemfica,Bellucci2,Banerjee2,Calmet,scholtz,Banerjee4,Bastos,scholtz2,Banerjee3,Gango,Gomes,Bertolami1,Murray}.
Also the Dirac equation has been considered in these scenarios \cite{Mikhail-4,Mik-6,Kupri,Panella}.

When studying these problems, the (phase) space noncommutativity has been implemented through the Moyal product of ordinary functions of coordinates \cite{mezincescu,Kupri-Dima,DouNek,Szabo} or, alternatively, by explicit realizations of dynamical variables as operators on the Hilbert space \cite{Poly,Bellucci,Scholtz-formulation} satisfying a deformed Heisenberg algebra  of the form
\begin{equation}\label{conmutadores}
 \left[\hat{X}_i , \hat{X}_j\right] = \imath \theta\, \epsilon_{ij} \,,\qquad
  \left[\hat{X}_i , \hat{P}_j\right] = \imath \hbar \,  \delta_{ij} \,,\qquad
  \left[\hat{P}_i , \hat{P}_j\right] = \imath \kappa \, \epsilon_{ij}\,.
\end{equation}
This can be achieved, for example, by taking linear combinations of ordinary canonically conjugated dynamical variables with conveniently chosen coefficients (Bopp's shift). It is worth noticing that the presence of \emph{boundaries} in these noncommutative spaces must then be implemented by imposing conditions on the Hilbert space expressed in terms of these operators \cite{scholtz,Boundaries}.

Several articles have pointed out that nonrelativistic quantum mechanical models on the noncommutative phase space present two \emph{quantum phases} \cite{Poly,Bellucci,Mikhail1}, according to the value of the dimensionless parameter $\kappa \theta/\hbar^2$, separated by a \emph{critical point} where the effective dimension of the system is reduced \cite{Poly,Bellucci,Mik-Hor-2}. A similar behavior has been found in noncommutative extensions of the Dirac Hamiltonian \cite{Mikhail-4,Mik-6,Panella}.


\smallskip

It is the aim of this article to consider nonrelativistic two-dimensional rotationally invariant Hamiltonians on the noncommutative phase space, where the operators representing coordinates and momenta satisfy the commutation relations in Eq.\ \eqref{conmutadores}. We show that, as a consequence of this commutator algebra, the rotationally invariant Hermitian quadratic forms in $\hat{X}_i, \hat{P}_i, i=1,2$ generate a nonabelian three-dimensional Lie algebra corresponding to $sl(2,\mathbb{R})$ or $su(2)$ according to $\kappa \theta< \hbar^2$ or $\kappa \theta>\hbar^2$.

In so doing, we construct the rotation generator in both the coordinate and momenta planes, $\hat{L}$, which is singular at the critical value $\kappa \theta = \hbar^2$, where a dimensional reduction takes place as noticed in \cite{Poly}. We also construct operators which perform the discrete transformations of time-reversal and parity in each region.

The most general rotationally invariant nonrelativistic Hamiltonian can then be expressed as a function of the generators of these Lie algebras and $\hat{L}$ itself, and then its characteristic subspaces are contained in unitary irreducible representations of the groups $SL(2,\mathbb{R})\otimes SO(2)$ or $SU(2)\otimes SO(2)$, according to the region. Moreover, we show that the quadratic Casimir of the representations is related with the eigenvalue of $\hat{L}$, relation which selects the physically sensible representations of these direct products.

In this framework, we consider the simple examples of the extensions to these noncommutative phase space of the Hamiltonians of the isotropic oscillator and the Landau model, {reproducing from this perspective their well known spectra}. Finally, we consider the case of cylindrical well potentials on the plane. {We are able to reproduce the results in \cite{scholtz}, where only coordinate noncommutativity has been considered and the state space is realized in terms of Hilbert-Schmidt class operators}, but we also solve the general case for the different parameters regions. In particular, for the case of infinite wells, we get the eigenvalues corresponding to each irreducible representation as the zeros of given polynomials.

Finally, three appendices are dedicated to clarify some intermediary steps.



\section{The  non-commutative two-dimensional phase space}\label{NC-2dim-space}

The non-commutative two-dimensional space with nonvanishing commutators among both coordinates and momenta \cite{Zhang-PLB-2004,Zhang-PRL-2004}
is characterized by the following commutation relations for the Hermitian operators representing position and momenta, $\hat{X}_i,\hat{P}_i, i=1,2$:
\begin{equation}\label{1}
      \left[\hat{X}_i , \hat{X}_j\right] = \imath \theta\, \epsilon_{ij} \,, \quad
      \left[\hat{X}_i , \hat{P}_j\right] = \imath \hbar \,  \delta_{ij} \,, \quad
      \left[\hat{P}_i , \hat{P}_j\right] = \imath \kappa \, \epsilon_{ij} \,,
\end{equation}
where $\theta$ and $\kappa$ are the (real) noncommutativity parameters. With no loss of generality, we can take $\theta\geq0$. These relations reduce to the Heisenberg algebra for ordinary dynamical variables in the $\theta,\kappa \rightarrow 0$ limit,
\begin{equation}\label{1p}
   \left[x_i , x_j\right] = 0 \,, \quad
      \left[x_i , p_j\right] = \imath \hbar\, \delta_{ij} \,, \quad
      \left[p_i , p_j\right] = 0 \,.
\end{equation}

From Eq.\ \eqref{1}, it is straightforward to verify that the generator of rotations on the NC plane of coordinates or momenta is given by \cite{Poly}
\begin{equation}\label{gen-rot}
    \hat{L}:=\frac{1}{\left(1-\frac{\theta\kappa}{\hbar^2}\right)}\left\{\left( \hat{X}_1 \hat{P}_2 - \hat{X}_2 \hat{P}_1\right)+
   {{  \frac{\theta}{2 \hbar } \left({\hat{P}_1}^2+{\hat{P}_2}^2\right)}+{ \frac{\kappa}{2 \hbar }
   \left({\hat{X}_1}^2+{\hat{X}_2}^2\right)}}\right\}\,,
\end{equation}
for $\frac{\kappa \theta}{\hbar ^2}\neq 1$.
Indeed, Eqs.\ \eqref{1} imply that $\hat{L}$ transforms $\hat{\mathbf{X}}$ and $\hat{\mathbf{P}}$ as vectors,
\begin{equation}\label{PQvec}
     \left[\hat{L},\hat{X}_i\right] =\imath \hbar\, \epsilon_{i j} \hat{X}_j\,, \quad
     \left[\hat{L},\hat{P}_i\right] =\imath \hbar\, \epsilon_{i j} \hat{P}_j \,.
\end{equation}

Therefore,
\begin{equation}\label{PQ2}
     \left[\hat{L},\hat{\mathbf{X}}^2\right] =0\,, \quad
     \left[\hat{L},\hat{\mathbf{P}}^2\right] =0
\end{equation}
and, consequently, the Hamiltonian of a particle living on this NC plane and subject to a \emph{central potential} $V(\hat{\mathbf{X}}^2)$, given by an expression of the form
\begin{equation}\label{Ham-P2-Q2}
    \hat{H}=\frac{1}{2\mu}\, \hat{\mathbf{P}}^2 + V(\hat{\mathbf{X}}^2)
\end{equation}
(where $\mu$ is a mass parameter), commutes with $\hat{L}$ and, then, presents an $SO(2)$ symmetry.

\smallskip

For the critical value $\kappa_c=\hbar^2/\theta$ (where $\hat{L}$ in Eq.\ \eqref{gen-rot} has no sense), the commutator algebra in Eq.\ \eqref{1} reduces to
\begin{equation}\label{1-critico}
      \left[\hat{X}_i , \hat{X}_j\right] = \imath \theta\, \epsilon_{ij} \,, \quad
      \left[\hat{X}_i , \hat{P}_j\right] = \imath \hbar \,  \delta_{ij} \,, \quad
      \left[\hat{P}_i , \hat{P}_j\right] = \imath \frac{\hbar^2}{\theta} \, \epsilon_{ij} \,.
\end{equation}
These relations can be satisfied by a single pair of dynamical variables as
$\hat{X}_1=-\frac{\theta}{\hbar}\,\hat{P}_2\,,  \hat{P}_1=\frac{\hbar}{\theta}\, \hat{X}_2$.
Therefore, for this particular value of $\kappa$ there occurs a kind of \emph{dimensional reduction} \cite{Poly,Bellucci}.

\smallskip

Let us mention that translation generators on the noncommutative coordinate plane can be defined as
\begin{equation}\label{trnaslations1}
    \hat{K}_i:=\frac{1}{\left(1-\frac{\theta\kappa}{\hbar^2}\right)}\left(
    \hat{P}_i-\frac{\kappa}{\hbar}\,\epsilon_{i j} \hat{X}_j \right)\,,
\end{equation}
which play a similar role to that of the \emph{magnetic translation generators} in the presence of an external magnetic field perpendicular to the plane \cite{Fradkin,Wen}. Indeed, we have
\begin{equation}\label{translation2}
\begin{array}{c}\displaystyle
    \left[\hat{K}_i,\hat{X}_j\right]= -\imath \hbar \delta_{i j}\,, \quad
    \left[\hat{K}_i,\hat{P}_j\right]=0\,,
   \\ \\ \displaystyle
    \left[\hat{K}_i,\hat{K}_j\right]=\frac{- \imath\kappa\,\epsilon_{i j}}
    {\left(1-\frac{\theta\kappa}{\hbar^2}\right)}\,, \quad
    \left[\hat{L},\hat{K}_i\right]= \imath \hbar \,  \epsilon_{ij} \hat{K}_j \,.
\end{array}
\end{equation}

\medskip

It is worth noting that $\hat{L}$ can be considered as one of the generators of an $su(2)$ or an $sl(2,\mathbb{R})$ Lie algebras \cite{Bellucci,Mik-Hor-2}, generated also by bilinear expressions in $\hat{P}_i$ and $\hat{K}_i$, according to $\kappa$ is less or greater than the critical value $\kappa_c$. For more details, see Appendix \ref{L-Lie}.

In the next Section we discuss the discrete symmetries of this noncommutative phase space and in Section \ref{algebraic-structure} we will consider the general algebraic structure of Hamiltonians with central potentials.

\section{Discrete symmetries for $\kappa<\kappa_c$}\label{discrete-menor}

\subsection{Time reversal}

In the normal (commutative) plane, the time-reversal transformation is realized as an antilinear unitary operator that leaves invariant the coordinates and change the signs of momenta. As a consequence, the angular momentum $L_0=x_1 p_2-x_2 p_1$ changes its sign. This is, of course, consistent with the Heisenberg algebra in Eq.\ (\ref{1p}).

In the case of the modified algebra in Eqs.\ (\ref{1}), an antilinear unitary transformation $\hat{\mathcal{T}}$ will change the sign of the right hand sides of these three equations, but this can not be compensated by just the change of sign of momenta.

We can construct a set of consistent \emph{time-reversal} transformed dynamical variables satisfying Eqs.\ (\ref{1}) with a minus sign in the right hand side as follows. Let us define ${\hat{\mathcal{T}}}$ by
\begin{equation}\label{time-reversal}
    \begin{array}{c}\displaystyle
      {\hat{X}_i}^{\mathcal{T}}:={\hat{\mathcal{T}}}{\hat{X}_i}{\hat{\mathcal{T}}}^\dagger=
       \frac{1}{\sqrt{1-\frac{\theta  \kappa }{\hbar^2}}}
      \left(\hat{X}_i+\frac{\theta  }{\hbar}\epsilon_{i j}\hat{P}_j\right)\,,
   \\ \\ \displaystyle
   {\hat{P}_i}^{\mathcal{T}}:={\hat{\mathcal{T}}}{\hat{P}_i}{\hat{\mathcal{T}}}^\dagger=
    \frac{1}{\sqrt{1-\frac{\theta  \kappa }{\hbar^2}}}
    \left(-\hat{P}_i+\frac{\kappa}{\hbar}\epsilon_{i j}\hat{X}_j\right)\,,
    \end{array}
\end{equation}
{expressions which reduce to those quoted in \cite{Scholtz-formulation} in the $\kappa\rightarrow 0$ limit.}
It is a straightforward exercise to verify that
\begin{equation}\label{1-Theta}
      \left[{\hat{X}_i}^{\mathcal{T}} , {\hat{X}_j}^{\mathcal{T}}\right] = -\imath \theta\, \epsilon_{ij} \,, \quad
      \left[{\hat{X}_i}^{\mathcal{T}} , {\hat{P}_j}^{\mathcal{T}}\right] = -\imath \hbar \,  \delta_{ij} \,, \quad
      \left[{\hat{P}_i}^{\mathcal{T}} , {\hat{P}_j}^{\mathcal{T}}\right] = -\imath \kappa \, \epsilon_{ij} \,.
\end{equation}
Notice also that Eqs.\ (\ref{time-reversal}), for $\theta,\kappa \rightarrow 0$, smoothly reduce to the usual time-reversal transformation of the dynamical variables in Eq.\ (\ref{1p}).


By direct calculation, it can be easily verified that the generator of rotations is transformed by time reversal as
\begin{equation}\label{L-Theta}
    {\hat{\mathcal{T}}} \hat{L} {\hat{\mathcal{T}}}^\dagger=-\hat{L}\,.
\end{equation}

For the squared radial distance on the noncommutative plane, $\mathbf{\hat{X}}^2$, we have
\begin{equation}\label{Theta-7}
   \begin{array}{c} \displaystyle
      {\hat{\mathcal{T}}} {{\mathbf{\hat{X}}}}^2 {\hat{\mathcal{T}}}^\dagger=
    \frac{1}{1-\frac{\theta  \kappa }{\hbar ^2}}\left\{ {{\mathbf{\hat{X}}}}^2+
     \frac{\theta ^2 }{\hbar ^2} {{\mathbf{\hat{P}}}}^2+
     \frac{2 \theta  }{\hbar }\left(\hat{X}_1 \hat{P}_2-\hat{X}_2\hat{P}_1\right)
     \right\}=
      \\ \\ \displaystyle
     ={{\mathbf{\hat{X}}}}^2+ \frac{2 \theta  }{\hbar }\, \hat{L}\,,
   \end{array}
\end{equation}
where Eq.\ (\ref{gen-rot}) has been used.

For the squared momentum $\mathbf{\hat{P}}^2$  we get
\begin{equation}\label{Theta-8}
   \begin{array}{c} \displaystyle
      {\hat{\mathcal{T}}} \mathbf{\hat{P}}^2 {\hat{\mathcal{T}}}^\dagger=
      \frac{1}{\left(1-\frac{\theta  \kappa }{\hbar ^2}\right)}\left\{ {{\mathbf{\hat{P}}}}^2+
    \frac{\kappa ^2 }{\hbar ^2}{{\mathbf{\hat{X}}}}^2+
    \frac{2 \kappa  }{\hbar }\left(\hat{X}_1 \hat{P}_2-\hat{X}_2\hat{P}_1\right)
    \right\} =
      \\ \\ \displaystyle
     ={{\mathbf{\hat{P}}}}^2+ \frac{2 \kappa  }{\hbar }\, \hat{L}
   \end{array}
\end{equation}
and for the scalar product of ${\mathbf{\hat{X}}}$ and ${\mathbf{\hat{P}}}$
\begin{equation}\label{Theta-Prod}
     {\hat{\mathcal{T}}} \left({\mathbf{\hat{X}}}\cdot {\mathbf{\hat{P}}}+{\mathbf{\hat{P}}}\cdot {\mathbf{\hat{X}}} \right)  {\hat{\mathcal{T}}}^\dagger
     = - \left({\mathbf{\hat{X}}}\cdot {\mathbf{\hat{P}}}+{\mathbf{\hat{P}}}\cdot {\mathbf{\hat{X}}} \right)\,.
\end{equation}

Consequently, $\hat{L}$ and $({\mathbf{\hat{X}}}\cdot {\mathbf{\hat{P}}}+{\mathbf{\hat{P}}}\cdot {\mathbf{\hat{X}}} )$ have the usual transformation properties, but neither $\mathbf{\hat{X}}^2$ nor $\mathbf{\hat{P}}^2$ are left invariant by the time-reversal transformation, although in both cases this symmetry is recovered in the commutative limit.


\subsection{Parity}

In the normal two-dimensional space, the \emph{parity}  transformation is realized by a unitary linear operator which changes (for example) the signs of $x_2$ and $p_2$, leaving invariant\footnote{Notice that the simultaneous change of signs of both coordinates and both momenta is equivalent to a rotation in $\pi$ on the plane.} $x_1$ and $p_1$. But in the noncommutative phase space, such transformation of the dynamical variables does not leave invariant the first and third commutators in Eqs.\ (\ref{1}).

Rather, we define
\begin{equation}\label{par-3}
    \begin{array}{c}\displaystyle
      {\hat{X}_1}^{\mathcal{P}}:={\hat{\mathcal{P}}}{\hat{X}_1}{\hat{\mathcal{P}}}^\dagger= {\hat{X}_1}^{\mathcal{T}}\,,
      \quad {\hat{X}_2}^{\mathcal{P}}:={\hat{\mathcal{P}}}{\hat{X}_2}{\hat{\mathcal{P}}}^\dagger= -{\hat{X}_2}^{\mathcal{T}}       \,,
   \\ \\ \displaystyle
   {\hat{P}_1}^{\mathcal{P}}:={\hat{\mathcal{P}}}{\hat{P}_1}{\hat{\mathcal{P}}}^\dagger=-{\hat{P}_1}^{\mathcal{T}}\,,  \quad
    {\hat{P}_2}^{\mathcal{P}}:={\hat{\mathcal{P}}}{\hat{P}_2}{\hat{\mathcal{P}}}^\dagger= {\hat{P}_2}^{\mathcal{T}}   \,,
    \end{array}
\end{equation}
where ${\hat{X}_i}^{\mathcal{T}}$ and ${\hat{P}_i}^{\mathcal{T}}$ are given in Eqs.\ (\ref{time-reversal}).


We can also analyze the behavior of (Hermitian) quadratic expressions in the dynamical variables against this parity transformation. We straightforwardly get
\begin{equation}\label{transf-parity}
   \begin{array}{c}\displaystyle
     {\hat{\mathcal{P}}} \hat{L} {\hat{\mathcal{P}}}^{\dagger}=   {\hat{\mathcal{T}}} \hat{L} {\hat{\mathcal{T}}}^\dagger =-\hat{L} \,,
     \\ \\ \displaystyle
     {\hat{\mathcal{P}}} {{\mathbf{\hat{X}}}}^2 {\hat{\mathcal{P}}}^\dagger=
      {\hat{\mathcal{T}}} {{\mathbf{\hat{X}}}}^2 {\hat{\mathcal{T}}}^\dagger=
      {{\mathbf{\hat{X}}}}^2+ \frac{2 \theta  }{\hbar }\, \hat{L}\,,
       \\ \\ \displaystyle
      {\hat{\mathcal{P}}} \mathbf{\hat{P}}^2 {\hat{\mathcal{P}}}^\dagger=
       {\hat{\mathcal{T}}} \mathbf{\hat{P}}^2 {\hat{\mathcal{T}}}^\dagger=
      {{\mathbf{\hat{P}}}}^2+ \frac{2 \kappa  }{\hbar }\, \hat{L}\,,
     \\ \\ \displaystyle
      {\hat{\mathcal{P}}}\left\{ \left({\mathbf{\hat{X}}}\cdot {\mathbf{\hat{P}}}+{\mathbf{\hat{P}}}\cdot {\mathbf{\hat{X}}} \right)\right\} {\hat{\mathcal{P}}}^\dagger= -
        {\hat{\mathcal{T}}}\left\{  \left({\mathbf{\hat{X}}}\cdot {\mathbf{\hat{P}}}+{\mathbf{\hat{P}}}\cdot {\mathbf{\hat{X}}} \right)\right\} {\hat{\mathcal{T}}}^\dagger=
       \left({\mathbf{\hat{X}}}\cdot {\mathbf{\hat{P}}}+{\mathbf{\hat{P}}}\cdot {\mathbf{\hat{X}}} \right)\,.
   \end{array}
\end{equation}

Here again, $\hat{L}$ and $({\mathbf{\hat{X}}}\cdot {\mathbf{\hat{P}}}+{\mathbf{\hat{P}}}\cdot {\mathbf{\hat{X}}})$ have the usual transformation properties, but neither $\mathbf{\hat{X}}^2$ nor $\mathbf{\hat{P}}^2$ are left invariant by the parity transformation, although this symmetry is recovered in both cases in the commutative limit.

\medskip

Finally, let us remark that the  time-reversal and parity transformations so defined commute, as follows from Eqs.\ (\ref{time-reversal}) and (\ref{par-3}). Indeed, we have
\begin{equation}\label{TP-var}
\begin{array}{c}\displaystyle
\left({\hat{X}_1}^{\mathcal{P}}\right)^\mathcal{T}={\hat{X}_1}=\left({\hat{X}_1}^{\mathcal{T}}\right)^\mathcal{P}\,, \quad
    \left( {\hat{X}_2}^{\mathcal{P}}\right)^\mathcal{T} =-{\hat{X}_2}=  \left( {\hat{X}_2}^{\mathcal{T}}\right)^\mathcal{P} \,,
   \\ \\ \displaystyle
   \left({\hat{P}_1}^{\mathcal{P}}\right)^\mathcal{T}=-{\hat{P}_1}=\left({\hat{P}_1}^{\mathcal{T}}\right)^\mathcal{P}\,, \quad
    \left( {\hat{P}_2}^{\mathcal{P}} \right)^\mathcal{T}={\hat{P}_2} = \left( {\hat{P}_2}^{\mathcal{T}} \right)^\mathcal{P}\,.
\end{array}
\end{equation}

Moreover, the combination ${\hat{\mathcal{P}}}{\hat{\mathcal{T}}}$ leaves invariant the first three quadratic forms considered in Eq.\ \eqref{transf-parity} and changes the sign of the fourth one,
\begin{equation}\label{par-8}
    \begin{array}{c} \displaystyle
      {\hat{\mathcal{P}}}{\hat{\mathcal{T}}}\hat{L}  {\hat{\mathcal{T}}}^\dagger {\hat{\mathcal{P}}}^\dagger=- {\hat{\mathcal{P}}}\hat{L}   {\hat{\mathcal{P}}}^\dagger=\hat{L}\,,
      \\ \\ \displaystyle
      {\hat{\mathcal{P}}}{\hat{\mathcal{T}}}  {\mathbf{\hat{X}}}^2  {\hat{\mathcal{T}}}^\dagger {\hat{\mathcal{P}}}^\dagger=
      {\hat{\mathcal{P}}}  \left( {\mathbf{\hat{X}}}^2  + \frac{2\theta}{\hbar}\, \hat{L} \right)   {\hat{\mathcal{P}}}^\dagger=
       {\mathbf{\hat{X}}}^2  + \frac{2\theta}{\hbar}\, \hat{L} -  \frac{2\theta}{\hbar}\, \hat{L}= {\mathbf{\hat{X}}}^2\,,
     \\ \\ \displaystyle
     {\hat{\mathcal{P}}}{\hat{\mathcal{T}}}  {\mathbf{\hat{P}}}^2  {\hat{\mathcal{T}}}^\dagger {\hat{\mathcal{P}}}^\dagger=
      {\hat{\mathcal{P}}}  \left( {\mathbf{\hat{P}}}^2  + \frac{2\kappa}{\hbar}\, \hat{L} \right)   {\hat{\mathcal{P}}}^\dagger=
       {\mathbf{\hat{P}}}^2  + \frac{2\kappa}{\hbar}\, \hat{L} -  \frac{2\kappa}{\hbar}\, \hat{L}= {\mathbf{\hat{P}}}^2\,.
        \\ \\ \displaystyle
     {\hat{\mathcal{P}}}{\hat{\mathcal{T}}} \left({\mathbf{\hat{X}}}\cdot {\mathbf{\hat{P}}}+{\mathbf{\hat{P}}}\cdot {\mathbf{\hat{X}}} \right) {\hat{\mathcal{T}}}^\dagger {\hat{\mathcal{P}}}^\dagger=
     -  {\hat{\mathcal{P}}}  \left({\mathbf{\hat{X}}}\cdot {\mathbf{\hat{P}}}+{\mathbf{\hat{P}}}\cdot {\mathbf{\hat{X}}} \right)   {\hat{\mathcal{P}}}^\dagger=-
      \left({\mathbf{\hat{X}}}\cdot {\mathbf{\hat{P}}}+{\mathbf{\hat{P}}}\cdot {\mathbf{\hat{X}}} \right)\,.
    \end{array}
\end{equation}

In particular, any \emph{nonrelativistic} Hamiltonian on the noncommutative phase space with a kinetic term proportional to ${\mathbf{\hat{P}}}^2$ and a central potential  is $\mathcal{P}\mathcal{T}$-invariant, as well as $\hat{L}$ is.

\subsection{Alternative definitions of discrete transformations}

Let us remark that these discrete transformations of dynamical variables are defined up to a linear \emph{canonical} transformation in $Sp(4,\mathbb{R})$.

If we call $\xi:=(\hat{X}_1,\hat{X}_2,\hat{P}_1,\hat{P}_2)^t$, the commutation relations of the deformed Heisenberg algebra in Eq.\ \eqref{1} can be written in a matricial form \cite{Poly} as $\left(  [\xi_i,\xi_j] \right)=\imath \hbar \hat{G}$, where
\begin{equation}\label{GconNCparameters}
      \hat{G}=
\left(
  \begin{array}{cc}
     \frac{\theta}{\hbar} \, \imath \sigma_2 & \mathbf{1}_2 \\  \\
    - \mathbf{1}_2 & \frac{\kappa}{\hbar} \, \imath \sigma_2 \\
  \end{array}
\right)
\end{equation}
with $\sigma_2$ the second Pauli matrix. Notice that the determinant
\begin{equation}\label{determinant}
    \det \left([\xi_i,\xi_j]\right)=\left(1-\frac{\kappa \theta}{\hbar^2}\right)^2 \hbar^4
\end{equation}
vanishes for the \emph{critical value} $\kappa_c=\hbar^2/\theta$ \cite{Poly,Bellucci}, where the operators $\hat{X}_i,\hat{P}_i, i=1,2$, do not represent independent dynamical variables, as previously discussed.

Now, these commutation relations are left invariant by linear transformations which preserve the Hermiticity of the dynamical variables and belong to (an equivalent representation of) the group $Sp(4,\mathbb{R})$. Indeed,  for $\kappa< \kappa_c$ we can write $\hat{G}=A G A^t$ where
\begin{equation}\label{GsinNCparameters}
    G:= \left(
  \begin{array}{cc}
    \mathbf{0}_2 & \mathbf{1}_2 \\ \\
    - \mathbf{1}_2 &  \mathbf{0}_2 \\
  \end{array}
\right)=  \mathbf{1}_2  \otimes \imath \sigma_2
\end{equation}
and
\begin{equation}\label{AGA}
    A=
    \left(
               \begin{array}{cc}
                 \lambda\, \mathbf{1}_2 & -\frac{\theta}{2\hbar\lambda} \, \imath \sigma_2 \\ \\
                  \frac{\kappa}{2\hbar\lambda} \, \imath \sigma_2  & \lambda\, \mathbf{1}_2  \\
               \end{array}
             \right) \quad {\rm with}\quad
              {\lambda}^2=\frac{1}{2}\left(1+ \sqrt{1-\frac{\kappa \theta}{\hbar^2}} \right)\,,
\end{equation}
as can be easily verified. Then, for a linear transformation $\xi \rightarrow U \xi$ which preserves the commutation relations we have
\begin{equation}\label{UinSP4-1}
   U  \hat{G} U^t= \hat{G} \quad \Rightarrow \quad U A G A^t U^t  = A G A^t \,.
\end{equation}
Then
\begin{equation}\label{UinSP4-2}
    \left( A^{-1}U A\right) G \left( A^{-1}U A\right)^t =G  \quad
    \Rightarrow \quad
    \left( A^{-1}U A\right) \in Sp(4,\mathbb{R})\,.
\end{equation}

Therefore, as previously stated,  the discrete transformations $\mathcal{T}$ and $\mathcal{P}$ in Eqs.\ \eqref{time-reversal} and \eqref{par-3} are defined up to an  $Sp(4,\mathbb{R})$ linear transformation of the noncommutative dynamical variables.


\section{Discrete symmetries for $\kappa>\kappa_c$}\label{discrete-mayor}

Since the dynamical variables must be represented by Hermitian operators, we see that the definition of time-reversal and parity transformations in Eqs.\ \eqref{time-reversal} and \eqref{par-3} are valid only for $\kappa<\kappa_c$. For the region where $\kappa>\kappa_c$ we must adopt another definitions compatible with the transformation rules. Notice also that the commutative limit can not be attained from this region.

So, we \emph{define} for $\kappa>\kappa_c$
\begin{equation}\label{discrete-k-mayor-1}
    \begin{array}{c}\displaystyle
      {\hat{X}_1}^{\mathcal{T}}= \frac{1}{\sqrt{1-\frac{\hbar^2}{\theta \kappa}}} \left( \hat{X}_1+\frac{\hbar}{\kappa}\, \hat{P}_2 \right) = {\hat{X}_1}^{\mathcal{P}}\,,
      \\ \\ \displaystyle
      {\hat{X}_2}^{\mathcal{T}}= \frac{1}{\sqrt{1-\frac{\hbar^2}{\theta \kappa}}} \left( -\hat{X}_2+\frac{\hbar}{\kappa}\, \hat{P}_1 \right) = - {\hat{X}_2}^{\mathcal{P}}\,,
      \\ \\ \displaystyle
      {\hat{P}_1}^{\mathcal{T}}= \frac{1}{\sqrt{1-\frac{\hbar^2}{\theta \kappa}}} \left(  \hat{P}_1- \frac{\hbar}{\theta}\, \hat{X}_2 \right) = - {\hat{P}_1}^{\mathcal{P}}\,,
      \\ \\ \displaystyle
      {\hat{P}_2}^{\mathcal{T}}=  \frac{1}{\sqrt{1-\frac{\hbar^2}{\theta \kappa}}} \left( - \hat{P}_2- \frac{\hbar}{\theta}\, \hat{X}_1 \right) ={\hat{P}_2}^{\mathcal{P}}\,, \\
    \end{array}
\end{equation}
which satisfy the commutation relations in Eq.\ \eqref{1} (with a $(-1)$ additional factor on the right hand side of these equations in the case of the antilinear time-reversal transformation) as can be easily verified.

Under these discrete transformations, the previously considered quadratic expressions in the variables  transform as
\begin{equation}\label{dkmayor-2}
  \begin{array}{c} \displaystyle
   {\hat{\mathcal{T}}} \hat{L} {\hat{\mathcal{T}}}^\dagger  =  {\hat{\mathcal{P}}} \hat{L} {\hat{\mathcal{P}}}^\dagger =  \hat{L}\,,
    \\ \\ \displaystyle
    {\hat{\mathcal{T}}} {{\mathbf{\hat{X}}}}^2 {\hat{\mathcal{T}}}^\dagger=
  {\hat{\mathcal{P}}} {{\mathbf{\hat{X}}}}^2 {\hat{\mathcal{P}}}^\dagger =- \frac{\theta}{\kappa}\, {{\mathbf{\hat{P}}}}^2 - \frac{2 \theta  }{\hbar }\, \hat{L}\,,
  \\ \\ \displaystyle
    {\hat{\mathcal{T}}} {{\mathbf{\hat{P}}}}^2 {\hat{\mathcal{T}}}^\dagger=
  {\hat{\mathcal{P}}} {{\mathbf{\hat{P}}}}^2 {\hat{\mathcal{P}}}^\dagger=- \frac{\kappa}{\theta}\, {{\mathbf{\hat{X}}}}^2 - \frac{2 \kappa  }{\hbar }\, \hat{L}\,,
    \\ \\ \displaystyle
    {\hat{\mathcal{T}}} \left({\mathbf{\hat{X}}}\cdot {\mathbf{\hat{P}}}+{\mathbf{\hat{P}}}\cdot {\mathbf{\hat{X}}} \right) {\hat{\mathcal{T}}}^\dagger=-
  {\hat{\mathcal{P}}} \left({\mathbf{\hat{X}}}\cdot {\mathbf{\hat{P}}}+{\mathbf{\hat{P}}}\cdot {\mathbf{\hat{X}}} \right){\hat{\mathcal{P}}}^\dagger
  =   \left({\mathbf{\hat{X}}}\cdot {\mathbf{\hat{P}}}+{\mathbf{\hat{P}}}\cdot {\mathbf{\hat{X}}} \right)\,,
  \end{array}
\end{equation}
(compare with Eq.\ \eqref{transf-parity}). These four quadratic forms are also $\mathcal{P T}$-invariant.

\smallskip

Notice that these transformation rules are different  from those in Eqs.\ \eqref{L-Theta} - \eqref{Theta-Prod} and \eqref{transf-parity}, although the relation between $\xi^{\mathcal{T}}$ and $\xi^{\mathcal{P}}$ is the same as in Eq.\ \eqref{par-3} and their definitions are consistent with the deformed algebra in Eq.\ \eqref{1}.

In this region of the noncommutative parameters, the $\mathcal{T}$ and $\mathcal{P}$ transformed dynamical variables are also determined up to a linear canonical transformation in $Sp(4,\mathbb{R})$. Indeed, for $\kappa>\kappa_c$ we can write $\hat{G}=B G B^t$ where
\begin{equation}\label{Sp4Rkmayor-1}
    B= \left({\frac{\hbar ^2}{\theta  \kappa }}\right)^{\frac{1}{4}} C  D
\end{equation}
with
\begin{equation}\label{Sp4Rkmayor-2}
    C=\left(
\begin{array}{cccc}
 1 & 0 & 0 & 0 \\
 0 & 0 & \sqrt{\frac{\theta }{\kappa }} & 0 \\
 0 & \sqrt{\frac{\kappa }{\theta }} & 0 & 0 \\
 0 & 0 & 0 & 1 \\
\end{array}
\right)\,, \quad
  D=
    \left(
    \begin{array}{cc}
    \gamma\, \mathbf{1}_2 & -\frac{\theta}{2\hbar\gamma} \, \imath \sigma_2 \\ \\
    \frac{\kappa}{2\hbar \gamma} \, \imath \sigma_2  & \gamma\, \mathbf{1}_2  \\
    \end{array}
    \right)\,,
\end{equation}
and
\begin{equation}\label{Sp4Rkmayor-3}
  \gamma= \sqrt{\frac{\theta  \kappa }{2 \hbar ^2}}
   \left({{1+\sqrt{1-\frac{\hbar ^2}{\theta  \kappa }}}}\right)^{\frac{1}{2}}\,,
\end{equation}
as can be easily verified.

Then,  a linear transformation $\xi \rightarrow U \xi$ which preserves the commutation relations must satisfy
\begin{equation}\label{Sp4Rkmayor-4}
     U B G B^t U^t  = B G B^t \quad \Rightarrow \quad  \left( B^{-1}U B\right) \in Sp(4,\mathbb{R})\,.
\end{equation}


\section{Algebraic structure of Hamiltonians with central potentials}\label{algebraic-structure}

This Section is devoted to the study of the algebraic structure of nonrelativistic Hamiltonians with central potential in the noncommutative two-dimensional phase space.

It will be shown that the Hamiltonian eigenvalue problem can be referred to the representation spaces of irreducible unitary representations of the groups $SL(2,\mathbb{R})$ or $SU(2)$, according to $\kappa<\kappa_c$ or $\kappa>\kappa_c$ respectively. {The existence of two \emph{quantum-mechanical phases} for these systems has been found in \cite{Bellucci} employing particular linear realizations of the noncommutative dynamical variables in terms of ordinary dynamical variables.} In the following we will show that this algebraic structure is independent of these realizations and are determined by just the commutation relations in Eq.\ (\ref{1}).


\subsection{The $0<\kappa<\kappa_c$ case}\label{discrete-kappa-menor}

Let us first consider the region with $0<\kappa<\kappa_c$ where we define
\begin{equation}\label{algeb1}
    \begin{array}{c} \displaystyle
      \mathcal{J}_0= \frac{1}{4} \sqrt{\frac{\theta  \kappa }{\hbar ^2}}
   \left(\frac{\mathbf{{\hat{{P}}}}^2}{\kappa
   }+\frac{\mathbf{{\hat{{X}}}}^2}{\theta }+
   \frac{2 \hat{L}}{\hbar }\right) = {\mathcal{J}_0}^\dagger\,,
    \\ \\ \displaystyle
      \mathcal{J}_\pm=\frac{1}{4 \sqrt{\frac{\hbar
   ^2}{\theta  \kappa }-1}} \left\{-\frac{\mathbf{{\hat{{P}}}}^2}{\kappa
   }+\frac{\mathbf{{\hat{{X}}}}^2}{\theta }  \mp
   \frac{i }{\sqrt{\theta  \kappa
   }}\left(\mathbf{\hat{P}}\cdot \mathbf{\hat{X}}+\mathbf{\hat{X}}\cdot \mathbf{\hat{P}}\right) \right\}=
   {\mathcal{J}_\mp}^\dagger \,.
    \end{array}
\end{equation}
It is a straightforward exercise to verify that these operators satisfy the commutation relations characteristics of an $sl(2,\mathbb{R})$ Lie algebra,
\begin{equation}\label{algeb2}
    \left[\mathcal{J}_0 , \mathcal{J}_\pm\right] = \pm \mathcal{J}_\pm\,, \quad
    \left[\mathcal{J}_+ , \mathcal{J}_-\right] = -2\mathcal{J}_0 \,,
\end{equation}
with  $\mathcal{J}_\pm=\mathcal{J}_1\pm \imath \mathcal{J}_2$ ($\mathcal{J}_i,i=0,1,2,$ Hermitian operators), and the quadratic Casimir invariant ${\mathcal{J}}^2=\mathcal{J}_0(\mathcal{J}_0\mp 1)-\mathcal{J}_\pm \mathcal{J}_\mp=\mathcal{J}_0^2-\mathcal{J}_1^2-\mathcal{J}_2^2$ is given by
\begin{equation}\label{algeb3}
   \begin{array}{c} \displaystyle
      {\mathcal{J}}^2=\frac{1}{16\left( \frac{\hbar^2}{\theta\kappa}-1 \right)}
    \left\{\left( 1-\frac{\theta\kappa}{\hbar^2}\right)
    \left( \frac{\mathbf{\hat{X}}^2}{\theta} +  \frac{\mathbf{\hat{P}}^2}{\kappa} + 2 \frac{\hat{L}}{\hbar}\right)^2  - \right.
    \\ \\ \displaystyle
     \left. -  \left( \frac{\mathbf{\hat{X}}^2}{\theta} -  \frac{\mathbf{\hat{P}}^2}{\kappa} \right)^2 -
     \frac{\left( \mathbf{\hat{P}}\cdot \mathbf{\hat{X}}+\mathbf{\hat{X}}\cdot \mathbf{\hat{P}} \right)^2}{\theta \kappa} \right\} \,,
   \end{array}
\end{equation}
which is not a positive definite operator \cite{Bargmann}.

Notice that Eqs.\ (\ref{PQvec}-\ref{PQ2})  imply that
\begin{equation}\label{algeb5}
    \left[ \hat{L} , \mathcal{J}_0 \right]=0\,, \quad \left[ \hat{L} , \mathcal{J}_\pm \right]=0
    \quad \Rightarrow \quad \left[ \hat{L} , {\mathcal{J}}^2 \right]=0 \,.
\end{equation}
In fact, {one can also verify that (See Appendix \ref{Jcuad-Lcuad})}
\begin{equation}\label{JcuadLcuad}
    \mathcal{J}^2=\frac{1}{4}\left\{\left( \frac{\hat{L}}{\hbar}\right)^2-1 \right\} \geq - \frac{1}{4}\,,
\end{equation}
which imposes a \emph{constraint} on the acceptable irreducible representations of the direct product of groups $SL(2,\mathbb{R}) \otimes SO(2)$ (where the second (Abelian) factor is generated by $\hat{L}$).
In particular, this relation implies that only the \emph{discrete classes} of irreducible unitary representations of $SL(2,\mathbb{R})$ \cite{Bargmann,Vega} will be of interest for our purposes.

\smallskip

Since a general Hamiltonian $\hat{H}( \mathbf{\hat{P}},\mathbf{\hat{X}} )$ which commutes with $\hat{L}$ can only be a function of ${\mathbf{\hat{P}}^2}$, ${\mathbf{\hat{X}}^2}$, $( \mathbf{\hat{P}}\cdot \mathbf{\hat{X}}+\mathbf{\hat{X}}\cdot \mathbf{\hat{P}})$ and $\hat{L}$ itself, and Eqs.\ (\ref{algeb1}) allow to write
\begin{equation}\label{algeb6}
    \begin{array}{c} \displaystyle
    \frac{\mathbf{\hat{X}}^2}{\theta} = 2 \sqrt{\frac{\hbar^2}{\theta\kappa}} \ \mathcal{J}_0 +
      \sqrt{\frac{\hbar^2}{\theta\kappa}-1} \left(\mathcal{J}_++\mathcal{J}_-\right) - \frac{\hat{L}}{\hbar} \,,
     \\ \\ \displaystyle
     \frac{\mathbf{\hat{P}}^2}{\kappa} = 2 \sqrt{\frac{\hbar^2}{\theta\kappa}} \ \mathcal{J}_0 -
     \sqrt{\frac{\hbar^2}{\theta\kappa}-1} \left(\mathcal{J}_++\mathcal{J}_-\right) - \frac{\hat{L}}{\hbar}\,,
      \\ \\ \displaystyle
      \frac{\left(\mathbf{\hat{P}}\cdot \mathbf{\hat{X}}+\mathbf{\hat{X}}\cdot \mathbf{\hat{P}}\right)}{\hbar}
      =2\imath \sqrt{1-\frac{\theta\kappa}{\hbar^2}}\left(\mathcal{J}_+-\mathcal{J}_-\right)\,,
    \end{array}
\end{equation}
it is clear that such $\hat{H}$ can be expressed in terms of $\mathcal{J}_0$, $\mathcal{J}_\pm$ and $\hat{L}$ only.  Therefore, the simultaneous characteristic subspaces of $\hat{H}$ and $\hat{L}$ are contained in representation spaces of irreducible unitary representations of the noncompact group $SL(2,{\mathbb{R}})$ (or, equivalently, $SU(1,1)$) which satisfy the constraint in Eq.\ \eqref{JcuadLcuad}.

These irreducible unitary representations in the discrete classes are characterized by the eigenvalue of the Casimir $\mathcal{J}^2$, which can take the values $\lambda=k(k-1)$ with $k=\frac{1}{2},1,\frac{3}{2},2,\cdots$. For a given $k$, the representation space is generated by the simultaneous eigenvectors of $\mathcal{J}^2$, $\mathcal{J}_0$ and $\hat{L}$,
\begin{equation}\label{eigen-sim}
    \begin{array}{c}\displaystyle
      \mathcal{J}^2  \left|k,m,l \right\rangle=k(k-1) \left|k,m,l \right\rangle\,,
      \\ \\ \displaystyle
      \mathcal{J}_0  \left|k,m,l \right\rangle = m \left|k,m,l \right\rangle\,, \  {\rm with} \  m= k,k+1, \cdots\ {\rm or}\ m= -k,-(k+1), \cdots\,,
      \\ \\ \displaystyle
      \hat{L}  \left|k,m,l \right\rangle= \hbar l \left|k,m,l \right\rangle\,, \quad {\rm with}\quad l^2= (2k-1)^2\,,
    \end{array}
\end{equation}
as follows from Eq.\ \eqref{JcuadLcuad}. Notice that these representations, which can be denoted by $\langle k,l \rangle$,  are not of finite dimension (See, for example, \cite{Bargmann,Vega}).

\smallskip

It is also worth mentioning that the time reversal and parity discrete transformations in Eqs.\ \eqref{transf-parity} leave invariant the extended $sl(2,\mathbb{R})$ algebra in Eqs.\ \eqref{algeb2} and \eqref{algeb5}. Indeed, while these transformations change the sign of the central element $\hat{L}$, one straightforwardly gets
\begin{equation}\label{discrete-J}
   \begin{array}{c}\displaystyle
      \mathcal{T} \mathcal{J}_0 \mathcal{T}^\dagger = \mathcal{P} \mathcal{J}_0 \mathcal{P}^\dagger = \mathcal{J}_0\,, \quad
    \mathcal{T} \mathcal{J}_\pm \mathcal{T}^\dagger =  \mathcal{P} \mathcal{J}_\pm \mathcal{P}^\dagger =  \mathcal{J}_\pm\,,
    \\ \\  \displaystyle
     \Rightarrow \quad   \mathcal{T} \mathcal{J}^2 \mathcal{T}^\dagger =  \mathcal{P} \mathcal{J}^2 \mathcal{P}^\dagger = \mathcal{J}^2\,.
   \end{array}
\end{equation}
We have, for example,
\begin{equation}\label{TP-autovector}
   \begin{array}{c} \displaystyle
      \hat{L} \mathcal{T} \left|k,m,l \right\rangle= - \left( \mathcal{T} \hat{L} \mathcal{T}^\dagger  \right) \mathcal{T} \left|k,m,l \right\rangle
    =- \hbar l \, \mathcal{T} \left|k,m,l \right\rangle
    \\ \\ \displaystyle
     \Rightarrow \quad  \mathcal{T} \left|k,m,l \right\rangle \sim  \left|k,m,-l \right\rangle\,.
   \end{array}
\end{equation}
Then, the discrete transformations map one irreducible representation $\langle k,l \rangle$ of $SL(2,\mathbb{R}) \otimes SO(2)$ satisfying Eq.\ \eqref{JcuadLcuad} onto the other one with the same $|l|$,
$\langle k,-l \rangle$, leaving invariant the direct sum $\langle k,l \rangle \oplus \langle k,-l \rangle$.




\smallskip

On the other hand, from Eq.\ (\ref{algeb2}) one gets
\begin{equation}\label{algeb66}
    \left\langle m,\lambda,l \right|
    \mathcal{J}_\pm \left|m,\lambda,l \right\rangle =0\,.
\end{equation}
Therefore, the positivity of $\mathbf{\hat{X}^2}$ (or $\mathbf{\hat{P}^2}$) implies that, for each irreducible representation,
\begin{equation}\label{algeb67}
    l\leq 2 m \sqrt{\frac{\hbar^2}{\theta\kappa}}
\end{equation}
for all values of $m$. In particular, this is possible only for the \emph{discrete class} of irreducible representations of $sl(2,\mathbb{R})$ with
integer or half-integer $m\geq k$ \cite{Bargmann,Vega}. Therefore, $ l\leq 2 k \sqrt{{\hbar^2}/{\theta\kappa}}$ (Notice that $\sqrt{{\hbar^2}/{\theta\kappa}}>1$).

\smallskip

We can establish a more stringent bound if we take into account Eq.\ \eqref{transf-4-ap} in Appendix \ref{rot-sl2r}, which establishes that, in any unitary representation of $sl(2,\mathbb{R})$, we have
\begin{equation}\label{transf-4}
e^{\displaystyle {\imath} \alpha \mathcal{J}_2}
    \left( A \mathcal{J}_0 + B \mathcal{J}_1 \right)
    e^{\displaystyle-{\imath} \alpha  \mathcal{J}_2}
    =A \sqrt{1-\frac{B^2}{A^2}} \, \mathcal{J}_0
\end{equation}
for real $A,B$ such that $|A|>|B|$, with $\tanh\alpha = B/A$. Then, $\mathbf{\hat{X}}^2$ in Eq.\ \eqref{algeb6} (or, equivalently, $\mathbf{\hat{P}}^2$) can be diagonalized by a unitary transformation as
\begin{equation}\label{Pcuad-diag}
     \frac{\mathbf{\hat{X}}^2}{\theta} \quad  \rightarrow \quad
      2 \sqrt{\frac{\hbar^2}{\theta\kappa}-\left( \frac{\hbar^2}{\theta\kappa}-1 \right)} \ \mathcal{J}_0
      - \frac{\hat{L}}{\hbar} =2 \mathcal{J}_0 - \frac{\hat{L}}{\hbar}\,.
\end{equation}

Therefore, the positivity of ${\mathbf{\hat{X}}^2}$ (or, similarly, $\mathbf{\hat{P}^2}$) implies that
\begin{equation}\label{Pcuad-diag-2}
    2m \geq l \,, \forall m \quad \Rightarrow \quad  2k \geq l\,.
\end{equation}
Since $ l^2= (2k-1)^2$, one concludes that
\begin{equation}\label{l-intermedio}
    l=\pm(2k-1)\,,
\end{equation}
being acceptable both signs.

\smallskip

Finally, since a general Hamiltonian with central potential $H( \hat{L}, \mathcal{J}_0 , \mathcal{J}_\pm)$ is transformed by $\mathcal{T}$ (or $\mathcal{P}$) into $\mathcal{T} H( \hat{L}, \mathcal{J}_0 , \mathcal{J}_\pm ) \mathcal{T}^\dagger= H(- \hat{L}, \mathcal{J}_0 , \mathcal{J}_\pm)$,  these discrete transformations are symmetries of the Hamiltonian (with the corresponding degeneracy of eigenvalues) if $H$ depends on $| \hat{L} |$ only.


\subsection{The $\kappa<0$ case}\label{discrete-kappa-negativo}

Let us now consider the region with $\kappa<0$. Here we can define the operators
\begin{equation}\label{algeb1-neg}
    \begin{array}{c} \displaystyle
      \mathcal{J}_0= \frac{1}{4\sqrt{1+\frac{\hbar ^2}{\theta  |\kappa| }}}
   \left(\frac{\mathbf{{\hat{{P}}}}^2}{|\kappa|
   }+\frac{\mathbf{{\hat{{X}}}}^2}{\theta }\right)\,,
    \\ \\ \displaystyle
      \mathcal{J}_\pm=\frac{1}{4} \left\{
      \frac{\left(\mathbf{\hat{P}}\cdot \mathbf{\hat{X}}+\mathbf{\hat{X}}\cdot \mathbf{\hat{P}}\right)}{\hbar{\sqrt{1+\frac{\theta |\kappa|}{\hbar^2}}}}
      \pm i {\sqrt{\frac{\theta |\kappa|}{\hbar^2}}}
      \left(-\frac{\mathbf{{\hat{{P}}}}^2}{|\kappa|}+\frac{\mathbf{{\hat{{X}}}}^2}{\theta } +\frac{2 \hat{L}}{\hbar} \right)
    \right\}=\mathcal{J}_1 \pm i \mathcal{J}_2 \,,
    \end{array}
\end{equation}
(with Hermitian $\mathcal{J}_i\,, i=0,1,2$) which satisfy the commutation relations characteristics of an $sl(2,\mathbb{R})$ Lie algebra (Eq.\ \eqref{algeb2}) and commute with $\hat{L}$. The quadratic Casimir invariant reduces also in this case to (See Appendix \ref{Jcuad-Lcuad})
\begin{equation}\label{algeb3-neg}
   \begin{array}{c} \displaystyle
      {\mathcal{J}}^2=\mathcal{J}_0^2-\mathcal{J}_1^2-\mathcal{J}_2^2=
     \frac{1}{4}\left\{\left( \frac{\hat{L}}{\hbar}\right)^2-1 \right\}  \,,
   \end{array}
\end{equation}
thus satisfying the same constraint with $\hat{L}$ as in Eq.\ \eqref{JcuadLcuad}.

Then, only the direct product of unitary irreducible representations of $SL(2,\mathbb{R})$ in the discrete classes \cite{Bargmann,Vega} with those of $SO(2)$ satisfying the constraint 
(subspaces of the Hilbert space generated by the simultaneous eigenvectors of $\mathcal{J}^2$, $\mathcal{J}_0$ and $\hat{L}$ as in Eq.\ \eqref{eigen-sim}) are relevant to describe the physical states of a system with a central potential.

Indeed, a general non-relativistic Hamiltonian $\hat{H}( \mathbf{\hat{P}},\mathbf{\hat{X}} )$ which commutes with $\hat{L}$ can be expressed in terms of the rotation invariants (compare with Eq.\ \eqref{algeb6})
\begin{equation}\label{algeb6-neg}
    \begin{array}{c} \displaystyle
    \frac{\mathbf{\hat{X}}^2}{\theta} = 2 {\sqrt{1+\frac{\hbar ^2}{\theta  |\kappa| }}} \ \mathcal{J}_0 - i
      \sqrt{\frac{\hbar^2}{\theta|\kappa|}} \left(\mathcal{J}_+-\mathcal{J}_-\right) - \frac{\hat{L}}{\hbar} \,,
     \\ \\ \displaystyle
     \frac{\mathbf{\hat{P}}^2}{|\kappa|} = 2 {\sqrt{1+\frac{\hbar ^2}{\theta  |\kappa| }}} \ \mathcal{J}_0 + i
      \sqrt{\frac{\hbar^2}{\theta|\kappa|}} \left(\mathcal{J}_+-\mathcal{J}_-\right) + \frac{\hat{L}}{\hbar} \,,
      \\ \\ \displaystyle
      \frac{\left(\mathbf{\hat{P}}\cdot \mathbf{\hat{X}}+\mathbf{\hat{X}}\cdot \mathbf{\hat{P}}\right)}{\hbar}
      =2 \sqrt{1+\frac{\theta|\kappa|}{\hbar^2}}\left(\mathcal{J}_++\mathcal{J}_-\right)\,,
    \end{array}
\end{equation}
and $\hat{L}$ itself, and then $\hat{H}$ is a function of $\mathcal{J}_0$, $\mathcal{J}_\pm$ and $\hat{L}$ only.

\smallskip

On the other hand, taking into account Eq.\ \eqref{transf-44-ap}, one can see that the positivity of $\mathbf{\hat{X}}^2$ (or $\mathbf{\hat{P}}^2$) requires that $2m-l\geq 0, \forall m$, which implies that $m=k,k+1,\cdots$ with $k$ a positive integer or half-integer, and $l\leq 2k$. Therefore, as in the previous Section,
\begin{equation}\label{l-kappa-negativo}
    l=\pm(2k-1)\,.
\end{equation}

Moreover, also in this region of parameters the time reversal and parity discrete transformations in Eqs.\ \eqref{transf-parity} leave invariant the extended $sl(2,\mathbb{R})$ algebra.
Indeed, these transformations change the sign of the central element $\hat{L}$ and for the $sl(2,\mathbb{R})$ generators we straightforwardly get
\begin{equation}\label{discrete-J-neg}
   \begin{array}{c}\displaystyle
      \mathcal{T} \mathcal{J}_0 \mathcal{T}^\dagger = \mathcal{P} \mathcal{J}_0 \mathcal{P}^\dagger = \mathcal{J}_0\,, \quad
    \mathcal{T} \mathcal{J}_\pm \mathcal{T}^\dagger = - \mathcal{J}_\pm = - \mathcal{P} \mathcal{J}_\pm \mathcal{P}^\dagger \,,
    \\ \\  \displaystyle
     \Rightarrow \quad   \mathcal{T} \mathcal{J}^2 \mathcal{T}^\dagger =  \mathcal{P} \mathcal{J}^2 \mathcal{P}^\dagger = \mathcal{J}^2\,,
   \end{array}
\end{equation}
and they leave invariant the direct sum of irreducible representations $\langle k,l \rangle \oplus \langle k,-l \rangle$ with $l=\pm(2k-1)$, as discussed in the previous Section (See Eq.\ \eqref{TP-autovector}).

\smallskip

Therefore, the region with $\kappa<0$ is completely similar to the one with $0<\kappa<\kappa_c$.


\subsection{The $\kappa>\kappa_c$ case}\label{algebraic-structure-kmayor}

Let us now consider the region with $\kappa>\kappa_c$. Here we define
\begin{equation}\label{algeb7}
    \begin{array}{c} \displaystyle
      J_3=\frac{1}{4} \sqrt{\frac{\theta  \kappa }{\hbar ^2}}
   \left(\frac{\mathbf{{\hat{{P}}}}^2}{\kappa
   }+\frac{\mathbf{{\hat{{X}}}}^2}{\theta }+
   \frac{2 \hat{L}}{\hbar }\right)\,,
    \\ \\ \displaystyle
      J_\pm=\frac{1}{4 \sqrt{1-\frac{\hbar
   ^2}{\theta  \kappa }}} \left\{-\frac{\mathbf{{\hat{{P}}}}^2}{\kappa
   }+\frac{\mathbf{{\hat{{X}}}}^2}{\theta }  \mp
   \frac{i }{\sqrt{\theta  \kappa
   }}\left(\mathbf{\hat{P}}\cdot \mathbf{\hat{X}}+\mathbf{\hat{X}}\cdot \mathbf{\hat{P}}\right) \right\} \,.
    \end{array}
\end{equation}
It can be straightforwardly verified that, as a consequence of Eq.\ \eqref{1}, these operators satisfy the commutation relations
\begin{equation}\label{algeb8}
    \left[{J}_3 , {J}_\pm\right] = \pm {J}_\pm\,, \quad
    \left[{J}_+ , {J}_-\right] = 2{J}_3 \,,
\end{equation}
which correspond to an $su(2)$ Lie algebra, with $J_\pm=J_1\pm\imath J_2$ ($J_i, i=1,2,3,$ Hermitian operators). In this case the quadratic Casimir operator, $\mathbf{J}^2=J_\pm J_\mp+J_3(J_3\mp1)$,  is given by  the same expression as in Eq.\ (\ref{algeb3}) which, in this region, can be written as the manifestly nonnegative operator
\begin{equation}\label{algeb9}
   \begin{array}{c} \displaystyle
      {{J}}^2=\frac{1}{16\left(1- \frac{\hbar^2}{\theta\kappa}\right)}
    \left\{\left(\frac{\theta\kappa}{\hbar^2}-1\right)
    \left( \frac{\mathbf{\hat{X}}^2}{\theta} +  \frac{\mathbf{\hat{P}}^2}{\kappa} + 2 \frac{\hat{L}}{\hbar}\right)^2  + \right.
    \\ \\ \displaystyle
     \left. +  \left( \frac{\mathbf{\hat{X}}^2}{\theta} -  \frac{\mathbf{\hat{P}}^2}{\kappa} \right)^2 +
     \frac{\left( \mathbf{\hat{P}}\cdot \mathbf{\hat{X}}+\mathbf{\hat{X}}\cdot \mathbf{\hat{P}} \right)^2}{\theta \kappa} \right\}
     \,.
   \end{array}
\end{equation}

Here again, the generator of rotations $\hat{L}$ can be incorporated as a central element of the extended algebra,
\begin{equation}\label{algeb10}
    \left[ \hat{L} , {J}_3 \right]=0\,, \quad \left[ \hat{L} , {J}_\pm \right]=0 \quad \Rightarrow
    \quad  \left[ \hat{L} , {J}^2 \right]=0 \,,
\end{equation}
which is related to the $su(2)$ Casimir operator by the constraint
\begin{equation}\label{JcuadLcuadSU2}
    {J}^2=\frac{1}{4}\left\{\left( \frac{\hat{L}}{\hbar}\right)^2-1 \right\}\,,
\end{equation}
which selects the admissible unitary irreducible representations of the direct product $SU(2) \otimes SO(2)$. In particular, notice that $l=0$ is excluded.

The Hermitian expressions quadratic in the dynamical variables can be written as
\begin{equation}\label{algeb11}
    \begin{array}{c} \displaystyle
    \frac{\mathbf{\hat{X}}^2}{\theta} = 2 \sqrt{\frac{\hbar^2}{\theta\kappa}} \ {J}_3 +
     \sqrt{1-\frac{\hbar^2}{\theta\kappa}} \left({J}_++{J}_-\right) - \frac{\hat{L}}{\hbar} \,,
     \\ \\ \displaystyle
     \frac{\mathbf{\hat{P}}^2}{\kappa} = 2 \sqrt{\frac{\hbar^2}{\theta\kappa}} \ {J}_3 -
     \sqrt{1-\frac{\hbar^2}{\theta\kappa}} \left({J}_++{J}_-\right) - \frac{\hat{L}}{\hbar}\,,
      \\ \\ \displaystyle
      \frac{\left(\mathbf{\hat{P}}\cdot \mathbf{\hat{X}}+\mathbf{\hat{X}}\cdot \mathbf{\hat{P}}\right)}{\hbar}
      =2 \imath \sqrt{\frac{\theta\kappa}{\hbar^2}-1}\left({J}_+-{J}_-\right)\,.
    \end{array}
\end{equation}
Consequently, any Hamiltonian $\hat{H}$ which commutes with $\hat{L}$ can be expressed as a function of $J_3$, $J_\pm$ and $\hat{L}$ itself. In this case, the simultaneous eigenvectors of $\hat{L}$ and $\hat{H}$  are contained in representation spaces of unitary irreducible representations of the \emph{compact} group $SU(2)$, which are of finite dimension.

The unitary irreducible representations of $SU(2) \otimes SO(2)$ can be characterized by the indices $\langle j , l \rangle$ where, as well known, j can take nonnegative integer or half-integer values, $j=0,\frac{1}{2},1,\frac{3}{2},\cdots$, and $l \in \mathbb{Z}$. Eq.\ \eqref{JcuadLcuadSU2} imposes the restriction $l^2 =(2j+1)^2$.   For a given $j$, the representation space is generated by the simultaneous eigenvectors of ${J}^2$, ${J}_3$ and $\hat{L}$,
\begin{equation}\label{eigen-SU2SO2}
    \begin{array}{c}\displaystyle
      {J}^2  \left|j,m,l \right\rangle=j(j+1) \left|j,m,l \right\rangle\,,
      \\ \\ \displaystyle
      {J}_3  \left|j,m,l \right\rangle = m \left|j,m,l \right\rangle\,, \  {\rm with} \  m= -j,-j+1, \cdots, j-1,j\,,
      \\ \\ \displaystyle
      \hat{L}  \left|j,m,l \right\rangle= \hbar l \left|j,m,l \right\rangle\,, \quad {\rm with}\quad l^2= (2j+1)^2\,.
    \end{array}
\end{equation}

\smallskip

According to Eqs.\ \eqref{dkmayor-2}, this $su(2)$ algebra is left invariant by the time-reversal and parity discrete transformations defined in Eq.\ \eqref{discrete-k-mayor-1}. Indeed, we straightforwardly get
\begin{equation}\label{tptransform-su2-1}
      \begin{array}{c}\displaystyle
        \mathcal{T} J_3 \mathcal{T}^\dagger =  \mathcal{P} J_3 \mathcal{P}^\dagger = -J_3\,,
    \quad   \mathcal{T} J_\pm \mathcal{T}^\dagger =  \mathcal{P} J_\pm \mathcal{P}^\dagger = J_\mp
    \\ \\ \displaystyle
       \left( \quad \Rightarrow \    \mathcal{T} J_{1,2} \mathcal{T}^\dagger=  J_{1,2} \quad \right) \,,
      \end{array}
\end{equation}
which preserve Eqs.\ \eqref{algeb8} and $\mathbf{J}^2$,
\begin{equation}\label{tptransform-su2-2}
    \mathcal{T} \mathbf{J}^2 \mathcal{T}^\dagger =  \mathcal{T} \left( J_\pm J_\mp+J_3(J_3\mp1) \right) \mathcal{T}^\dagger
     =J_\mp J_\pm+J_3(J_3\pm 1) =  \mathbf{J}^2 \,.
\end{equation}
Moreover, since in this region the discrete transformations do not change $\hat{L}$, we conclude that the unitary irreducible representations $\langle j , l \rangle$ of the direct product $SU(2) \otimes SO(2)$, of dimension $(2j+1)$, are also $\mathcal{T}$ and $\mathcal{P}$-invariant.

For example,
\begin{equation}\label{TeigenJ3}
   \begin{array}{c} \displaystyle
      J_3 \mathcal{T} \left|j,m,l \right\rangle = - \left( \mathcal{T }J_3 \mathcal{T}^\dagger\right)\mathcal{T} \left|j,m,l \right\rangle
    = - m \mathcal{T} \left|j,m,l \right\rangle
    \\ \\ \displaystyle
     \Rightarrow \quad \mathcal{T} \left|j,m,l \right\rangle \sim \left|j,-m,l \right\rangle\,, \quad {\rm for} \quad m=-j, -j+1,\cdots,j-1,j \,,
   \end{array}
\end{equation}
and the transformed vector belongs to the same irreducible representation.

Even though the Hamiltonians in the class we are considering are not, in general, $\mathcal{T}$ or $\mathcal{P}$-invariant, they have its characteristic subspaces contained in such irreducible representations of $SU(2) \otimes SO(2)$.

\medskip

On the other hand, from Eqs.\ \eqref{algeb11} one can see that the positivity of $\mathbf{\hat{X}^2}$ (or $\mathbf{\hat{P}^2}$) implies, for the unitary irreducible representation  $\langle j , l \rangle$, that
\begin{equation}\label{algeb67-kappa-mayor}
    l\leq - 2 j \sqrt{\frac{\hbar^2}{\theta\kappa}} \leq 0\,.
\end{equation}

\smallskip

But, $\mathbf{\hat{X}}^2$ (or $\mathbf{\hat{P}}^2$) can also be diagonalized by a unitary transformation in $SU(2)$ as in Eq.\ \eqref{transf-6-ap}. Indeed, in any unitary representations of $SU(2)$, a linear combination of generators of the form $A J_3 + B J_1$, with $A,B\in \mathbb{R}$, can be transformed into
\begin{equation}\label{diag-SU2}
    e^{\imath \varphi J_2} \left( A J_3 + B J_1 \right) e^{- \imath \varphi J_2} = A \sqrt{1+\frac{B^2}{A^2}} J_3\,,
\end{equation}
where $\varphi=\arctan \left( \frac{B}{A} \right)$ (See Appendix \ref{rot-sl2r}). Then,
\begin{equation}\label{diag-x2-su2}
    \mathbf{\hat{X}}^2 \quad \rightarrow \quad 2 \sqrt{\frac{\hbar^2}{\kappa\theta}+\left(1-\frac{\hbar^2}{\kappa\theta}\right)} J_3
    -\frac{\hat{L}}{\hbar} = 2 J_3 - \frac{\hat{L}}{\hbar}\,.
\end{equation}
Therefore, the positivity of $\mathbf{\hat{X}}^2$ (or, equivalently, $\mathbf{\hat{P}}^2$) requires that
\begin{equation}\label{positivity-x2-su2}
    2m\geq l\,, \  \forall m=-j,-j+1,\cdots,j-1,j\,, \quad \Rightarrow \quad l\leq -2j \leq 0\,.
\end{equation}
And since $l^2= (2j+1)^2$, one concludes that
\begin{equation}\label{l-kappa-mayor}
    l= -(2j+1)\leq -1
\end{equation}
for the unitary irreducible representation $\langle j,l \rangle$ which satisfy the constraint in Eq.\ \eqref{JcuadLcuadSU2}.

\smallskip

\medskip

In conclusion, for any Hamiltonian with a central potential as in Eq.\ (\ref{Ham-P2-Q2}) (also dependent on $\hat{L}$, possibly) there exists another conserved quantity corresponding to the quadratic Casimir invariant in Eq.\ \eqref{algeb3} for $\kappa<\kappa_c$ or in Eq.\ (\ref{algeb9}) for $\kappa>\kappa_c$, whose eigenvalues determine irreducible representations of $SL(2,\mathbb{R})$  \cite{Bargmann,Vega} or
$SU(2)$ respectively.  There is also a constraint relating this Casimir invariant with the square of the generator of rotations on the noncommutative plane, $\hat{L}$.
Moreover, the positivity of $\mathbf{\hat{X}}^2$ (or $\mathbf{\hat{P}}^2$), implied by the Hermiticity of the dynamical variables, imposes an additional constraint which determines the possible eigenvalues of $\hat{L}$ for each irreducible representation.

In these conditions, the eigenvalue problem for $\hat{H}$ reduces, for $\kappa>\kappa_c$, to a finite-dimensional one in the $\langle j, l \rangle$ unitary irreducible representation of  $SU(2) \otimes SO(2)$ with $ l=-(2j+1)$.

 For $\kappa<\kappa_c$, only the (infinite dimensional) unitary irreducible representations $\langle k,l\rangle$ of $SL(2,\mathbb{R})$ with $\mathcal{J}_0$ bounded below satisfy the constraints. In this case, $ l=\pm(2k-1) $ and both signs are related by a time-reversal (or parity) transformation.

 \smallskip

Notice also that, from  Eqs.\ (\ref{algeb1}) or \eqref{algeb1-neg}, neither $\mathcal{J}_0$ nor $\mathcal{J}_1$ have a well defined double limit for $\theta,\kappa \rightarrow 0$. In particular, they can not be defined neither in the $\kappa\rightarrow 0$  nor in the $\theta\rightarrow 0$ limits, even though the double limit of $\mathbf{\mathcal{J}}^2$ in Eq.\ (\ref{algeb3}) exists. Indeed,  if we
define $L_0=X_1P_2-X_2P_1$ (the commutative limit of $\hat{L}$), we get
\begin{equation}\label{algeb12}
   \begin{array}{c}\displaystyle
     \lim_{\theta,\kappa \rightarrow 0} \mathbf{\mathcal{J}}^2 = \left. \frac{1}{16 \hbar^2}\left\{ 2 {\mathbf{\hat{X}}^2}  {\mathbf{\hat{P}}^2} +
   2 {\mathbf{\hat{P}}^2}  {\mathbf{\hat{X}}^2} - \left({\mathbf{\hat{X}}}\cdot{\mathbf{\hat{P}}}+{\mathbf{\hat{P}}}\cdot{\mathbf{\hat{X}}} \right)^2\right\}\right|_{\theta,\kappa \rightarrow 0} =
   \\ \\ \displaystyle
     = \left. \frac{1}{4\hbar^2} \left( L_0^2-\hbar^2 \right)  \right|_{\theta,\kappa \rightarrow 0}
   \end{array}
\end{equation}
(which clearly commutes with ${\mathbf{\hat{X}}^2}$ and ${\mathbf{\hat{P}}^2}$ in this limit).

Therefore, the algebraic structure described above is a consequence of the noncommutativity present in both coordinates and momenta.

\smallskip

In the following Section we apply these results to the resolution of some simple examples.


\section{The isotropic oscillator}

A simple example is offered by the \emph{isotropic oscillator} in this noncommutative phase spaces, since in this case the Hamiltonian \cite{Poly,Bellucci}
\begin{equation}\label{Hosc-isotropo}
     H_{osc}:= \frac{\mathbf{\hat{P}}^2}{2\mu}  +\frac{1}{2}\mu \omega^2 {\mathbf{\hat{X}}^2}
\end{equation}
itself is an element of the Lie algebra generated by $\left\{\mathbf{\hat{P}}^2,\mathbf{\hat{X}}^2,
\left(\mathbf{\hat{X}}\cdot\mathbf{\hat{P}}+\mathbf{\hat{P}}\cdot\mathbf{\hat{X}}\right),\hat{L}\right\}$.


\subsection{The isotropic harmonic oscillator for $\kappa>\kappa_c$}

Let us first consider this Hamiltonian in the region  where $\kappa>\kappa_c$,
\begin{equation}\label{algeb13}
    \begin{array}{c} \displaystyle
      H_{osc}     =\frac{\kappa+\theta\mu^2\omega^2}{2\mu}\left(  2 \sqrt{\frac{\hbar^2}{\theta\kappa}} \ {J}_3 -\frac{\hat{L}}{\hbar} \right)
     -\frac{\kappa-\theta\mu^2\omega^2}{2\mu}   \sqrt{1-\frac{\hbar^2}{\theta\kappa}} \left({J}_++{J}_-\right) \,,
    \end{array}
\end{equation}
where we have employed Eq.\ (\ref{algeb11}).



According to the discussion in Section \ref{algebraic-structure-kmayor}, the Hamiltonian in Eq.\ \eqref{algeb13} can be diagonalized through a \emph{rotation} generated by $J_2$.
Indeed, if
\begin{equation}\label{rotJ2-1}
    \varphi = \arctan \left\{ -\left( \frac{\kappa-\theta \mu^2 \omega^2}{\kappa+\theta \mu^2 \omega^2} \right)
    \sqrt{\frac{\kappa\theta}{\hbar^2}-1} \right\}\,,
\end{equation}
the unitary transformation $e^{\imath \varphi J_2}$ brings $H_{osc}$ to the diagonal form (See Eq.\ \eqref{transf-6-ap})
\begin{equation}\label{rotJ2-2}
    \begin{array}{c} \displaystyle
      e^{\imath \varphi J_2} H_{osc} \, e^{-\imath \varphi J_2} =
      \\ \\ \displaystyle
     =  \frac{1}{\mu}\sqrt{\frac{\hbar^2}{\theta\kappa}} \left\{ \left(\kappa+\theta \mu^2 \omega^2 \right)^2 +\left[\frac{\kappa\theta}{\hbar^2}-1 \right]\left(\kappa-\theta \mu^2 \omega^2 \right)^2 \right\}^{1/2} J_3
      -\frac{\kappa+\theta\mu^2\omega^2}{2\mu} \frac{\hat{L}}{\hbar} \,,
    \end{array}
\end{equation}
from which one concludes that the eigenvectors of $H_{osc}$ in the irreducible representation $\langle j,l \rangle$ (with $l=-(2j+1)$), of dimension $2j+1$, are
\begin{equation}\label{eigenvectors-su2}
    \left| \psi_{j,m,l} \right\rangle  =  e^{-\imath \varphi J_2}  \left|j,m,l \right\rangle\,.
\end{equation}
The corresponding energy eigenvalues are given by
\begin{equation}\label{rotJ2-3}
    \begin{array}{c} \displaystyle
      E_m^{(j)}
       = \frac{1}{\mu}\sqrt{\frac{\hbar^2}{\theta\kappa}} \left\{ \left(\kappa+\theta \mu^2 \omega^2 \right)^2 +\left[\frac{\kappa\theta}{\hbar^2}-1 \right]\left(\kappa-\theta \mu^2 \omega^2 \right)^2 \right\}^{1/2}  m +
         \\ \\ \displaystyle
         +\frac{\kappa+\theta\mu^2\omega^2}{2\mu} \, (2j+1) =
                \frac{m}{\mu} \sqrt{\left(\kappa -\theta  \mu ^2 \omega ^2\right)^2+4
   \mu ^2 \omega ^2 \hbar ^2}  + \frac{\kappa+\theta\mu^2\omega^2}{2\mu}\, (2j+1) \,,
    \end{array}
\end{equation}
where $m=-j,-j+1,\cdots, j-1, j$.

\smallskip

Taking into account that, in this region,
\begin{equation}\label{positivity-1}
       \begin{array}{c}\displaystyle
          \sqrt{\left(\kappa -\theta  \mu ^2 \omega ^2\right)^2+4 \mu ^2 \omega ^2 \hbar ^2} =
        \\ \\ \displaystyle
         =\sqrt{\left(\kappa +\theta  \mu ^2 \omega ^2\right)^2+4 \mu ^2 \omega ^2 \hbar ^2\left(1-\frac{\kappa\theta}{\hbar^2}\right)}
          <\left( {\kappa+\theta\mu^2\omega^2}\right)\,,
       \end{array}
\end{equation}
we see that
\begin{equation}\label{positivity-2}
   E_m^{(j)}\geq E_{-j}^{(j)} \geq \left( \frac{\kappa+\theta\mu^2\omega^2}{2\mu}\right)  \geq 0\,,
\end{equation}
in agreement with the positivity of $H_{osc}$ in Eq.\ \eqref{Hosc-isotropo}.

Notice that the spectrum of $H_{osc}$ is left invariant by the change $m\rightarrow-m$, even though the Hamiltonian in Eq.\ \eqref{algeb13} is not time-reversal or parity invariant (remember that $\mathcal{T}$ and $\mathcal{P}$ change the sign of $J_3$ while leave invariant $J_1$ and $\hat{L}$). Indeed, for the \emph{antilinear} transformation $\mathcal{T}$ we have (see Eq.\ \eqref{TeigenJ3})
\begin{equation}\label{spectrTR-SU2}
    \left| \psi_{j,-m,l} \right\rangle  \sim  e^{-\imath \varphi J_2}  \mathcal{T} \left|j,m,l \right\rangle =e^{-\imath \varphi J_2}  \mathcal{T}    e^{\imath \varphi J_2}  \left| \psi_{j,m,l} \right\rangle \,,
\end{equation}
thus mapping biunivocally  $E_m^{(j)} \leftrightarrow E_{-m}^{(j)}$.

\smallskip

The eigenvalues in Eq.\ \eqref{rotJ2-3} can also be written as
\begin{equation}\label{positivity-3}
    E_m^{(j)} = \hbar \Omega_+ \left(n_+ +\frac{1}{2}\right) + \hbar \Omega_- \left(n_- +\frac{1}{2}\right)\,,
\end{equation}
where we have defined the (positive) frequencies
\begin{equation}\label{positivity-4}
    \Omega_\pm=\frac{1}{2\mu \hbar}\left\{
    \left( {\kappa+\theta\mu^2\omega^2}\right)\pm
     \sqrt{\left(\kappa -\theta  \mu ^2 \omega ^2\right)^2+4 \mu ^2 \omega ^2 \hbar ^2}
    \right\}
\end{equation}
and $n_+,n_-$  are univocally determined by $j$ and $m$ through the relations
\begin{equation}\label{positivity-5}
   \begin{array}{c}\displaystyle
      n_++n_-+1=2j+1\,, \quad n_+-n_-=2m\,, \quad \Rightarrow
    \\ \\ \displaystyle
          n_\pm=j \pm m \geq  0\,.
   \end{array}
\end{equation}
Then, $n_+,n_-$ are both nonnegative integers.

Conversely, given the nonnegative integers $n_\pm$, Eq.\  \eqref{positivity-5} univocally determines the nonnegative integer or half-integer $j$ and $m$ satisfying $-j \leq m \leq j$. Therefore, the spectrum of $H_{osc}$ coincides with that of two uncoupled ordinary harmonic oscillators of frequencies $\Omega_\pm$, in agreement with the results in \cite{Poly}.

\subsection{The isotropic harmonic oscillator for $0<\kappa<\kappa_c$}

We can give a similar solution for the case $\kappa<\kappa_c$, where the Hamiltonian in Eq.\ \eqref{Hosc-isotropo} can be written as
\begin{equation}\label{algeb13-menor}
    \begin{array}{c} \displaystyle
      H_{osc}:=
    \frac{\kappa+\theta\mu^2\omega^2}{2\mu}\left(  2 \sqrt{\frac{\hbar^2}{\theta\kappa}} \ {\mathcal{J}}_0 -\frac{\hat{L}}{\hbar} \right)
     -\frac{\kappa-\theta\mu^2\omega^2}{2\mu}   \sqrt{\frac{\hbar^2}{\theta\kappa}-1} \left({\mathcal{J}}_++{\mathcal{J}}_-\right) \,,
    \end{array}
\end{equation}
expression which can be diagonalized through a unitary transformation in $SL(2,\mathbb{R})$, as explained in Appendix \ref{rot-sl2r}.

Indeed, if we take
\begin{equation}\label{AyBsl2R}
   \begin{array}{c} \displaystyle
      A=  \frac{\kappa+\theta\mu^2\omega^2}{\mu}   \sqrt{\frac{\hbar^2}{\theta\kappa}}  \,, \quad
    B= - \frac{\kappa-\theta\mu^2\omega^2}{\mu}   \sqrt{\frac{\hbar^2}{\theta\kappa}-1} \,,
    \\ \\ \displaystyle
     {\rm and} \quad \tanh \alpha = \frac{B}{A} \,,
   \end{array}
\end{equation}
applying the result quoted in Eq.\ \eqref{transf-4-ap} to $H_{osc}$ in Eq.\ \eqref{algeb13-menor} we get the operator
\begin{equation}\label{transf-5}
    \begin{array}{c} \displaystyle
      e^{\imath \alpha  \mathcal{J}_2} H_{osc}  \,  e^{-\imath \alpha  \mathcal{J}_2}=
      \\ \\ \displaystyle
     \frac{1}{\mu}\sqrt{\frac{\hbar^2}{\theta\kappa}} \left\{ \left(\kappa+\theta \mu^2 \omega^2 \right)^2 -\left[1-\frac{\kappa\theta}{\hbar^2}\right]\left(\kappa-\theta \mu^2 \omega^2 \right)^2 \right\}^{1/2} \mathcal{J}_0
      -\frac{\kappa+\theta\mu^2\omega^2}{2\mu} \frac{\hat{L}}{\hbar} \,,
    \end{array}
\end{equation}
diagonal in any unitary irreducible representation of $SL(2,\mathbb{R}) \otimes SO(2)$.

From Eq.\ \eqref{transf-5} one concludes that the Hamiltonian's eigenvectors are given by
\begin{equation}\label{eigenvectors-kappa-menor}
    \left| \phi_{k,m,l} \right\rangle =  e^{-\imath \alpha  \mathcal{J}_2}
    \left| {k,m,l} \right\rangle \,,
\end{equation}
where $\left| {k,m,l} \right\rangle$ belongs to the unitary irreducible representation $\langle k, l\rangle$, where $k$ is some positive integer or half-integer, $m=k,k+1,\cdots$ and $l=(2k-1)$ or $l=-(2k-1)$ (see Section \ref{discrete-kappa-menor}). The corresponding energy eigenvalues are
 \begin{equation}\label{transf-6}
        \begin{array}{c} \displaystyle
            E_m^{(k,\pm)}=
            \frac{1}{\mu}\sqrt{\frac{\hbar^2}{\theta\kappa}} \left\{ \left(\kappa+\theta \mu^2 \omega^2 \right)^2 -\left[1-\frac{\kappa\theta}{\hbar^2} \right]\left(\kappa-\theta \mu^2 \omega^2 \right)^2 \right\}^{1/2} m -
                  \\ \\ \displaystyle
      \mp \frac{\kappa+\theta\mu^2\omega^2}{2\mu}\, (2k-1) =
     \\ \\ \displaystyle =
\frac{m}{\mu} \sqrt{\left(\kappa +\theta  \mu ^2 \omega ^2\right)^2+4
   \mu ^2 \omega^2 \hbar ^2 \left(1 -\frac{\kappa \theta}{\hbar ^2} \right)}  \mp \frac{\kappa+\theta\mu^2\omega^2}{2\mu}\, (2k-1)\,.
        \end{array}
 \end{equation}
Notice that
 \begin{equation}\label{E-pos-kappa-menor}
     E_m^{(k,\pm)}\geq  E_k^{(k,\pm)} \geq \frac{\kappa+\theta\mu^2\omega^2}{2\mu}\left( 2k\mp(2k-1) \right) \geq
      \frac{\kappa+\theta\mu^2\omega^2}{2\mu} \,,
 \end{equation}
 in agreement with the positivity of $H_{osc}$.

Moreover, notice that, even though $H_{osc}$ in Eq.\ \eqref{algeb13-menor} is not time-reversal  (or parity) invariant, the constraint relating $\hat{L}$ to $\mathcal{J}^2$ and the condition of positivity of $\mathbf{\hat{X}}^2$ (or $\mathbf{\hat{P}}^2$) make the energy spectrum $\mathcal{T}$-invariant (In this region, the discrete transformations change the sign of $\hat{L}$ and $\mathcal{J}_2$ leaving invariant $\mathcal{J}_{0,1}$ - see Eq.\ \eqref{discrete-J}), since
\begin{equation}\label{spectTR}
    \mathcal{T}\left| \phi_{k,m,l} \right\rangle =  e^{-\imath \alpha  \mathcal{J}_2}  \mathcal{T} \left| {k,m,l} \right\rangle
    \sim  e^{-\imath \alpha  \mathcal{J}_2}   \left| {k,m,-l} \right\rangle=\left| \phi_{k,m,-l} \right\rangle \,,
\end{equation}
with $l=\pm(2k-1)$, thus mapping biunivocally  $ E_m^{(k,\pm)} \leftrightarrow  E_m^{(k,\mp)}$.

\smallskip

 The energy eigenvalues can also be written as
 \begin{equation}\label{transf-6-1}
            E_m^{(k,s)}=  \hbar \Omega_+ \left( n_++\frac{1}{2}\right) + \hbar \Omega_- \left( n_-+\frac{1}{2}\right)\,,
 \end{equation}
 with $s=\pm 1$,  where the (positive) frequencies are
 \begin{equation}\label{omegamas-omegamenos}
    \Omega_\pm=\frac{1}{2 \mu  \hbar }\left\{ { \sqrt{\left(\kappa +\theta  \mu ^2 \omega ^2\right)^2+4
   \mu ^2 \omega^2 \hbar ^2 \left(1 -\frac{\kappa \theta}{\hbar ^2} \right)}
    \pm (\kappa+\theta  \mu ^2
   \omega ^2) } \right\}
 \end{equation}
 and $n_+,n_-$ are univocally determined by $m$ and $l=s(2k-1)$ through the relations
 \begin{equation}\label{omegamas-omegamenos-1}
   \begin{array}{c} \displaystyle
      n_++n_-+1=2m\,,\quad n_+-n_-=-l\,,
      \\ \\ \displaystyle
    \Rightarrow \quad  n_\pm=m\mp \frac{l}{2}-\frac{1}{2} \geq \left(1 \mp s\right) \left(k-\frac{1}{2}\right) \geq 0 \,.
   \end{array}
 \end{equation}
Therefore, also in this case, $n_\pm$ are both nonnegative integers.

Conversely, given the nonnegative integers $n_\pm$, Eq.\ \eqref{omegamas-omegamenos-1} univocally determines $m$ and $l$, with $k=(|l|+1)/2$. Therefore, also in this region the spectrum of $H_{osc}$ coincides with that of two uncoupled harmonic oscillators of frequencies $\Omega_\pm$, {in agreement  with \cite{Poly}.}


\subsection{The isotropic harmonic oscillator for $\kappa<0$}

In this region, the Hamiltonian in Eq.\ \eqref{Hosc-isotropo} can be written as
\begin{equation}\label{algeb13-kappa-neg}
      H_{osc}:= A \mathcal{J}_0+B\mathcal{J}_2+\left(\frac{|\kappa|-\theta\mu^2\omega^2}{2\mu}\right) \frac{\hat{L}}{\hbar}
\end{equation}
with
\begin{equation}\label{AyBsl2R-menor}
   \begin{array}{c} \displaystyle
      A=  \frac{|\kappa|+\theta\mu^2\omega^2}{\mu}   \sqrt{1+\frac{\hbar^2}{\theta|\kappa|}}  \,, \quad
    B= \frac{\theta\mu^2\omega^2-|\kappa|}{\mu}   \sqrt{\frac{\hbar^2}{\theta|\kappa|}} \,.
   \end{array}
\end{equation}
Applying the result quoted in Eq.\ \eqref{transf-44-ap} with $\tanh \beta =-{B}/{A}$ to $H_{osc}$ in Eq.\ \eqref{algeb13-kappa-neg} we get the diagonal operator
\begin{equation}\label{transf-5-kappa-neg}
    \begin{array}{c} \displaystyle
      e^{\imath \beta  \mathcal{J}_1} H_{osc}  \,  e^{-\imath \beta  \mathcal{J}_1}=
      \\ \\ \displaystyle
     \frac{1}{\mu} \left\{ \left(|\kappa|+\theta \mu^2 \omega^2 \right)^2 + 4 \hbar^2 \mu^2 \omega^2 \right\}^{1/2} \mathcal{J}_0
      +\left(\frac{|\kappa|-\theta\mu^2\omega^2}{2\mu}\right) \frac{\hat{L}}{\hbar} \,.
    \end{array}
\end{equation}

This shows that the Hamiltonian's eigenvectors are given by
\begin{equation}\label{eigenvectors-kappa-menor-neg}
    \left| \phi_{k,m,l} \right\rangle =  e^{-\imath \beta  \mathcal{J}_1}
    \left| {k,m,l} \right\rangle \,,
\end{equation}
where $\left| {k,m,l} \right\rangle$ belongs to the unitary irreducible representation $\langle k, l\rangle$, with $k$ a positive integer or half-integer, $m=k,k+1,\cdots$ and $l=\pm(2k-1)$ (see Section \ref{discrete-kappa-menor}), corresponding to the energy eigenvalues
 \begin{equation}\label{transf-6-2}
        \begin{array}{c} \displaystyle
            E_m^{(k,\pm)}=
     \frac{m}{\mu} \left\{ \left(|\kappa|-\theta \mu^2 \omega^2 \right)^2 + \left(1+\frac{\hbar^2}{\theta|\kappa|}\right) 4 |\kappa|\theta \mu^2 \omega^2 \right\}^{1/2}
     \\ \\ \displaystyle
      \pm \left(\frac{|\kappa|-\theta\mu^2\omega^2}{2\mu}\right) (2k-1) \,.
        \end{array}
 \end{equation}
Here again
\begin{equation}\label{E-pos-kappa-menor-neg}
     E_m^{(k,\pm)}\geq  E_k^{(k,\pm)} \geq \frac{\left||\kappa|-\theta\mu^2\omega^2\right|}{2\mu}\left( 2k- |2k-1| \right) =
      \frac{\left||\kappa|-\theta\mu^2\omega^2\right|}{2\mu} \,,
 \end{equation}
 in agreement with the positivity of $H_{osc}$.

\smallskip

 The energy eigenvalues can also be written as in Eq.\ \eqref{transf-6-1}, where the frequencies are given in this region by
 \begin{equation}\label{omegamas-omegamenos-2}
    \Omega_\pm=\frac{1}{2 \mu  \hbar }\left\{ { \sqrt{\left(|\kappa| -\theta  \mu ^2 \omega ^2\right)^2+
    \left(1 +\frac{|\kappa| \theta}{\hbar ^2} \right) 4 \hbar ^2  \mu ^2 \omega^2 }
    \pm \left||\kappa| -\theta  \mu ^2 \omega ^2 \right| } \right\}
 \end{equation}
 and, for $\mathfrak{s}= {\rm sign}\left(|\kappa| -\theta  \mu ^2 \omega ^2\right)$,
 \begin{equation}\label{omegamas-omegamenos-12}
   \begin{array}{c} \displaystyle
      n_++n_-+1=2m\,,\quad n_+-n_-=\mathfrak{s}\,  l\,,
      \\ \\ \displaystyle
     \Rightarrow \quad  n_\pm=m\pm \mathfrak{s}\,  \frac{l}{2} - \frac{1}{2} \geq
    \left(k-\frac{1}{2}\right) (1\pm \mathfrak{s} \, {\rm sign}(l)) \geq 0  \,.
   \end{array}
 \end{equation}
Then, $n_\pm$ are both nonnegative integers univocally related to $m$ and $l$, with $k=(|l|+1)/2$. Therefore, also in this case we find that the spectrum of $H_{osc}$ is equivalent to the spectrum of two uncoupled harmonic oscillators of frequencies $\Omega_\pm$ \cite{Poly}.


\section{The Landau problem}

Let us now consider the extension of the Landau Hamiltonian to the noncommutative phase space \cite{jcm2,Hor-Mik-1},
\begin{equation}\label{Landau-1}
      \begin{array}{c} \displaystyle
        2\mu H_L:=\left( \hat{P}_1+\frac{e B}{2}\, \hat{X}_2 \right)^2 +\left( \hat{P}_2-\frac{e B}{2}\, \hat{X}_1 \right)^2=
        \\ \\ \displaystyle
        =\left(1+\frac{e B \theta }{2 \hbar } \right) \mathbf{\hat{P}}^2+
         \frac{e B }{4 \hbar } (e B  \hbar +2 \kappa )  \mathbf{\hat{X}}^2 -
        e B \left(1-\frac{\kappa\theta}{\hbar^2}\right) \hat{L}\,,
      \end{array}
\end{equation}
where $B$ stands for the magnetic field perpendicular to the plane. For definiteness, we take $e B > 0$.

This Hamiltonian should be compared with $H_{osc}$ in Eq.\ (\ref{Hosc-isotropo}). It is evident that its spectrum can be determined in a similar way.

\subsection{The Landau problem for $\kappa>\kappa_c$} From Eqs.\ \eqref{algeb11}, we can write the Hamiltonian as
\begin{equation}\label{Landau-1-mayor}
    \begin{array}{c} \displaystyle
      2 \mu H_L = \frac{
    \left(B^2 e^2
   \theta  \hbar +4 B e \theta  \kappa +4 \kappa  \hbar
   \right)}{2 \sqrt{{\theta  \kappa }} }\, {J_3}+
   \frac{1}{2}  \sqrt{1-\frac{\hbar ^2}{\theta  \kappa
   }} \left(B^2 e^2 \theta -4 \kappa \right) \, {J_1}-
   \\ \\ \displaystyle
      - \frac{ \left(B^2 e^2 \theta +4 B e
   \hbar +4 \kappa \right)}{4 \hbar }\, \hat{L}\,.
    \end{array}
\end{equation}
Taking into account Eq.\ \eqref{transf-6-ap}, one can see that a $J_2$-generated rotation in an angle
\begin{equation}\label{Landau-2}
    \varphi=\arctan\left(
    \frac{\sqrt{\theta  \kappa -\hbar ^2} \left(B^2 e^2
   \theta -4 \kappa \right)}{B^2 e^2 \theta  \hbar +4
   B e \theta  \kappa +4 \kappa  \hbar }
   \right)
\end{equation}
transforms the Hamiltonian into
\begin{equation}\label{Landau-3}
    e^{\imath \varphi J_2}\left( 2 \mu H_L \right) e^{-\imath \varphi J_2} =
    \frac{ \left(4  B e \hbar + B^2 e^2 \theta +4 \kappa \right)}{2}\left({J_3} -\frac{L}{2\hbar}\right)\,.
\end{equation}

Therefore, the eigenvectors of $H_L$ in the irreducible representation $\langle j , l\rangle$ are given by
\begin{equation}\label{Landau-4}
    \left| \psi_{j,m,l} \right\rangle  =  e^{-\imath \varphi J_2}  \left|j,m,l \right\rangle
\end{equation}
with $l=-(2j+1)$, for integer or half-integer $j\geq 0$ and $-j\leq m\leq j$. The corresponding eigenvalue is
\begin{equation}\label{Landau-5}
    E_m^{(j)}=
     \frac{ \left(4  B e \hbar + B^2 e^2 \theta +4 \kappa \right)}{4\mu}\left(m + j +\frac{1}{2}\right)
     \geq
     \frac{ \left(4  B e \hbar + B^2 e^2 \theta +4 \kappa \right)}{8\mu}\,.
\end{equation}

Notice that these eigenvalues depend only on the nonnegative integer $n_a=(j+m)$. Therefore, given $n_a$, each irreducible representation with $j \geq n_a/2$ contains a state with energy ${ \left(4  B e \hbar + B^2 e^2 \theta +4 \kappa \right)}\left(n_a +\frac{1}{2}\right)/{4\mu}$ which, then, has (a countable) infinite degeneracy. The vectors in this sequence can be identified by the nonnegative integer $n_b=j-m$. Compared with the Landau problem on the normal commutative plane, one can see that $\left(  B^2 e^2 \theta +4 \kappa  \right)/4e\hbar$ appears as an additional contribution to the \emph{effective} external magnetic field. Remember that only negative integer values of $l\leq -1$ are possible in this region.

\smallskip

Taking into account that each irreducible representation contributes with only one state to each Landau level, we can now evaluate the density of states. From Eq.\ \eqref{transf-su2-gen} with $\varphi$ given in Eq.\  \eqref{Landau-2} we get for the mean value of $\mathbf{\hat{x}}^2$ in a given Hamiltonian eigenvector (See ec.\ \eqref{algeb11})
\begin{equation}\label{densidad-1}
    \left \langle j,m,l  \right|  e^{-\imath \varphi J_2} \, \mathbf{\hat{x}}^2 e^{\imath \varphi J_2}  \left|j,m,l \right\rangle
    =m \left(\frac{4 \left(B^2 e^2 \theta ^2+4 B e \theta \hbar
   +4\hbar^2\right)}{B^2 e^2 \theta +4 B e \hbar+4 \kappa }-2
   \theta \right)+\theta  (2 j+1)\,,
\end{equation}
where we have taken into account that the mean value of $J_1$ vanishes, $ \langle j,m,l  |  J_1 | j,m,l \rangle =0$.
Then, with $m=n_a - j$, we see that $\langle  \mathbf{\hat{x}}^2 \rangle$ linearly grows with $j$ in each Landau level (fixed $n_a$),
\begin{equation}\label{densidad-2}
    \langle  \mathbf{\hat{x}}^2 \rangle
     = \frac{16 j (\theta  \kappa -\hbar^2)}{B^2 e^2 \theta +4 B
   e\hbar+4 \kappa } + \frac{2 n_a \left(B^2 e^2 \theta ^2+4 B
   e \theta \hbar - 4 \theta  \kappa +8\hbar^2 \right)}{B^2 e^2
   \theta +4 B e\hbar + 4 \kappa }+\theta   \,.
\end{equation}
If we assume that $\left| \psi_{j,m,l} \right\rangle$ represents a probability distribution which is essentially concentrated in a \emph{circle} of area $\pi  \langle  \mathbf{\hat{x}}^2 \rangle$, we can assume that this circle \emph{contains} one state for each integer of half-integer $0\leq j' \leq j$. This number is $\sum_{r=0}^{2j} 1= 2j+1$.

Therefore, the density of states in the thermodynamic limit ($j \rightarrow \infty$) can be approximated by
\begin{equation}\label{densidad-3}
    \rho_{\kappa>\kappa_c}\simeq\frac{2j+1}{ \pi  \langle  \mathbf{\hat{x}}^2 \rangle }=
    \frac{ B^2 e^2 \theta +4 B e\hbar+4 \kappa }{8 \pi  (\theta \kappa -\hbar^2)}
    + O(j^{-1})\,.
\end{equation}
Notice that this density is the same for all the Landau levels and diverges for $\kappa \rightarrow \kappa_c^+$.

\subsection{The Landau problem for $0<\kappa<\kappa_c$} From Eqs.\ \eqref{algeb6}, one can see that the Hamiltonian can be written as
\begin{equation}\label{Landau-1-1}
    \begin{array}{c} \displaystyle
      2 \mu H_L = \frac{
    \left(B^2 e^2
   \theta  \hbar +4 B e \theta  \kappa +4 \kappa  \hbar
   \right)}{2 \sqrt{{\theta  \kappa }} }\, {\mathcal{J}_0}+
   \frac{1}{2}  \sqrt{\frac{\hbar ^2}{\theta  \kappa
   }-1} \left(B^2 e^2 \theta -4 \kappa \right) \, {\mathcal{J}_1}-
   \\ \\ \displaystyle
      - \frac{ \left(B^2 e^2 \theta +4 B e
   \hbar +4 \kappa \right)}{4 \hbar }\, \hat{L}\,.
    \end{array}
\end{equation}
Then, employing Eq.\ \eqref{transf-4-ap} with
\begin{equation}\label{Landau-2-1}
    \tanh \alpha=\left(
    \frac{\sqrt{\hbar^2-\theta  \kappa} \left(B^2 e^2
   \theta -4 \kappa \right)}{B^2 e^2 \theta  \hbar +4
   B e \theta  \kappa +4 \kappa  \hbar }
   \right)\,,
\end{equation}
the Hamiltonian can be transformed into
\begin{equation}\label{Landau-3-1}
    e^{\imath \alpha \mathcal{J}_2}\left( 2 \mu H_L \right) e^{-\imath \alpha \mathcal{J}_2} =
    \frac{ \left(4  B e \hbar + B^2 e^2 \theta +4 \kappa \right)}{2}\left({\mathcal{J}_0} -\frac{L}{2\hbar}\right)\,.
\end{equation}

Therefore, the eigenvectors of $H_L$ in the unitary irreducible representation $\langle k , l\rangle$ of 
$SL(2,\mathbb{R})\otimes SO(2)$ with $k$ a positive integer or half-integer and $l=\pm(2k-1)$ (See Section \ref{discrete-kappa-menor})
are given by
\begin{equation}\label{Landau-4-1}
    \left| \psi_{k,m,s} \right\rangle  =  e^{-\imath \alpha \mathcal{J}_2}  \left|k,m,s(2k-1) \right\rangle\,,
\end{equation}
with $s=\pm 1$. The corresponding eigenvalue is
\begin{equation}\label{Landau-5-1}
    E_m^{(k,s)}=
     \frac{ \left(4  B e \hbar + B^2 e^2 \theta +4 \kappa \right)}{4\mu}\left[m -s \left(k -\frac{1}{2}\right)\right]
     \geq
     \frac{ \left(4  B e \hbar + B^2 e^2 \theta +4 \kappa \right)}{8\mu}\,.
\end{equation}

Let us first consider the case with $s=+1$ (corresponding to  $k\geq \frac{1}{2}$ with non-negative $l=(2k-1)\geq 0$). Writing $m=k+n$ with $n=0,1,2,\cdots$, we get
\begin{equation}\label{Landau-5-1-1}
    E_m^{(k,+1)}=
     \frac{ \left(4  B e \hbar + B^2 e^2 \theta +4 \kappa \right)}{4\mu}\left(n  +\frac{1}{2}\right)\,,
\end{equation}
which is independent of $k$ (a countable infinite degeneracy).

On the other hand, for $s=-1$ (states with $k\geq 1$ and negative $ l=-(2k-1)\leq -1$), we get
\begin{equation}\label{Landau-5-1-2}
    E_m^{(k,-1)}=
     \frac{ \left(4  B e \hbar + B^2 e^2 \theta +4 \kappa \right)}{4\mu}\left[n+(2k-1)  +\frac{1}{2}\right]\,.
\end{equation}
Since $n':=n+(2k-1)\in \mathbb{N}$, for a given $n'$ we have $2k-1=n'-n\geq 1$, and then $n=0,1,\cdots,n'-1$. Therefore, these representations of the group with negative $l$ contribute to the characteristic subspace of $H_L$ with energy $E_m^{(k,-1)}= \frac{ \left(4  B e \hbar + B^2 e^2 \theta +4 \kappa \right)}{4\mu}\left(n'  +\frac{1}{2}\right)$ with a finite number $n'$ of additional linearly independent eigenvectors.

Then we see that, all together, one gets a spectrum equivalent to that of the Landau model on the usual commutative plane with an effective magnetic field given by $B_{eff}=B+\left(  B^2 e^2 \theta +4 \kappa  \right)/4e\hbar$. {This spectrum, which is exact, agrees with the one obtained in \cite{Gango2} up to first order in the noncommutativity parameters (Notice the different convention for the commutators between dynamical variables adopted in this reference).}

\smallskip

Let us now consider the density of states in each Landau level. In this case it is sufficient to consider the contributions of irreducible representations with $l=(2k-1)\geq 0$, since those with $l\leq-1$ (finite in number) do not contribute in the thermodynamic limit.

With $\mathbf{\hat{x}}^2$ written as in  Eq.\ \eqref{algeb6}, from Eq.\ \eqref{transf-5-ap} with $\alpha$ given in Eq.\ \eqref{Landau-2-1} we get
\begin{equation}\label{densidad-5}
   \begin{array}{c} \displaystyle
      \left\langle k,m,(2k-1) \right| e^{\imath \alpha \mathcal{J}_2} \,
    \mathbf{\hat{x}}^2
    e^{-\imath \alpha \mathcal{J}_2}  \left|k,m,(2k-1) \right\rangle=
    \\ \\ \displaystyle
     = k \, \frac{16  (\hbar^2-\theta  \kappa )}{B^2 e^2 \theta +4 B
   e\hbar+4 \kappa }+2 n\, \frac{   (B e \theta  (B e
   \theta +4\hbar)-4 \theta  \kappa +8\hbar^2)}{B e   (B e
   \theta +4\hbar)+4   \kappa }+\theta\,,
   \end{array}
\end{equation}
where we have written $m=k+n$ and taken into account that  $\langle k,m,(2k-1) |  \mathcal{J}_1|k,m,(2k-1) \rangle=0$. Then we see that, in a given Landau level (fixed $n$), the mean value $\langle \mathbf{\hat{x}}^2 \rangle$ linearly grows with $k$.

We will again assume that $\left| \psi_{k,m,1} \right\rangle$ corresponds to a probability distribution which is essentially concentrated in a \emph{circle} of area $\pi \langle \mathbf{\hat{x}}^2 \rangle$, with $\langle \mathbf{\hat{x}}^2 \rangle$ given in ec.\ \eqref{densidad-5}. And, taking into account that each irreducible representation with $l=2k-1\geq 0$ contribute with only one state to each Landau level, we can assume that this \emph{circle} contains one state for each integer or half integer $\frac{1}{2}\leq k' \leq k$, whose number is given by $\sum_{r=1}^{2k} 1= 2k$.

Therefore, the density of states in the thermodynamic limit ($k \rightarrow \infty$) can be approximated as
\begin{equation}\label{densidad-6}
    \rho_{0<\kappa<\kappa_c} \simeq \frac{2k}{ \pi \langle \mathbf{\hat{x}}^2 \rangle }=
    \frac{B^2 e^2 \theta +4 B e\hbar+4 \kappa }{8 \pi
   (\hbar^2-\theta  \kappa )} +O(k^{-1})\,.
\end{equation}
The leading term is independent of the Landau level (compare with Eq.\ \eqref{densidad-3}), smoothly reduces to the well known result in the commutative plane when $\kappa,\theta\rightarrow 0$, $eB/2\pi\hbar$, and also diverges in the limit $\kappa \theta/\hbar^2 \rightarrow {1}^-$.

\subsection{The Landau problem for $\kappa<0$} Employing  Eqs.\ \eqref{algeb6-neg}, one gets
\begin{equation}\label{Landau-1-2}
    \begin{array}{c} \displaystyle
      2 \mu H_L = \frac{1}{2}  \sqrt{1+\frac{\hbar ^2}{\theta  |\kappa|
   }} \left(B^2 e^2 \theta +4 |\kappa| \right) \, {\mathcal{J}_0}+
   \frac{     \left(B^2 e^2
   \theta  \hbar - 4 B e \theta  |\kappa| - 4 |\kappa|  \hbar
   \right)}{2 \sqrt{{\theta  |\kappa| }} }\, {\mathcal{J}_2}-
   \\ \\ \displaystyle
      - \frac{ \left(B^2 e^2 \theta +4 B e
   \hbar - 4 |\kappa| \right)}{4 \hbar }\, \hat{L}\,.
    \end{array}
\end{equation}
Then, employing Eq.\ \eqref{transf-44-ap} with
\begin{equation}\label{Landau-2-2}
    \tanh \beta=-\left(     \frac{B^2 e^2 \theta  \hbar - 4
   B e \theta  |\kappa| - 4 |\kappa|  \hbar }{\sqrt{\hbar^2+\theta  |\kappa|} \left(B^2 e^2
   \theta + 4 |\kappa| \right)}
   \right)\,,
\end{equation}
the Hamiltonian can be unitarily transformed into
\begin{equation}\label{Landau-3-2}
    e^{\imath \beta \mathcal{J}_1}\left( 2 \mu H_L \right) e^{-\imath \beta \mathcal{J}_1} =
    \frac{ \left|4  B e \hbar + B^2 e^2 \theta - 4 |\kappa| \right|}{2}\left({\mathcal{J}_0} -t \frac{L}{2\hbar}\right)\,,
\end{equation}
where $t={\rm sign}\left(B^2 e^2 \theta +4 B e \hbar - 4 |\kappa| \right)$ (the sign of the effective magnetic field).

Also in this case $k$ is a positive integer or half-integer and $l=s(2k-1)$ with $s=\pm1$ (See Section \ref{discrete-kappa-negativo}). So, the eigenvectors of $H_L$ are
\begin{equation}\label{Landau-4-2}
    \left| \psi_{k,m,s} \right\rangle  =  e^{-\imath \beta \mathcal{J}_1}  \left|k,m,s(2k-1) \right\rangle\,,
\end{equation}
corresponding to the eigenvalue
\begin{equation}\label{Landau-5-2}
    E_m^{(k,s)}=
     \frac{ \left|4  B e \hbar + B^2 e^2 \theta - 4 |\kappa| \right|}{4\mu}\left[m -s\, t \left(k -\frac{1}{2}\right)\right]
     \geq
     \frac{ \left|4  B e \hbar + B^2 e^2 \theta - 4 |\kappa| \right|}{8\mu}\,.
\end{equation}

{From this point, the analysis continues as in the previous section with the replacement $s \rightarrow s\, t$. So, one concludes that this system is also equivalent to the Landau model on the usual commutative plane with an effective magnetic field $B_{eff}=B+\left(  B^2 e^2 \theta - 4 |\kappa|  \right)/4e\hbar$.}

\smallskip

Notice that only Hamiltonian eigenvectors with $s t=1$ contribute to the density of states in a given Landau level in the thermodynamic limit. {So, through a similar reasoning as that in the previous Section, from Eqs.\ \eqref{algeb6-neg}, \eqref{trans-sl2r-general} and \eqref{Landau-2-2} we get for this density}
\begin{equation}\label{densidad-7}
     \rho_{\kappa<0}\simeq\frac{2k}{ \pi \langle \mathbf{\hat{x}}^2 \rangle }=\frac{\left| B e (B e \theta +4\hbar)-4 |\kappa|\right| }{8\pi ( \theta  |\kappa|+\hbar^2 )} + O(k^{-1})
\end{equation}
which, for $\kappa\rightarrow 0^-$ and $\theta\rightarrow 0$, smoothly reduces to $e B/2 \pi\hbar$.

Comparing this result with those in Eqs.\ \eqref{densidad-3} y \eqref{densidad-6}, we see that we can express the density of states in the thermodynamic limit by a single expression as
\begin{equation}\label{densidad-8}
    \rho=\left| \frac{4 B e \hbar+B^2 e^2 \theta +4 \kappa }{8 \pi    (\hbar^2-\theta  \kappa )} \right|\,,
\end{equation}
which presents a singularity for $\kappa \rightarrow \kappa_c$ from both sides\footnote{{A singular behavior of the density of states at the critical value $\kappa_c$ had been described in \cite{Poly}. Compare Eq.\ \eqref{densidad-8} with Eq.\ (24) in this reference, with the identifications $B\rightarrow0, \kappa \rightarrow B, \theta\rightarrow\theta, \hbar\rightarrow1$.}}.

\medskip

{Finally, let us remark that the contributions to the energy eigenvalues coming from $\theta$ depend on $B$ while those coming from $\kappa$ are independent of the external magnetic field. The same is true for the density of states in the Landau levels for small  $\kappa$ and $\theta$. This suggests the possibility that an experiment involving these elements and performed at different values of $B$ could establish some bounds on  the noncommutativity parameters.}


\section{Hamiltonian with a central potential}

Le us now consider a non-relativistic Hamiltonian with a central potential as in Eq.\ \eqref{Ham-P2-Q2}.
For any $\kappa\neq 0$, it is possible to diagonalize $\mathbf{\hat{X}}^2 $ (and, consequently, $V( \mathbf{\hat{X}}^2 )$) expressed in terms of the generators of the group through a unitary transformation as discussed in the previous Sections, transformation which must also be applied to the kinetic term as we describe in the following. {We show that the eigenvalue problem for such Hamiltonians reduces to a three-term recursion relation, similarly to the results obtained in \cite{Bellucci} through a different formulation.} Later, we will apply these results to the example of the cylindrical well potential.

\subsection{\underline{$\kappa > \kappa_c$} :}

With $\varphi=\arctan \sqrt{\frac{\kappa\theta}{\hbar^2}-1}$, from Eqs.\ \eqref{algeb11} and \eqref{transf-su2-gen} we get
\begin{equation}\label{CP2-1}
    \begin{array}{c}\displaystyle
    e^{ i \varphi J_2} \left(\frac{\mathbf{\hat{X}}^2}{\theta} \right) e^{-i \varphi J_2}
    = 2 J_3 - \frac{\hat{L}}{\hbar}\,,
       \\ \\  \displaystyle
    e^{i \varphi J_2} \left( \frac{\mathbf{\hat{P}}^2}{\kappa} \right) e^{-i \varphi J_2}  =
    -2\left(1-\frac{2 \hbar^2}{\kappa\theta} \right) J_3 - \frac{\hat{L}}{\hbar} -
    \frac{2 \hbar^2}{\kappa\theta} \sqrt{\frac{\kappa\theta}{\hbar^2}-1} \left(J_+ + J_- \right)\,,
    \end{array}
\end{equation}
which, taking into account that $l=-(2j+1)$ (Eq.\ \eqref{l-kappa-mayor}), implies that the non-vanishing matrix elements of the Hamiltonian are
\begin{equation}\label{CP3-1}
    \begin{array}{c} \displaystyle
     2 \mu  \left\langle j,m,l \right| e^{ i \varphi J_2} \hat{H} e^{- i \varphi J_2} \left| j,m',l\right\rangle=
      \\ \\ \displaystyle
      =\left\{-2\kappa\left(1-\frac{2 \hbar^2}{\kappa\theta} \right) m + \kappa (2j+1) +2\mu V\big(\theta(2m+2j+1)\big) \right\} \delta_{m,m'}-
      \\ \\ \displaystyle
      -\frac{2 \hbar^2}{\theta} \sqrt{\frac{\kappa\theta}{\hbar^2}-1}
      \left\{ \sqrt{(j-m+1)(j+m)}\,  \delta_{m-1,m'} +
      \sqrt{(j+m+1)(j-m)}\,  \delta_{m+1,m'}  \right\}\,.
    \end{array}
\end{equation}
Notice that the eigenvalue problem for the Hamiltonian in this region is reduced to a matricial problem, since the irreducible unitary representations of $SU(2)$ are of finite dimension.


\subsection{\underline{$0<\kappa < \kappa_c$} :}

If we take $\tanh \alpha=\sqrt{1-\frac{\kappa\theta}{\hbar^2}}$, from Eqs.\ \eqref{algeb6} and \eqref{transf-5-ap} we get
\begin{equation}\label{CP2-2}
    \begin{array}{c}\displaystyle
    e^{ i \alpha \mathcal{J}_2} \left(\frac{\mathbf{\hat{X}}^2}{\theta} \right) e^{ -i \alpha \mathcal{J}_2}
    = 2 \mathcal{J}_0 - \frac{\hat{L}}{\hbar}\,,
       \\ \\  \displaystyle
    e^{ i \alpha \mathcal{J}_2} \left( \frac{\mathbf{\hat{P}}^2}{\kappa} \right) e^{ -i \alpha \mathcal{J}_2}  =
    2\left(\frac{2 \hbar^2}{\kappa\theta} -1\right) \mathcal{J}_0 - \frac{\hat{L}}{\hbar} -
    \frac{2 \hbar^2}{\kappa\theta} \sqrt{1-\frac{\kappa\theta}{\hbar^2}} \left(\mathcal{J}_+ + \mathcal{J}_- \right)\,,
    \end{array}
\end{equation}
which, taking into account Eq.\ \eqref{l-intermedio}, implies that the non-vanishing matrix elements of the Hamiltonian are
\begin{equation}\label{CP3-2}
    \begin{array}{c} \displaystyle
     2 \mu  \left\langle k,m,l \right| e^{ i \alpha \mathcal{J}_2} \hat{H} e^{ -i \alpha \mathcal{J}_2} \left| k,m',l\right\rangle=
      \\ \\ \displaystyle
      =\left\{2\kappa\left(\frac{2 \hbar^2}{\kappa\theta} -1\right) m - \kappa l +2\mu V\big(\theta(2m-l)\big) \right\} \delta_{m,m'}-
      \\ \\ \displaystyle
      -\frac{2 \hbar^2}{\theta} \sqrt{1-\frac{\kappa\theta}{\hbar^2}}
      \left\{ \sqrt{\left(m-\frac{1}{2}\right)^2-\left(k-\frac{1}{2}\right)^2}\,  \delta_{m-1,m'} + \right.
      \\ \\ \displaystyle
      \left. + \sqrt{\left(m+\frac{1}{2}\right)^2-\left(k-\frac{1}{2}\right)^2}\,  \delta_{m+1,m'}  \right\}\,,
    \end{array}
\end{equation}
where $k$ is a positive integer or half-integer, $m=k, k+1,k+2, \cdots$ and $l=\pm(2k-1)$. Notice that in this case the Hamiltonian eigenvectors are contained in infinite-dimensional irreducible representations and, for both signs, the argument of the potential is $\theta[2m\mp(2k-1)]\geq \theta$.


\subsection{\underline{$\kappa<0$} :}
From Eqs.\ \eqref{algeb6-neg} and \eqref{trans-sl2r-general}, taking $\tanh \beta=\frac{-1}{\sqrt{1+\frac{|\kappa|\theta}{\hbar^2}}}$, we get
\begin{equation}\label{CP2-3}
    \begin{array}{c}\displaystyle
    e^{ i \beta \mathcal{J}_1} \left(\frac{\mathbf{\hat{X}}^2}{\theta} \right) e^{ -i \beta \mathcal{J}_1}
    = 2 \mathcal{J}_0 - \frac{\hat{L}}{\hbar}\,,
       \\ \\  \displaystyle
    e^{ i \beta \mathcal{J}_1} \left( \frac{\mathbf{\hat{P}}^2}{|\kappa|} \right) e^{ -i \beta \mathcal{J}_1}  =
    2\left(1+\frac{2 \hbar^2}{|\kappa|\theta}\right) \mathcal{J}_0 + \frac{\hat{L}}{\hbar} +
    \frac{2 i \hbar^2}{|\kappa|\theta} \sqrt{1+\frac{|\kappa|\theta}{\hbar^2}} \left(\mathcal{J}_+ - \mathcal{J}_- \right)\,.
    \end{array}
\end{equation}
Taking into account Eq.\ \eqref{l-kappa-negativo}, we conclude that the non-vanishing matrix elements of the Hamiltonian are in this case
\begin{equation}\label{CP3-3}
    \begin{array}{c} \displaystyle
     2 \mu  \left\langle k,m,l \right| e^{ i \beta \mathcal{J}_1} \hat{H} e^{ -i \beta \mathcal{J}_1} \left| k,m',l\right\rangle=
      \\ \\ \displaystyle
      =\left\{2|\kappa|\left(1+\frac{2 \hbar^2}{|\kappa|\theta}\right) m + |\kappa| l +2\mu V\big(\theta(2m-l)\big) \right\} \delta_{m,m'} +
      \\ \\ \displaystyle
      +\frac{2 i \hbar^2}{\theta} \sqrt{1+\frac{|\kappa|\theta}{\hbar^2}}
      \left\{ \sqrt{\left(m-\frac{1}{2}\right)^2-\left(k-\frac{1}{2}\right)^2}\,  \delta_{m-1,m'} - \right.
      \\ \\ \displaystyle
      \left. - \sqrt{\left(m+\frac{1}{2}\right)^2-\left(k-\frac{1}{2}\right)^2}\,  \delta_{m+1,m'}  \right\}\,,
    \end{array}
\end{equation}
where again $k$ is a positive integer or half-integer, $m=k, k+1,k+2, \cdots$ and $l=\pm(2k-1)$.

\medskip

{Therefore, irrespective to the region of parameters, we see that the eigenvalue problem of a nonrelativistic Hamiltonian with central potential can always be reduced to a three-term recursion relation.}


\section{The cylindrical well potential}

In this Section we apply the results summarized in the previous Section to the simple example of a cylindrical well potential on the noncommutative plane, for which
\begin{equation}\label{CWP-1}
    V(\mathbf{\hat{X}}^2):=V_0 \, \Theta(\mathbf{\hat{X}}^2 - A^2)\,,
\end{equation}
where $V_0$ is a constant with units of energy, $A$ is the \emph{radius} of the well and we take the step function as
\begin{equation}\label{CWP-2}
    \Theta(z):=\left\{ \begin{array}{cc}
                         0\,, & z\leq 0\,, \\
                         1\,, & z>0 \,.
                       \end{array}
    \right.
\end{equation}

In each region of the $\kappa$ parameter, the operator $\Theta(\mathbf{\hat{X}}^2 - A^2)$ is defined through its spectral decomposition, as implicitly done for a general central potential in the previous Section.


\subsection{\underline{$0<\kappa<\kappa_c$} :}

From Eq.\ \eqref{CP2-2} we know that we can write
\begin{equation}\label{CWP-13}
    \begin{array}{c} \displaystyle
      e^{ i \alpha \mathcal{J}_2}\,  \Theta(\mathbf{\hat{X}}^2 - A^2) \, e^{- i \alpha \mathcal{J}_2} := \Theta\left(2 \mathcal{J}_0 - \frac{\hat{L}}{\hbar} - \frac{A^2}{\theta}\right) =
      \\ \\ \displaystyle
     =\sum_{k\geq \frac{1}{2}} \  \sum_{l=\pm (2k-1)}  \  \sum_{m=k}^{\infty}  \left|k,m,l \right\rangle
     \Theta\left( 2m-l- \frac{A^2}{\theta}\right)
     \left\langle k,m,l \right|
    \end{array}
\end{equation}
for real $ \alpha=\tanh^{-1} \sqrt{1-\frac{\kappa\theta}{\hbar^2}}$.

If we write $m=k+n$ with $n=0,1,2,\cdots$ and $l=s(2k-1)$ with $s=\pm1$, we see that the step function in the right hand side of Eq.\  \eqref{CWP-13} vanishes for
\begin{equation}\label{def-n0}
   0\leq n\leq \frac{A^2}{2\theta}-\frac{s}{2}+k(s-1)\,.
\end{equation}
Then, we define $ n_{0,k}^{(s)}$ as the maximum integer in this set,
\begin{equation}\label{CWP-14}
    n_{0,k}^{(s)} =\left\lfloor \frac{A^2}{2\theta} - \frac{s}{2}\right\rfloor +k(s-1) \geq 0
\end{equation}
(where $\lfloor x \rfloor$ means the highest integer less than or equal to $x$)
and write
\begin{equation}\label{CWP-15}
    \begin{array}{c} \displaystyle
      e^{ i \alpha \mathcal{J}_2}\,  \Theta(\mathbf{\hat{X}}^2 - A^2) \,  e^{ - i \alpha \mathcal{J}_2}
     = \\ \\ \displaystyle
     =\sum_{k\geq \frac{1}{2}} \  \sum_{s=\pm}  \  \sum_{m=k}^{\infty}
     \left|k,m,s  (2k-1) \right\rangle  \Theta\left(m-k-n_{0,k}^{(s)}\right)
     \left\langle k,m, s  (2k-1) \right|\,.
    \end{array}
\end{equation}
This is the spectral resolution of an operator which, in each irreducible representation, reduces to the orthogonal proyector onto the subspace characterized by the condition $m>k+n_{0,k}^{(\pm)}$. Indeed, the sum on the right hand side excludes those vectors $\left|k,m,\pm(2k-1)  \right\rangle$ for which the \emph{mean square radius} is less than or equal to the square of the cylindrical well radius (See Eq.\ \eqref{CP2-1}), $2m\mp(2k-1)\leq A^2/\theta$.

Writing $\left|\psi_{k,\pm}\right\rangle=  e^{- i \alpha \mathcal{J}_2}  \sum_{m\geq k}^{\infty}C_m \left|k,m,\pm(2k-1) \right\rangle$, we get for the eigenvectors of $\hat{H}$ in the irreducible representation $\left\langle k,s (2k-1)\right\rangle$ the following three-term recursion relation,
\begin{equation}\label{CWP-16}
    \begin{array}{c} \displaystyle
     2 \mu  \left\langle k,m,s (2k-1)\right| e^{ i \alpha \mathcal{J}_2} \left( \hat{H} - E \right) \left|\psi_{k,s}\right\rangle=
      \\ \\ \displaystyle
      =\left\{2\kappa\left(\frac{2 \hbar^2}{\kappa\theta} -1\right) m -s \kappa (2k-1) +2\mu V_0 \Theta\left(m-k-n_{0,k}^{(s)}\right) -2 \mu E\right\} C_m -
      \\ \\ \displaystyle
      -\frac{2 \hbar^2}{\theta} \sqrt{1-\frac{\kappa\theta}{\hbar^2}}
      \left\{ \sqrt{\left(m-\frac{1}{2}\right)^2-\left(k-\frac{1}{2}\right)^2}\,  C_{m-1} + \right.
      \\ \\ \displaystyle
      \left. + \sqrt{\left(m+\frac{1}{2}\right)^2-\left(k-\frac{1}{2}\right)^2}\,  C_{m+1}  \right\}=0\,.
    \end{array}
\end{equation}

\smallskip

Let us first remark that, even though the generators in Eq.\ \eqref{algeb1} do not have a well-defined $\kappa\rightarrow 0$-limit, the recursion in the previous equation does,
\begin{equation}\label{CWP-17}
    \begin{array}{c} \displaystyle
      \left\{2m +\frac{\mu \theta}{\hbar^2} \left[V_0 \Theta\left(m-k-n_{0,k}^{(s)}\right) -E\right]\right\} C_m -
      \\ \\ \displaystyle
      -  \left\{ \sqrt{(m-k)(m+k-1)}\,  C_{m-1}  + \sqrt{(m-k+1)(m+k)}\,  C_{m+1}  \right\}=0\,.
    \end{array}
\end{equation}
{This relation has two linearly independent solutions which have been found in \cite{scholtz}, where the Hilbert space of states of the non-commutative well potential (with $\kappa=0$) has been realized in terms of Hilbert-Schmidt class operators defined on an auxiliary Hilbert space}. Indeed, the general solution for $C_m$ in Eq.\ \eqref{CWP-17}, for $m\leq (k+n_{0,k}^{(s)}+1)$ or $m \geq( k+n_{0,k}^{(s)}+2)$, can be written as a linear combination of the form $C_{k+n}=a \psi_{n,k}(z) + b \phi_{n,k}(z)$, where
\begin{equation}\label{CWP-18}
    \begin{array}{c}\displaystyle
    \psi_{n,k}(z) = \sqrt{\frac{n!(2k-1)!}{(n+2 k-1)!}}\  L_n^{(2 k-1)}(z)= \sqrt{\frac{(n+2 k-1)!}{n!(2k-1)!}}\ M(-n,2k,z)\,,
      \\ \\ \displaystyle
      \phi_{n,k}(z)= \sqrt{\frac{n! (n+2 k-1)!}{(2k-1)!}} \  U(n+1,2-2 k,-z)\,.
    \end{array}
\end{equation}
In these definitions, $ L_n^{(\alpha)}(x)$ is the associated Laguerre polynomial, $M(a,b,x)$ and $U(a,b,x)$ are the Kummer's confluent hypergeometric functions and  $z=$ \linebreak  $\frac{\mu \theta}{\hbar^2} \left[E-V_0 \Theta\left(n-n_{0,k}^{(s)}\right)\right]$ is a constant for $n\leq n_{0,k}^{(s)}$ and for $n> n_{0,k}^{(s)}$.

\smallskip
Let us first consider the case with $m\leq k+n_{0,k}^{(s)}$, for which $z$ in the recursion equation takes the value $z=\frac{\mu \theta}{\hbar^2} \,E>0$. The first step in Eq.\ \eqref{CWP-17} (with $m=k$) reduces to
\begin{equation}\label{CWP-19}
    \left(2k -z\right)C_k - \sqrt{2k}\,  C_{k+1}=0\,.
\end{equation}
It is evident that, in order to get a nontrivial solution, one must take $C_k\neq 0$, and that linearity allows to choose $C_k=1$.

Moreover, it is easy to verify that only $\psi_{n,k}(z)$ satisfy this first equation,
\begin{equation}\label{CWP-20}
    \begin{array}{c}\displaystyle
      \left(2k -z\right)\psi_{0,k}(z) - \sqrt{2k}\,  \psi_{1,k}(z)=0\,,
      \\ \\ \displaystyle
      \left(2k -z\right)\phi_{0,k}(z) - \sqrt{2k}\,  \phi_{1,k}(z)=1\,.
    \end{array}
\end{equation}
Then, since $\psi_{0,k}(z)=1$, we have
\begin{equation}\label{CWP-21}
    C_{k+n}=\psi_{n,k}(z)=\sqrt{\frac{n!(2k-1)!}{(n+2 k-1)!}}\  L_n^{(2 k-1)}\left(\frac{\mu\theta E}{\hbar^2} \right)\,,\quad
    0\leq n \leq n_{0,k}^{(s)}\,.
\end{equation}

\smallskip
If we are looking for bound states, a normalizability condition must be imposed since the coefficients must satisfy the Bessel's inequality,
\begin{equation}
    \sum_{n=0}^{\infty} \left|C_{k+n} \right|^2<\infty\,.
\end{equation}
So, the behavior of $\psi_{n,k}(z)$ and $\phi_{n,k}(z)$ for large $n$ must be taken into account when considering the case $m \geq( k+n_{0,k}^{(s)}+1)$, where $z=\frac{\mu \theta}{\hbar^2} \left[E-V_0\right]$.

For the two factors in $\psi_{n,k}(z)$ we have \cite{A-S,Math6}
\begin{equation}\label{CWP-22}
    \begin{array}{c} \displaystyle
      \sqrt{\frac{n!(2k-1)!}{(n+2 k-1)!}} \approx \sqrt{(2k-1)!} \ n^{\frac{1}{2}-k}  \left(1+O(n^{-1}) \right)\,,
      \\ \\ \displaystyle
       L_n^{(2 k-1)}(z) \approx \frac{e^{z/2} n^{k-\frac{3}{4}} z^{\frac{1}{4}-k}}{\sqrt{\pi }}
       \left\{ \sin \left[\pi  \left(k+\frac{1}{4}\right)-2   \sqrt{z (k+n)}\right]  + O\left(n^{-1/2}\right) \right\}\,.
    \end{array}
\end{equation}
Then, for $E<V_0$ ($ z<0$), this behavior leads to exponentially growing coefficients, $C_{k+n}\sim n^{-1/4}\, e^{2\sqrt{n} \sqrt{\frac{\mu \theta}{\hbar^2} \left[V_0-E\right]}}$, and must be discarded. For scattering states ($E>V_0 \Rightarrow z>0$), $\psi_{n,k}(z)$ behaves as $n^{-1/4}$ times an oscillatory function, giving rise to bounded coefficients (which do not lead to normalizable solutions).

On the other hand, from Eq.\ (9) in page 279 of Ref.\ \cite{Bateman1} we have that
\begin{equation}\label{Bateman-U-1}
    \begin{array}{c} \displaystyle
      U(a,b,x)=
      \\ \\ \displaystyle
      \frac{(1+\imath)}{2}\,\gamma^{ -\frac{1}{4}+\gamma}
    x^{ \frac{1}{4}-\frac{b}{2}}
    \exp{\left[\displaystyle -\gamma+\frac{x}{2} - \imath \gamma \pi +2\imath \sqrt{\gamma x}\right]} \left\{
    1+O\left( \frac{1}{\sqrt{|\gamma|}} \right) \right\}\,,
    \end{array}
\end{equation}
for $a\rightarrow \infty$, with $\gamma=\frac{b}{2}-a$, $ \varepsilon < {\rm arg} \, \gamma < 3\pi- \varepsilon <$, and $\left|{\rm arg} \, x - {\rm arg} \, \gamma \right|\leq \pi$. Moreover,  the Stirling's approximation gives
\begin{equation}\label{}
    \sqrt{\frac{n!(n+2 k-1)!}{(2k-1)!}} \approx  \frac{\sqrt{2 \pi}}{\sqrt{(2k-1)!}}\ e^{-n} n^{n+k}  \left(1+O(n^{-1}) \right)
\end{equation}
Then,
\begin{equation}\label{CWP-23}
    \phi_{n,k}(z) \approx \frac{\sqrt{\pi } e^{\displaystyle -z/2} (-z)^{k-\frac{3}{4}}}{\sqrt{(2 k-1)!}}\  {n^{-1/4}}{e^{\displaystyle -2 \sqrt{n} \sqrt{-z}}} \left( 1+ O(n^{-1/2}) \right)
\end{equation}
which, for bound states ($z<0$), has an exponentially vanishing behavior while, for scattering states ($z>0$) behaves as $n^{-1/4}$ times an oscillatory function.

\smallskip

Therefore, to construct (normalizable) bound states we must write $C_{k+n}=N \, \phi_{n,k}(z)$ for $n \geq( n_{0,k}^{(s)}+1)$. The proportionality constant $N$ and the energy eigenvalues are determined by two \emph{matching conditions} followed from Eq.\ \eqref{CWP-17} with $m=k+n_{0,k}^{(s)}$ and $m=k+n_{0,k}^{(s)}+1$, equations in which both kinds of coefficients apear:
\begin{equation}\label{CWP-24}
    \begin{array}{c} 
          \left\{2\left(k+n_{0,k}^{(s)}\right) -\frac{\mu \theta}{\hbar^2} E\right\} C_{k+n_{0,k}^{(s)}} -
          \left\{ \sqrt{n_{0,k}^{(s)}\left(2k-1+n_{0,k}^{(s)}\right)}\,  C_{k+n_{0,k}^{(s)}-1}  + \right.
      \\ \\ 
      \left. + \sqrt{\left(n_{0,k}^{(s)}+1\right)\left(2k+n_{0,k}^{(s)}\right)}\,  C_{k+n_{0,k}^{(s)}+1}  \right\}=0\,,
       \\ \\ 
             \left\{2\left(k+n_{0,k}^{(s)}+1\right) +\frac{\mu \theta}{\hbar^2} \left[V_0-E\right]\right\} C_{k+n_{0,k}^{(s)}+1} -
             \left\{ \sqrt{\left(n_{0,k}^{(s)}+1\right)\left(2k+n_{0,k}^{(s)}\right)}\,  C_{k+n_{0,k}^{(s)}}  + \right.
      \\ \\ 
      \left. + \sqrt{\left(n_{0,k}^{(s)}+2\right)\left(2k+n_{0,k}^{(s)}+1\right)}\,  C_{k+n_{0,k}^{(s)}+2}  \right\}=0\,,
    \end{array}
\end{equation}
where
\begin{equation}\label{CWP-25}
    \begin{array}{c} 
    C_{k+n_{0,k}^{(s)}-1}=\psi_{n_{0,k}^{(s)}-1,k}\left(\frac{\mu \theta E}{\hbar^2}\right) \,, \quad
    C_{k+n_{0,k}^{(s)}}= \psi_{n_{0,k}^{(s)},k}\left(\frac{\mu \theta E}{\hbar^2}\right) \,,
      \\ \\ 
          C_{k+n_{0,k}^{(s)}+1}= N \, \phi_{n_{0,k}^{(s)}+1,k}\left(\frac{\mu \theta (E-V_0)}{\hbar^2}\right) \,, \ \,
          C_{k+n_{0,k}^{(s)}+2} = N \, \phi_{n_{0,k}^{(s)}+2,k}\left(\frac{\mu \theta (E-V_0)}{\hbar^2}\right)  \,.
    \end{array}
\end{equation}

\smallskip

For scattering states, with $z=\frac{\mu \theta}{\hbar^2} \left[E-V_0\right]>0$, the coefficients for $n\geq n_{0,k}^{(s)}+1$ must be taken as a linear combination of the form $C_{k+n}=N_1 \, \psi_{n,k}(z)+ N_2 \, \phi_{n,k}(z)$, since both functions behave similarly. In this case the constants $N_{1,2}$ are determined by the matching conditions in Eq.\ \eqref{CWP-24} as functions of $E$.

\smallskip

{This is in agreement with the results in \cite{scholtz}}. In particular, for the infinite cylindrical well potential $(V_0 \rightarrow \infty)$, from  Eq.\ \eqref{CWP-17} one concludes that $C_m=O\left({V_0}^{-1}\right)$ for $m\geq k+n_{0,k}^{(s)}+1$. Then, the first matching condition in Eq.\ \eqref{CWP-24} requires that
\begin{equation}\label{CWP-24-1}
    \left\{2\left(k+n_{0,k}^{(s)}\right) -\frac{\mu \theta}{\hbar^2} E\right\} C_{k+n_{0,k}^{(s)}} -
          \sqrt{n_{0,k}^{(s)}\left(2k-1+n_{0,k}^{(s)}\right)}\,  C_{k+n_{0,k}^{(s)}-1} \begin{array}{c}
                                                                                           \\
                                                                                         \longrightarrow \\
                                                                                         {V_0 \rightarrow \infty}
                                                                                       \end{array} 0\,.
\end{equation}
In this limit, this imposes a condition which determines the spectrum,
\begin{equation}\label{CWP-24-2}
   \begin{array}{c} \displaystyle
      \left\{2\left(k+n_{0,k}^{(s)}\right) -\frac{\mu \theta}{\hbar^2} E\right\} \psi_{n_{0,k}^{(s)},k}\left(\frac{\mu \theta E}{\hbar^2}\right)-
          \sqrt{n_{0,k}^{(s)}\left(2k-1+n_{0,k}^{(s)}\right)}\, \psi_{n_{0,k}^{(s)}-1,k}\left(\frac{\mu \theta E}{\hbar^2}\right)
          \\ \\ \displaystyle
        = \sqrt{\left(n_{0,k}^{(s)}+1\right)\left(2k+n_{0,k}^{(s)}\right)}\,  \psi_{n_{0,k}^{(s)}+1,k} = 0\,,
   \end{array}
\end{equation}
where the recursion relation satisfied by the $\psi_{n,k}(z)$ has been employed.

Therefore, from Eq.\ \eqref{CWP-18} we conclude that the energy eigenvalues are determined by the $n_{0,k}^{(s)}+1$ roots of the associated Laguerre polynomial $ L_{n_{0,k}^{(s)}+1}^{(2 k-1)}\left(z\right)$,
\begin{equation}\label{CWP-26}
    L_{n_{0,k}^{(s)}+1}^{(2 k-1)}\left(z_r\right)=0\,, \quad {\rm with} \ z_r= \frac{\mu\theta }{\hbar^2}\, E_r\,,
    \quad r=1,2,\cdots,n_{0,k}^{(s)}+1 \,.
\end{equation}

Notice that, for a given $k$ and $s=+1$ ($l\geq 0$), the number of linearly independent eigenvectors of the Hamiltonian in this irreducible representation is
\begin{equation}\label{CWP-27}
    n_{0,k}^{(+)}+1 =\left\lfloor \frac{A^2}{2\theta} - \frac{1}{2}\right\rfloor +1  \,,
\end{equation}
which is finite and independent of $k$ (and at least one - we assume that $\frac{A^2}{\theta}\geq 1$). On the other hand, for $s=-1$ ($l<0$),  the number of linearly independent eigenvectors is
\begin{equation}\label{CWP-28}
     n_{0,k}^{(-)} +1 =\left\lfloor \frac{A^2}{2\theta} +\frac{1}{2}\right\rfloor -(2k-1)
\end{equation}
which, for a given $A$, decreases linearly with $k$. In particular, there are no nontrivial solutions for
\begin{equation}\label{CWP-29}
     k>\frac{1}{2}\left\lfloor \frac{A^2}{2\theta} +\frac{1}{2}\right\rfloor + \frac{1}{2}\,.
\end{equation}
Therefore, for each non-negative $l=2k-1$, the Hamiltonian of the infinite cylindrical well on the noncommutative plane has a finite number of eigenvectors which is independent of $k=\frac{1}{2},1,\frac{3}{2},2,\cdots$, while for negative $l=-(2k-1)$, there is only a finite number of irreducible representation where the Hamiltonian has eigenvectors, and this number decreases linearly with $k$.  Notice that in the $\theta \rightarrow 0$ limit one recovers an infinite number of states for each $l\in \mathbb{Z}$, as corresponds to this system on the usual commutative plane.

{As previously stated, these results are in complete agreement with those in \cite{scholtz}, where a completely different formulation of the coordinate noncommutativity has been implemented.}

\medskip

Let us now return to the case with $0<\kappa<\kappa_c$, Eq.\ \eqref{CWP-16}. One possibility consists in the expansion in powers of $\kappa$ of the different elements and perturbatively determine the corrections to the coefficients $C_m$ and the energy eigenvalues.

Alternatively, one can transform the recursion equations into an ordinary differential equation whose solution is the generating function of the coefficients. Indeed, let us write Eq.\ \eqref{CWP-16} as
\begin{equation}\label{CWP-30}
    (a n-w) {C}_{k+n}-\sqrt{n (2 k+n-1)}  {C}_{k+n-1}-\sqrt{(n+1) (2 k+n)} {C}_{k+n+1}=0\,,
\end{equation}
where
\begin{equation}\label{CWP-31}
    \begin{array}{c}\displaystyle
      a=b+\frac{1}{b} >2\,, \quad b=\sqrt{1-\frac{\theta  \kappa }{\hbar^2}} <1 \,,
      \\ \\ \displaystyle
      w = \frac{\theta  \mu }{\hbar^2 \sqrt{1-\frac{\theta\kappa }{\hbar^2}}} \left[ {E}- {V_0} \Theta\left(n-{n_{0,k}^{(s)}}\right)\right]
      -    \frac{2 k}{\sqrt{1-\frac{\theta  \kappa  }{\hbar ^2}}} +
      \frac{\theta \kappa  \left(2   k (s+1)-  s\right)}{2 \hbar ^2 \sqrt{1-\frac{\theta  \kappa
   }{\hbar ^2}}} \,,
    \end{array}
\end{equation}
and write
\begin{equation}\label{CWP-32}
    {C}_{k+n}=\frac{\sqrt{(2 k-1)!} \sqrt{n!} }{\sqrt{(2 k+n-1)!}} \,  {K}_{n}\,.
\end{equation}
Then, we get for the coefficients $K_n$
\begin{equation}\label{CWP-33}
    (w-a n){K}_{n} +(2 k+n-1) {K}_{n-1}+(n+1)
   {K}_{n+1}=0
\end{equation}
for $n\geq1$ and
\begin{equation}\label{CWP-34}
    w {K}_{0} + {K}_{1}=0
\end{equation}
for $n=0$.

We now define
\begin{equation}\label{CWP-35}
    f(x):= \sum_{n=0}^\infty K_n x^n\,.
\end{equation}
By multiplying both sides of Eq.\ \eqref{CWP-33} by $x^n$, rearranging  terms and taking $w$ as a constant we get for $f(x)$ the differential equation
\begin{equation}\label{CWP-36}
    \left(1-a x+x^2\right) f'(x)+(2 k x+w) f(x) =0\,,
\end{equation}
whose solution is
\begin{equation}\label{CWP-37}
    \begin{array}{c}\displaystyle
       f(x)
   =(1-b x)^{-\left(\frac{b w+2 k}{1-b^2}\right)}
   \left(1-\frac{x}{b}\right)^{\left(\frac{b w+2 k}{1-b^2}\right)-2k}
      \,.
    \end{array}
\end{equation}
The coefficients are finally given by
\begin{equation}\label{CWP-38}
    K_n=\frac{1}{n!}\, f^{(n)}(0)\,.
\end{equation}
In particular, $f(0)=1$, $f'(0)=-w$ and Eq.\ \eqref{CWP-34} is satisfied. One can easily verify that $K_n$ so obtained satisfy Eq.\ \eqref{CWP-33} for $n\geq 1$. So, one can write
\begin{equation}\label{CWP-39}
    C_{n+k}=\Psi_{n,k}(w):=\frac{\sqrt{(2 k-1)!}  }{\sqrt{(2 k+n-1)!}\sqrt{n!}} \,  f^{(n)}(0)\,, \quad {\rm for}\ n=0,1,2,\cdots, {{n_{0,k}^{(s)}}}\,,
\end{equation}
where $\Psi_{n,k}(w)$ are some polynomials in $w$ of degree $n$ with $b$-dependent coefficients, which can be expressed in terms of Jacobi Polynomials\footnote{The Jacobi polynomials are related with the Gaussian  hypergeometric function through
\begin{equation}\label{Jacobi-Hyper-1}
    P_n^{(\alpha,\beta)}(y)=\frac{(\alpha +1)_n }{n!}
    \   _2F_1\left(-n,n+\alpha +\beta +1;\alpha +1;\frac{1-y}{2}\right)
    \,,
\end{equation}
where $(\alpha)_n:= \Gamma(\alpha+n)/\Gamma(\alpha)$ is the Pochhammer symbol.
Then, the recursion satisfied by these polynomials can be mapped onto the relation
\begin{equation}\label{Jacobi-Hyper-2}
    \begin{array}{c} \displaystyle
      2 (\alpha )_{n+1} \,    _2F_1\left(-n-1,n+\alpha +\beta    +1;\alpha ;\frac{1-y}{2}\right)-
   \\ \\ \displaystyle
    (\alpha +1)_n (\alpha -\beta +2 n (y-1)+y (\alpha +\beta +1)-1) \,  _2F_1\left(-n,n+\alpha +\beta +1;\alpha +1;\frac{1-y}{2}\right)-
   \\ \\ \displaystyle
    n (y-1)   (\beta +n) (\alpha +2)_{n-1} \, _2F_1\left(1-n,n+\alpha +\beta +1;\alpha +2;\frac{1-y}{2}\right)=0\,,
    \end{array}
\end{equation}
satisfied by the hypergeometric function \cite{A-S,MathWorld}.} \cite{A-S,MathWorld} $P_n^{(\alpha,\beta)}(t)$ as
\begin{equation}\label{Jacobi}
    \Psi_{n,k}(w):=\frac{\sqrt{(2 k-1)!} \sqrt{n!} }{\sqrt{(2 k+n-1)!}} \,
    \left(-\frac{1}{b}\right)^n P_n^{\left(\frac{b w+2
   k}{1-b^2}-2 k-n,2 k-1\right)}\left(1-2 b^2\right)\,.
\end{equation}
{Notice that $f(x)$ reduces to the generating function of associated Laguerre polynomials  $L_n^{(2k-1)}(w+2k)$  \cite{A-S} in the $b \rightarrow 1$ ($\kappa \rightarrow 0$) limit: 
\begin{equation}\label{CWP-40}
    f_0(x):=\lim_{b \rightarrow 1} f(x)= \frac{e^{\displaystyle -\frac{x(w+2k)}{1-x}}}{\left(1-x\right)^{2k}}\,,
\end{equation}
with
\begin{equation}\label{Laguerre-1}
    \frac{1}{n!}\, f_0^{(n)}(0)= L_n^{2k-1}(w+2k)\,.
\end{equation}}

A second solution to the recursion system is given by the solution of the \emph{inhomogeneous} differential equation (with $w$ shifted by the $V_0$-dependent term)
\begin{equation}\label{CWP-41}
    \left(1-a x+x^2\right) g'(x)+(2 k x+w) g(x) =1\,,
\end{equation}
given by
\begin{equation}\label{CWP-42}
    g(x)=
    \frac{b }{b w+2 k} \ \,  _2F_1\left(1,2 k;1+\frac{2 k+b    w}{1-b^2};\frac{1-b x}{1-b^2}\right)\,.
\end{equation}
Indeed, by construction, the coefficients
\begin{equation}\label{CWP-43}
    K'_n=\frac{1}{n!}\, g^{(n)}(0)
\end{equation}
satisfy the relations in Eq.\ \eqref{CWP-33} for $n\geq 1$, but not the first one with $n=0$: It is clear from Eq.\ \eqref{CWP-41} that
\begin{equation}\label{CWP-44}
    K'_1 + w K'_0 =1\,.
\end{equation}

These coefficients can be employed to express the $C_{n+k}$ for $n \geq {{n_{0,k}^{(s)}}}+1$ and, as before, Eq.\ \eqref{CWP-30} for $n={{n_{0,k}^{(s)}}}, {{n_{0,k}^{(s)}}}+1$ provides the two matching conditions between both behaviors necessary to determine the spectrum.

\smallskip

For concreteness, let us consider the infinite cylindrical well.  From Eqs.\ \eqref{CWP-30} and \eqref{CWP-31} we conclude that  $C_{k+n} =O\left({V_0}^{-1}\right)$ for $n\geq {n_{0,k}^{(s)}}+1$. Then Eq.\  \eqref{CWP-30} implies that
\begin{equation}\label{CWP-44-1}
    \left(a {n_{0,k}^{(s)}}-w\right) {C}_{k+{n_{0,k}^{(s)}}}-\sqrt{{n_{0,k}^{(s)}} \left(2 k+{n_{0,k}^{(s)}}-1\right)}  {C}_{k+{n_{0,k}^{(s)}}-1}
    \begin{array}{c}
        \\
      \longrightarrow \\
      V_0 \rightarrow \infty
    \end{array}
    0\,.
\end{equation}
In terms of the polynomials $\Psi_{n,k}(w)$, this condition reads as
\begin{equation}\label{CWP-44-1-1}
   \begin{array}{c} \displaystyle
      \left(a {n_{0,k}^{(s)}}-w\right) \Psi_{{n_{0,k}^{(s)}},k}(w)-\sqrt{{n_{0,k}^{(s)}} \left(2 k+{n_{0,k}^{(s)}}-1\right)}  \Psi_{{n_{0,k}^{(s)}}-1,k}(w)=
    \\ \\ \displaystyle
     = \sqrt{\left({{n_{0,k}^{(s)}}}+1\right) \left(2 k+{{n_{0,k}^{(s)}}}\right)}\Psi_{{n_{0,k}^{(s)}}+1,k}(w) = 0\,,
   \end{array}
\end{equation}
where the recursion relation satisfied by the $\Psi_{n,k}(w)$ has been used.

Therefore, the spectrum is determined by the  ${{n_{0,k}^{(s)}}}+1$ roots of the polynomial ${\Psi}_{{n_{0,k}^{(s)}}+1,k}(w)$,
\begin{equation}\label{CWP-45}
    \Psi_{{n_{0,k}^{(s)}}+1,k}(w_r) \sim P_{n_{0,k}^{(s)}+1}^{\left(\frac{b w_r+2
   k}{1-b^2}-2 k-{n_{0,k}^{(s)}}-1,2 k-1\right)}\left(1-2 b^2\right) = 0
\end{equation}
for $r=1,2,\cdots,{{n_{0,k}^{(s)}}}+1$, as
\begin{equation}\label{CWP-46}
    {E}_r=\frac{\hbar^2}{\theta  \mu }
    \left\{      w_r \sqrt{1-\frac{\theta  \kappa}{\hbar ^2}}+2 k
    - \frac{\theta  \kappa}{\hbar ^2} \left( k (1+s)-\frac{s}{2}\right)
    \right\}\,.
\end{equation}
The analysis of the number of linearly independent Hamiltonian eigenvectors in each irreducible representation $\langle k,l\rangle$ goes as in the $\kappa=0$ case previously discussed and will not be repeated here. In particular, the spectrum smoothly converges to that of Eq.\ \eqref{CWP-26} when $\kappa \rightarrow 0^+$.


\subsection{\underline{$\kappa<0$} :}

From Eq.\ \eqref{CP2-3} we get the operator
\begin{equation}\label{CWP-47}
    \begin{array}{c} \displaystyle
        e^{ i \beta \mathcal{J}_1} \,  \Theta(\mathbf{\hat{X}}^2 - A^2) \,  e^{ -i \beta \mathcal{J}_1} := \Theta\left(2 \mathcal{J}_0 - \frac{\hat{L}}{\hbar} - \frac{A^2}{\theta}\right) =
      \\ \\ \displaystyle
    =\sum_{k\geq \frac{1}{2}} \  \sum_{s=\pm 1}  \  \sum_{m=k}^{\infty}
     \left|k,m,s  (2k-1) \right\rangle  \Theta\left(m-k-n_{0,k}^{(s)}\right)
     \left\langle k,m, s  (2k-1) \right|
    \end{array}
\end{equation}
where $ \beta=- \coth^{-1} {\sqrt{1+\frac{|\kappa|\theta}{\hbar^2}}}\,$ and $n_{0,k}^{(s)}$ given in Eq.\ \eqref{CWP-14},
which has the interpretation as the spectral resolution of the orthogonal proyector onto subspaces characterized by the condition $m>k+n_{0,k}^{(\pm)}$ (states with {mean square radius} grater than the cylindrical well squared radius).

We write $\left|\psi_{k,\pm}\right\rangle=  e^{- i \beta \mathcal{J}_1}  \sum_{m\geq k}^{\infty}C_m \left|k,m,\pm(2k-1) \right\rangle$ to get from Eq.\ \eqref{CP3-3}
\begin{equation}\label{CWP-48}
    \begin{array}{c} \displaystyle
     2 \mu  \left\langle k,m,s (2k-1)\right|  e^{ i \beta \mathcal{J}_1}  \left( \hat{H} - E \right) \left|\psi_{k,s}\right\rangle=
           \\ \\ \displaystyle
      =\left\{2|\kappa|\left(1+\frac{2 \hbar^2}{|\kappa|\theta}\right) m +s |\kappa| (2k-1) +2\mu V_0 \Theta\left(m-k-n_{0,k}^{(s)}\right) -2 \mu E\right\} C_m +
      \\ \\ \displaystyle
      +\frac{2 i \hbar^2}{\theta} \sqrt{1+\frac{|\kappa|\theta}{\hbar^2}}
      \left\{ \sqrt{\left(m-\frac{1}{2}\right)^2-\left(k-\frac{1}{2}\right)^2}\,  C_{m-1} - \right.
      \\ \\ \displaystyle
      \left. - \sqrt{\left(m+\frac{1}{2}\right)^2-\left(k-\frac{1}{2}\right)^2}\,  C_{m+1}  \right\}=0\,,
    \end{array}
\end{equation}
{which can be written as the three-term recursion
\begin{equation}\label{CWP-49}
    -i(a n-w) {C}_{k+n}+\sqrt{n (2 k+n-1)}  {C}_{k+n-1}-\sqrt{(n+1) (2 k+n)} {C}_{k+n+1}=0\,,
\end{equation}
with the same definitions of $a,b$ and $w$ as in Eq.\ \eqref{CWP-31}, with $\kappa=-|\kappa|$ and $b>1$ this time.}

{If we write
\begin{equation}\label{CWP-51}
    {C}_{k+n}=(-i)^n \frac{\sqrt{(2 k-1)!} \sqrt{n!} }{\sqrt{(2 k+n-1)!}} \,  {K}_{n}\,.
\end{equation}
we get for $K_n$ the same recursion as in Eqs.\ \eqref{CWP-33} and \eqref{CWP-34}. So, the generating function in this case is also given in Eq.\ \eqref{CWP-37}}
%
{and these coefficients are obtained as in Eq.\ \eqref{CWP-38}.}

{Therefore, we finally  get for the coefficients in Eq.\ \eqref{CWP-49} the polynomial in $w$ of degree $n$}
{\begin{equation}\label{Jacobi-2}
\begin{array}{c} \displaystyle
  C_{k+n}=(-i)^n {\Psi}_{n,k}(w)=
  \\ \\ \displaystyle
   =i^n \frac{\sqrt{(2 k-1)!} \sqrt{n!} }{\sqrt{(2 k+n-1)!}} \,
    \left(\frac{1}{b}\right)^n P_n^{\left(-\frac{2 k+b w}{b^2-1}-2 k-n,2 k-1\right)}\left(1-2 b^2\right)\,,
\end{array}
\end{equation}
for $n=0,1,2,\cdots, n_{0,k}^{(s)}.$}


{For larger $n$, we need a second solution for the coefficients, which can be derived from $g(x)$ in Eq.\ \eqref{CWP-42} as in Eq.\ \eqref{CWP-43}. Arguments about the matching conditions similar to those developed in the previous Section apply here and will not be repeated.}


\smallskip

{Rather, let us consider again the infinite cylindrical well in this region of parameters.
By an entirely similar argument as in the previous Section,
one concludes that the energy eigenvalues of the system are again determined by the roots $w_r$ of the polynomials ${\Psi}_{{n_{0,k}^{(s)}}+1,k}(w)$}
{as
\begin{equation}\label{CWP-46-kappa-neg}
    {E}_r=\frac{\hbar^2}{\theta  \mu }
    \left\{ {w_r \sqrt{1+\frac{\theta  |\kappa| }{\hbar ^2}}} +2 k
    + \frac{\theta  |\kappa|}{\hbar^2}  \left( k (s+1)-\frac{s}{2}\right)
    \right\}
\end{equation}
for  $r=1,2,\cdots,{{n_{0,k}^{(s)}}+1}$ (which coincides with the expression of the eigenvalues in Eq.\ \eqref{CWP-46} with $\kappa<0$).}

From here on, the analysis continues as in the previous Sections: There is a finite number $n_{0,k}^{(s)}+1$ of linearly independent Hamiltonian eigenvectors in each irreducible representation $\langle k,l\rangle$, the same for all positive $l=2k-1$ and a linearly decreasing number for negative $l=-(2k-1)$. The spectrum smoothly converges to that of Eq.\ \eqref{CWP-26} when $\kappa \rightarrow 0^-$.



\subsection{\underline{$\kappa>\kappa_c$} :}

In this region of parameters, from \eqref{CP2-1} we get
\begin{equation}\label{CWP-3}
    \begin{array}{c} \displaystyle
      e^{ i \varphi J_2}\,  \Theta(\mathbf{\hat{X}}^2 - A^2) \, e^{- i \varphi J_2} := \Theta\left(2 J_3 - \frac{\hat{L}}{\hbar} - \frac{A^2}{\theta}\right) =
      \\ \\ \displaystyle
     =\sum_{j} \sum_{m=-j}^{j} \left|j,m,l \right\rangle
     \Theta\left( 2m+(2j+1)- \frac{A^2}{\theta}\right)
     \left\langle j,m,l \right|
    \end{array}
\end{equation}
where the sum extends to all the irreducible representations $\left\langle j, l\right\rangle$ of $SU(2)\otimes SO(2)$ with $l= -(2j+1)$.

Let us define the integer or half-integer $m_{0,j}$ by
\begin{equation}\label{CWP-4}
    2 m_{0,j} =\left\lfloor \frac{A^2}{\theta}  \right\rfloor- (2j+1) \geq -(2j+1)  \,.
\end{equation}
Then,
\begin{equation}\label{CWP-5}
    \begin{array}{c} \displaystyle
      e^{ i \varphi J_2}\,  \Theta(\mathbf{\hat{X}}^2 - A^2) \, e^{- i \varphi J_2}
     =\sum_{j} \sum_{m>-j}^{j} \left|j,m,l \right\rangle  \Theta\left(m-m_{0,j}\right)
     \left\langle j,m,l \right|\,,
    \end{array}
\end{equation}
which is an orthogonal proyector on a (finite dimensional - possibly trivial) subspace of each irreducible representation. Indeed, the sum on the right hand side excludes those vectors $\left|j,m,l  \right\rangle$ for which the \emph{mean square radius} is less that or equal to the square of the cylindrical well radius (See Eq.\ \eqref{CP2-1}), $2m+2j+1\leq A^2/\theta$.

Writing $\left|\psi_{j}\right\rangle= e^{- i \varphi J_2} \sum_{m=-j}^{j}C_m \left|j,m,l \right\rangle$, we get for the eigenvectors of $\hat{H}$ in the irreducible representation $\left\langle j,l\right\rangle$
\begin{equation}\label{CWP-6}
    \begin{array}{c} \displaystyle
      2 \mu \left\langle j,m,l\right| e^{ i \varphi J_2}\left( \hat{H} - E \right)  \left| \psi_j \right\rangle=
      \\ \\ \displaystyle
    =\left\{-2\kappa\left(1-\frac{2 \hbar^2}{\kappa\theta} \right) m + \kappa (2j+1) +2\mu V_0
    \Theta\big(m-{m_{0,j}}\big)-2\mu E \right\} C_m -
      \\ \\ \displaystyle
      -\frac{2 \hbar^2}{\theta} \sqrt{\frac{\kappa\theta}{\hbar^2}-1}
      \left\{ \sqrt{(j-m+1)(j+m)}\, C_{m-1} +
      \sqrt{(j+m+1)(j-m)}\,  C_{m+1}  \right\}=0\,,
    \end{array}
\end{equation}
which is a linear three-term recursion relation for the Fourier coefficients $C_m$ with $-j \leq m \leq j$.

Notice that the first equation of this recursion ($m=-j$) reduces to
\begin{equation}\label{CWP-7}
   \begin{array}{c} \displaystyle
      \left\{2\kappa\left(1-\frac{2 \hbar^2}{\kappa\theta} \right) j + \kappa (2j+1) +2\mu V_0
    \Theta\big(-j-{m_{0,j}}\big)-2\mu E \right\} C_{-j} \, -
      \\ \\ \displaystyle
      -\frac{2 \hbar^2}{\theta} \sqrt{\frac{\kappa\theta}{\hbar^2}-1}
      \sqrt{2j}\,  C_{1-j}  =0\,.
   \end{array}
\end{equation}
Then, if $C_{-j}=0 \Rightarrow C_{1-j}=0$, and we get the trivial solution. Therefore, $C_{-j} \neq 0$. Linearity allows to take $C_{-j} =1$, for example, and Eq.\ \eqref{CWP-7} determines $C_{-j+1}$ as a linear function of $E$.

One then applies Eq.\ \eqref{CWP-6} $(2j-1)$ times to recursively construct the remaining $(2j-1)$ Fourier coefficients as polynomials of $E$ of growing degree. In so doing one gets $C_j$ as a polynomial of degree $2j$.
Finally, the last equation ($m=j$),
\begin{equation}\label{CWP-8}
   \begin{array}{c} \displaystyle
      \left\{-2\kappa\left(1-\frac{2 \hbar^2}{\kappa\theta} \right) j + \kappa (2j+1) +2\mu V_0
    \Theta\big(j-{m_{0,j}}\big)-2\mu E \right\} C_j -
      \\ \\ \displaystyle
      -\frac{2 \hbar^2}{\theta} \sqrt{\frac{\kappa\theta}{\hbar^2}-1} \sqrt{2j}\, C_{j-1} =0\,,
   \end{array}
\end{equation}
determines the eigenvalues as the roots of a $(2j+1)$-degree polynomial.

\smallskip

For example, for $j=0$ we have just one equation,
\begin{equation}\label{CWP-9}
    2\mu E =\kappa  +2\mu V_0 \Theta\left(-\frac{1}{2}\left\lfloor \frac{A^2}{\theta} - 1 \right\rfloor\right)\,,
\end{equation}
which determines the energy of the unique state in this irreducible representation. Notice that the particle in this state feels the potential only if the radius of the well is $A<\sqrt{\theta}$.

\smallskip

For $j=1/2$, with $C_{-1/2}=1$, we have
\begin{equation}\label{CWP-10}
    C_{1/2}=\frac{\theta\left[2    \mu E -3   \kappa - 2   \mu
   V_0 \Theta \left(-\frac{1}{2} -m_{0,{1}/{2}}\right)\right]+2 \hbar ^2}{
   2 \hbar ^2\sqrt{\frac{\theta  \kappa }{\hbar ^2}-1}}
\end{equation}
where $m_{0,{1}/{2}}=\frac{1}{2}\left\lfloor \frac{A^2}{\theta}-2 \right\rfloor$ and, from the condition
\begin{equation}\label{CWP-11}
   \begin{array}{c} \displaystyle
      \left\{-\kappa\left(1-\frac{2 \hbar^2}{\kappa\theta} \right)  +2 \kappa  +2\mu V_0
    \Theta\left(\frac{1}{2}-{m_{0,{1}/{2}}}\right)-2\mu E \right\} C_{{1}/{2}} -
      \\ \\ \displaystyle
      -\frac{2 \hbar^2}{\theta} \sqrt{\frac{\kappa\theta}{\hbar^2}-1}\, C_{-{1}/{2}} =0\,,
   \end{array}
\end{equation}
we get a degree-2 polynomial whose roots,
\begin{equation}\label{CWP-12}
    \begin{array}{c} 
      E_\pm=
      \frac{\kappa  }{\mu }
      \left[1\pm \frac{1}{2} \sqrt{\frac{\mu
   {V_0}  }{\kappa }\left(\frac{4 \hbar ^2}{\theta
   \kappa }+\frac{\mu  {V_0}}{\kappa
   }-2\right)\left[\Theta
   \left(\frac{1}{2}-{m_{0,1/2}}\right)-\Theta
   \left(-\frac{1}{2}-{m_{0,1/2}}
   \right)\right]+1}\right]
   \\ \\ 
   + \frac{1}{2} {V_0} \left[\Theta
   \left(-\frac{1}{2}-{m_{0,1/2}}
   \right)+\Theta
   \left(\frac{1}{2}-{m_{0,1/2}}\right)\right] \,,
    \end{array}
\end{equation}
correspond to the two eigenvectors of $\hat{H}$ in this irreducible representation, determined by the coefficient in Eq.\ \eqref{CWP-10}.
Notice also that, in this representation, the particle can feel the potential only if $A<\sqrt{3\theta}$.

\smallskip

For the general case, one can introduce the generating function of the coefficients and transform the recursion into a differential equation. Let us call
\begin{equation}\label{KM-1}
    \begin{array}{c} \displaystyle
      a=b- \frac{1}{b}\,, \quad {\rm with} \ b=\sqrt{\frac{\theta  \kappa }{\hbar ^2}-1}\,,
      \\ \\ \displaystyle
      z=\frac{\theta  \mu  \left[E-{V_0} \Theta\left(m-{m_0}\right)\right] }{\hbar ^2 \sqrt{\frac{\theta
    \kappa }{\hbar ^2}-1}}  -  \frac{\theta  \kappa (2 j+1)  }{2 \hbar ^2
   \sqrt{\frac{\theta  \kappa }{\hbar ^2}-1}}\,,
    \end{array}
\end{equation}
and write
\begin{equation}\label{KM-2}
    C_m=\frac{\sqrt{(j-m)!} \sqrt{(j+m)!}}{\sqrt{(2 j)!}}\ K_{m+j}\,.
\end{equation}
Then, from Eq.\ \eqref{CWP-6} we get
\begin{equation}\label{KM-3}
    \left(a (n-j)+z\right)K_n+(2 j-n+1) K_{n-1}+(n+1) K_{n+1} =0\,,
\end{equation}
for $n=1,2,\cdots,2j-1$ and
\begin{equation}\label{KM-4}
    \begin{array}{c}\displaystyle
    \left(-a j+z\right)K_0+ K_{1}=0\,,
      \\ \\ \displaystyle
      \left(a j+z\right)K_{2j}+ K_{2j-1}=0\,,
    \end{array}
\end{equation}
for $n=0$ and $n=2j$ respectively.

We define the generating function of the coefficients $K_n$ as
\begin{equation}\label{KM-5}
    f(x):= \sum_{n=0}^{2j} K_n x^n
\end{equation}
and, for constant $z$, transform the recursion relation in Eq.\ \eqref{KM-3} into the linear differential equation
\begin{equation}\label{KM-6}
    (2 j x+z-a j) f(x)+\left(a x-x^2+1\right) f'(x)=0\,,
\end{equation}
whose solution can be written as
\begin{equation}\label{KM-7}
    f(x)=\left(1-\frac{x}{b}\right)^{j+\frac{b
   z}{b^2+1}} (1+b\, x)^{j-\frac{b z}{b^2+1}}\,.
\end{equation}

The coefficients derived from $f(x)$ can be expressed in terms of Jacobi polynomials,
\begin{equation}\label{KM-8}
     K_n= \frac{1}{n!}\, f^{(n)}(0) = \frac{(-1)^n}{ b^{n}} P_n^{\left(\frac{b z}{b^2+1}+j-n,-2 j-1\right)}\left(1+2 b^2\right)\,.
\end{equation}
These coefficients are polynomial in $z$ of degree $n$ which satisfy Eq.\ \eqref{KM-3} for any $n\geq 0$, as can be easily verified.

\smallskip

Let us first consider those ($2j+1$-dimensional) irreducible representations with $j\leq m_{0,j}$. In this case, $\Theta\big(m-{m_{0,j}}\big)=0$ for all $-j \leq m \leq j$ and the particle does not feel the step potential. Therefore, the corresponding Hamiltonian eigenvalues are determined by the second condition in Eq.\ \eqref{KM-4},
\begin{equation}\label{KM-9}
   \begin{array}{c} \displaystyle
      \left(a j+z\right)
     \frac{(-1)^{2j}}{ b^{2j}} P_{2j}^{\left(\frac{b z}{b^2+1}-j,-2 j-1\right)}\left(1+2 b^2\right)
    +\frac{(-1)^{2j-1}}{ b^{2j-1}} P_{2j-1}^{\left(\frac{b z}{b^2+1}-j+1,-2 j-1\right)}\left(1+2 b^2\right)
    \\  \\ \displaystyle
     = - (2j+1) \frac{(-1)^{2j+1}}{ b^{2j+1}} P_{2j+1}^{\left(\frac{b z}{b^2+1}-j-1,-2 j-1\right)}\left(1+2 b^2\right)= 0\,,
   \end{array}
\end{equation}
where the recursion for these polynomials for $n=2j$ has been employed.
Then, the $2j+1$ roots of the polynomial in $z$
\begin{equation}\label{KM-10}
    P_{2j+1}^{\left(\frac{b z}{b^2+1}-j-1,-2 j-1\right)}\left(1+2 b^2\right)= 0\,, \quad
    z_r\,, r=1,2,\cdots,2j+1\,,
\end{equation}
give the eigenvalues as
\begin{equation}\label{KM-11}
    E_r= z_r \, \frac{ \hbar ^2 \sqrt{\frac{\theta  \kappa }{\hbar^2}-1}}{\theta  \mu }
    +\frac{(2 j+1) \kappa }{2 \mu } \,.
\end{equation}

\smallskip

On the other hand, for those irreducible representations with $j > m_{0,j}$, one must construct a second set of coefficients satisfying the recursion with the shifted value of $z$ as well as the second line in Eq.\  \eqref{KM-4}, and then determine the spectrum by imposing the two \emph{matching conditions} corresponding to Eq.\ \eqref{KM-3} with $m=m_{0,j}$ and $m=m_{0,j}+1$.

For definiteness, let us refer to the infinite cylindrical well potential. From Eqs.\ \eqref{KM-3} and \eqref{KM-1} one sees that $C_m \sim O\left({V_0}^{-1}\right)$ for any $m\geq m_{0,j}+1$. Therefore, in the $V_0 \rightarrow \infty$ limit we must have
\begin{equation}\label{KM-11-1}
     \left(a m_{0,j}+z\right)K_{m_{0,j}+j}+(j-m_{0,j}+1) K_{m_{0,j}+j-1} =0\,,
\end{equation}
which, according to the recursion satisfied by the polynomials in Eq.\ \eqref{KM-8}, implies that the eigenvalues are determined by the roots of  the $\left({{m_{0,j}+j+1}}\right)$-degree polynomial in $z$
\begin{equation}\label{KM-12}
    P_{{m_{0,j}+j+1}}^{\left(\frac{b z}{b^2+1}-m_{0,j}-1,-2 j-1\right)}\left(1+2 b^2\right)=0\,.
\end{equation}

Consequently, in this irreducible representation there are $\left({{m_{0,j}+j+1}}\right)$ linearly independent Hamiltonian eigenvectors, while those states with $m\geq m_{0,j}+1$ (mean squared radius grater than the well squared radius) are not accesibles to the particle.



\section{Conclusions and discussion}

In the framework of Quantum Mechanics on the noncommutative phase space, we have studied in this article the algebraic structure of the extension to such spaces of two-dimensional nonrelativistic rotationally invariant Hamiltonians.

In so doing, we have considered nonvanishing commutators not only for coordinates but also for momenta, as in Eq.\ \eqref{1}. With no loose of generality, we have taken $\theta>0$. In this phase space, we have constructed the generators of coordinate translations, $\hat{K}_i$ (playing a role similar to that of the magnetic translations), and the rotations generator on both the coordinates and momenta planes, $\hat{L}$, as well as operators which realize the discrete transformations of parity and time-reversal, $\hat{\mathcal{P}}$ and $\hat{\mathcal{T}}$ (defined up to an $Sp(4,\mathbb{R})$ transformation).  These operators reduce to their ordinary counterpart in the $\kappa,\theta \rightarrow 0$ limit, but they are singular at the critical value of the momentum noncommutativity parameter, $\kappa \rightarrow \kappa_c=\hbar^2/\theta$. At this point, a dimensional reduction takes place, since the system can there be described in terms of only one pair of conjugate dynamical variables. This fact requires separate descriptions for the two regions $\kappa>\kappa_c$ and $\kappa<\kappa_c$ and then, as remarked in \cite{Bellucci,Poly}, these systems present two quantum phases, of which only the second one can be connected with the ordinary (commutative) case.

Irrespective of the region, we have found that the rotationally invariant Hermitian quadratic forms in the noncommutative dynamical variables generate a three-dimensional Lie algebra, which corresponds to the compact algebra $su(2)$ for  $\kappa>\kappa_c$ and to the noncompact algebra $sl(2,\mathbb{R})$ for $\kappa<\kappa_c$. In particular, in the last case the realization of the standard generators are somewhat different for $\kappa<0$ than for $0<\kappa<\kappa_c$. Therefore, we have had to consider three non overlapping parameter regions.

We have also shown that, in all cases, the quadratic Casimir operator is given in terms of ${\hat{L}}^2$, relation which imposes a constraint on the physically sensible unitary irreducible representations of the direct product of $SU(2)$ or $SL(2,\mathbb{R})$ with the rotations group $SO(2)$. Moreover, even though the rotationally invariant quadratic forms and $\hat{L}$ do not transform in general under $\hat{\mathcal{P}}$ or $\hat{\mathcal{T}}$ as their ordinary counterparts do and their transformation rules differ from one region to the other, they are all $\hat{\mathcal{P}}\hat{\mathcal{T}}$ invariant.

It is worth noticing that this algebraic structure is a consequence of the noncommutativity properties of both coordinates and momenta, since the generators of these nonabelian Lie algebras do not have a well defined double limit for $\kappa,\theta \rightarrow 0$. Only the Casimir operator has a well defined behavior since $\hat{L}$ reduces in this limit to the ordinary angular momentum.

\hfill\break

The noncommutative extension of the most general nonrelativistic rotationally invariant Hamiltonian is a function of the Hermitian invariants $\hat{\bf P}^2,\hat{\bf X}^2$, $(\hat{\bf P}\cdot\hat{\bf X} +\hat{\bf X}\cdot\hat{\bf P})$ and $\hat{L}$ itself (So, it is also $\hat{\mathcal{P}}\hat{\mathcal{T}}$ invariant). Therefore, it can be expressed in terms of the generators of the corresponding three-dimensional Lie algebra and $\hat{L}$ only, and the Hamiltonian characteristic subspaces are necessarily contained in the representation spaces of unitary irreducible representations of the direct product $SU(2) \otimes SO(2)$ or $SL(2,\mathbb{R})\otimes SO(2)$ (according to $\kappa>\kappa_c$ or $\kappa<\kappa_c$ respectively) which also satisfy the constraint between the quadratic Casimir and the (integer) eigenvalues of $\hat{L}$.

It is worth to remark at this point that, while the unitary irreducible representations of $SU(2)$ are of finite dimension, those of  $SL(2,\mathbb{R})$ are not. Moreover, the positivity of $\hat{\mathbf{X}^2}$ (or, equivalently, $\hat{\mathbf{P}^2}$), followed from the assumed Hermiticity of the dynamical variables, selects those irreducible representations of $SL(2,\mathbb{R})$ in the \emph{discrete class} \cite{Bargmann,Vega} in which the representations are characterized by an integer or half-integer $k$ which bounds from below the eigenvalues of $\mathcal{J}_0$.

From this algebraic point of view, we have analyzed the simple cases of the (noncommutative extensions) of the isotropic harmonic oscillator and the Landau model, getting results completely consistent with the well known ones obtained from \emph{Bopp' s  shifts} of ordinary dynamical variables. Indeed, by considering the relevant unitary irreducible representations of the group $SU(2) \otimes SO(2)$ for the region $\kappa > \kappa_c$ and  those of $SL(2,\mathbb{R})\otimes SO(2)$ for  $0<\kappa<\kappa_c$ or $\kappa<0$,  we have obtained the expression of the Hamiltonian eigenvectors of the extended isotropic oscillator and shown that its spectrum (in the three regions) is equivalent to that of two uncoupled harmonic oscillators with frequencies dependent on both noncommutativity parameters, in agreement with the results obtained in \cite{Poly,Bellucci}, for example. This spectrum is time-reversal invariant, even though the Hamiltonian has not this symmetry, which is consistent with the mapping onto a pair of (time-reversal invariant) ordinary harmonic oscillators.

Similarly, for the Landau model in the different parameter regions we have obtained the eigenvectors and shown that the spectrum corresponds to Landau levels with infinite degeneracy and a constant gap between them, which depends on an \emph{effective magnetic field} receiving corrections from both noncommutativity parameters, also in agreement with known results. In particular, for this system one can study the mean \emph{square radius} in the Hamiltonian eigenvectors and define a \emph{density of states} on the noncommutative plane which reduces to the ordinary one in the commutative limit and diverges for $\kappa \rightarrow  \kappa_c$ from both sides, where the dimensional reduction takes place.

{We have also remarked that the contributions to the spectrum and density coming from $\theta$ and $\kappa$ have different dependencies on $B$, which gives the possibility that an experiment performed at different values of the external magnetic field could establish some bounds on  the noncommutativity parameters.}

\hfill\break

We have also considered the case of a general central potential $V(\mathbf{\hat{X}}^2)$ within this algebraic approach, showing that the Hamiltonian eigenvalue problem, when expressed in relation with the representation space of a unitary irreducible representation of the corresponding group, is reduced to a \emph{linear three-term recursion relation} for the Fourier coefficients of the eigenvectors. Indeed, for $\kappa>\kappa_c$, a unitary $SU(2)$ transformation (which preserves the spectrum) diagonalizes $\mathbf{\hat{X}}^2$ and, then, the central potential. Since $\mathbf{\hat{P}}^2$ is also expressed in terms of the generators, it has definite transformation rules which finally reduce the eigenvalue equation to a recursion involving three consecutive coefficients of the solution expansion. Linearity allows to arbitrarily fix the first coefficient, the first recursion step determines the second one and the successive coefficients are recursively obtained as polynomials in the eigenvalue of growing degree. Since the irreducible representations of $SU(2)$ are of finite dimension, there is a \emph{last} equation which determines the energies as the roots of a polynomial whose degree depends on the dimension of the representation.

Similarly, for $\kappa<\kappa_c$ there is a unitary transformation in $SL(2,\mathbb{R})$ which diagonalizes the central potential and transforms in a definite way the kinetic term. Expressing the eigenvector in terms of the basis of a unitary irreducible representation of this group in the discrete class, one also gets a linear three-term recursion relation for the Fourier coefficients. Since $\mathcal{J}_0$ is bounded below, there is a first equation determining the second coefficient in terms of the first one (which can be chosen equal to 1), from which one can recursively determine the remaining coefficients. Since in this case the irreducible representations are not finite-dimensional, to get the bound states one must impose a normalizability condition (Bessel's inequality), which finally determines the spectrum.

These ideas were applied to the case of a \emph{cylindrical well potencial}. Since the coordinates are represented by Hermitian operators on a Hilbert space, the \emph{boundary} of the well, separating the \emph{interior} from the \emph{exterior}, is implemented by means of an orthogonal projector onto a subspace of the representation space defined by a condition on the mean square radius. This allows to reduce the eigenvalue equation to a linear three-term recursion for the Fourier coefficients, in which the potential term changes when going from the interior to the exterior. This occurs for a particular eigenvalue of $\mathcal{J}_0$, and the equations involving the corresponding coefficient imposes a set of \emph{matching conditions} which appears as an \emph{effective boundary condition}. These kind of recursion relations (with constant potential) admits two linearly independent solutions, which gives one the necessary freedom to satisfy the first equation and the matching conditions, to finally determine the eigenvalues by imposing the normalizability condition.

Following this scheme we have been able to solve first the case of the cylindrical well potential with $\kappa=0$. We have found the two linearly independent solutions of the recursion for the interior and the exterior of the well (which can be expressed in terms of Laguerre polynomial and Kummer's confluent hypergeometric functions), and taken linear combinations which satisfy the first equation in the recursion and the normalizability condition. The eigenvalues are then determined by (the imposition of) the matching conditions. This problem simplifies in the case of an infinite well potential, for which the eigenvectors belong to the interior subspace and the eigenvalues are determined by the zeroes of a Laguerre polynomial. These results are in complete agreement with those in \cite{scholtz}, where this problem has been considered in a formulation of the noncommutative plane with the space of quantum states of the system realized in terms of Hilbert-Schmidt class operators acting on an auxiliary Hilbert space.

\hfill\break

We have also studied the cylindrical well potential for the different regions of the $\kappa$ parameter. In so doing, for each irreducible representation, we have translated the recursion relation for the Fourier coefficients of the eigenvector expansion corresponding to each subspace with constant potential (the \emph{interior} and the \emph{exterior} of the well) into a homogeneous differential equation for the \emph{generating function} of these coefficients. This is a first order differential equation which can be solved employing the first equation in the recursion as an initial condition, so that the resulting coefficients are expressed in terms of Jacobi polynomials (or, equivalently, in terms of Gaussian hypergeometric functions). A linearly independent solution of the recursion (except for the first equation) can be derived from a second generating function which solves the same differential equation with a constant inhomogeneity.

Once more, for the cases $0<\kappa<\kappa_c$ and $\kappa<0$, the matching conditions and the Bessel inequality determine the spectrum. The problem simplifies for the infinite well potential, for which the Fourier coefficients corresponding to the exterior subspace vanish and the recursion relations for the Jacobi polynomials determine the eigenvalues in terms of the roots of a polynomial expression in the energy, condition which can also be expressed in terms of a hypergeometric function whose parameters depend on the energy eigenvalue. It is worth noticing that the $\kappa \rightarrow 0$ limit of the generating function is smooth and reduces to the generating function of Laguerre polynomials, thus reproducing the results previously described for this particular case.

The analysis for the case $\kappa>\kappa_c$ is similar, except that the normalizability condition is replaced by the last recursion equation, since in this case the irreducible representations are of finite dimension. One can also introduce here a generating function, obtained as the solution of a first order differential equation with an initial condition, to get (for constant potential) the coefficients expressed in terms of Jacobi polynomials. The spectrum of the infinite well potential is determined by the zeroes of a polynomial expression which can also be related to a hypergeometric function, similarly to the previous case. As far as we know, these results have not been previously obtained.

\hfill\break

In conclusion, we have shown that the spectrum of rotationally invariant nonrelativistic Hamiltonians extended to the noncommutative phase space with nonvanishing constant noncommutativity in both coordinates and momenta can be derived directly from the properties of the nonabelian Lie algebra generated by the rotationally invariant quadratic forms in the noncommutative dynamical variables, with no need of an explicit realization in terms of ordinary dynamical variables. Indeed, in each region of parameters, these Hamiltonians can be expressed as functions of the generators of these three-dimensional algebras and the generator of rotations on the coordinates and momenta planes (which is related to the quadratic Casimir), and the analysis can be reduced to the relevant unitary irreducible representations of these groups.

\smallskip
\hfill \break

\noindent \textbf{Acknowledgments:}  We are indebted with Jorge Gamboa for useful discussions. H.F. and P.A.G.P. thank ANPCyT, CONICET  and UNLP, Argentina, for partial support through grants \emph{PICT-2011-0605}, \emph{PIP-112-2011-01-681} and \emph{Proy.\ Nro.\ 11/X615} respectively. F.V. thanks support from CONICET, Argentina. D.C.\  thanks support from FONDECyT, Chile. F.M. thanks FONDECyT, Chile, for partial support though grant \emph{FONDECyT 1140243}. M.L. thanks FONDECyT, Chile, for partial support though grants \emph{FONDECyT 1130056} and \emph{1120770}.

\medskip

\appendix


\section{Unitary transformations in $SL(2,\mathbb{R})$ and $SU(2)$}\label{rot-sl2r}

Let us first recall that the generators in the fundamental (nonunitary  irreducible) representation of $SL(2,\mathbb{R})$ can be chosen as \cite{Vega}
\begin{equation}\label{transf-2}
    X_0=-\frac{1}{2}\, \sigma_2 \,, \quad X_1= \frac{\imath}{2}\, \sigma_1\,, \quad X_2= \frac{\imath}{2}\, \sigma_3\,,
\end{equation}
where $\sigma_i\,, i=1,2,3,$  are the Pauli matrices. These generators satisfy the commutation relations
\begin{equation}\label{transf-3}
    \left[X_\mu , X_\nu \right]= -\imath \epsilon_{\mu,\nu,\lambda} X^\lambda\,,
\end{equation}
where $X^\mu=\eta^{\mu\nu}X_\nu$, with $\left( \eta^{\mu\nu} \right)= {\rm diag}\left( 1,-1,-1\right)$.

It is straightforward to verify by direct computation that
\begin{equation}\label{transf-1}
    e^{\displaystyle {\imath} \alpha X_2}
    \left( A X_0 + B X_1 \right)
    e^{\displaystyle-{\imath} \alpha  X_2}
    =A \sqrt{1-\frac{B^2}{A^2}} X_0
\end{equation}
for $A,B\in \mathbb{R}$ with $|A|>|B|$, if we take $\tanh \alpha= B/A$.

Then, in any unitary representation of $SL(2,\mathbb{R})$ (with Hermitian generators $\mathcal{J}_i$) we also have
\begin{equation}\label{transf-4-ap}
    e^{\displaystyle{\imath} \alpha \mathcal{J}_2}
    \left( A \mathcal{J}_0 + B \mathcal{J}_1 \right)
    e^{\displaystyle-{\imath} \alpha \mathcal{J}_2}
    = A \sqrt{1-\frac{B^2}{A^2}} \mathcal{J}_0\,.
\end{equation}
Indeed, the left hand side of Eq.\ \eqref{transf-4-ap} can be written as
\begin{equation}\label{transf-5-ap}
    \begin{array}{c}\displaystyle
       \sum_{k=0}^\infty \frac{\left(i \alpha\right) ^k}{k!}\, \left( \left[ \mathcal{J}_2, \cdot \right] ^k \left( A \mathcal{J}_0 + B \mathcal{J}_1 \right)  \right)=
       \\ \\ \displaystyle
      = \left(A \cosh \alpha -B \sinh \alpha\right) \mathcal{J}_0 +
       \left(B \cosh \alpha -A \sinh \alpha\right) \mathcal{J}_1\,.
      \end{array}
\end{equation}
The coefficient of  $\mathcal{J}_1$ on the right hand side of this equation vanishes if we choose $\tanh \alpha= B/A$ (for $|B/A|<1$), in which case one gets the right hand side of Eq.\ \eqref{transf-4-ap}.

\smallskip

Similarly,
\begin{equation}\label{trans-sl2r-general}
   \begin{array}{c} \displaystyle
      e^{\displaystyle{\imath} \beta \mathcal{J}_1}
    \left( A \mathcal{J}_0 + B \mathcal{J}_2 \right)
    e^{\displaystyle-{\imath} \beta \mathcal{J}_1}=
    \\ \\ \displaystyle
     =\left(A \cosh \beta + B \sinh \beta \right) \mathcal{J}_0 +
     \left(A \sinh \beta + B \cosh \beta \right) \mathcal{J}_2\,.
   \end{array}
\end{equation}
Therefore, by choosing $\tanh \beta=-B/A$, for real $A,B$ with $|A|>|B|$, one gets
\begin{equation}\label{transf-44-ap}
    e^{\displaystyle{\imath} \beta \mathcal{J}_1}
    \left( A \mathcal{J}_0 + B \mathcal{J}_2 \right)
    e^{\displaystyle-{\imath} \beta \mathcal{J}_1}
    = A \sqrt{1-\frac{B^2}{A^2}} \mathcal{J}_0\,.
\end{equation}
.

\medskip

On the other hand, for any unitary representation of $SU(2)$, with Hermitian generators $J_i\,, i=1,2,3$ and $A,B\in \mathbb{R}$, we immediately obtain
\begin{equation}\label{transf-su2-gen}
    \begin{array}{c} \displaystyle
      e^{i \varphi J_2} \left(A J_3 +B J_1 \right)
    e^{-i \varphi J_2}=
    \\ \\ \displaystyle
      =  \left( A \cos \varphi + B \sin \varphi \right) J_3
    + \left(- A \sin \varphi + B \cos \varphi \right) J_1\,.
    \end{array}
\end{equation}
Then, taking  $\tan\varphi=B/A$ with $\varphi\in \left(- \frac{\pi}{2}, \frac{\pi}{2}\right)$ we get
\begin{equation}\label{transf-6-ap}
    e^{\displaystyle{\imath} \varphi {J}_2}
    \left( A {J}_3 + B {J}_1 \right)
    e^{\displaystyle-{\imath} \varphi {J}_2}
    = A \sqrt{1+\frac{B^2}{A^2}} {J}_3\,.
\end{equation}


\section{Relation between $J^2$ and $\hat{L}^2$}\label{Jcuad-Lcuad}

Let us define the operators
\begin{equation}\label{A1}
    \mathbf{A} = \frac{\mathbf{X}}{\sqrt{\theta}}+i \, \frac{\mathbf{P}}{\sqrt{\kappa}}\,,
    \quad  \mathbf{A}^\dagger = \frac{\mathbf{X}}{\sqrt{\theta}} - i \, \frac{\mathbf{P}}{\sqrt{\kappa}}\,.
\end{equation}
We have
\begin{equation}\label{A2}
    \begin{array}{c} \displaystyle
      \left[A_k , A_l\right] =0\,,\quad
    \left[A_k^\dagger , A_l^\dagger\right] =0\,,\quad
     \left[A_k , A_l^\dagger \right] = 2 i \epsilon_{k l} +2\sqrt{\frac{\hbar^2}{\kappa\theta}} \, \delta_{k l}\,,
     \end{array}
\end{equation}
which imply that
\begin{equation}\label{A22}
    \begin{array}{c}\displaystyle
     \left[ 2i \epsilon_{k l} A_k^\dagger   A_l  ,  A_i^\dagger   A_i    \right] = 0\,,
    \end{array}
\end{equation}
and
\begin{equation}\label{A222}
    \begin{array}{c}\displaystyle
      \left[A_k A_k , A_l^\dagger A_l^\dagger \right]=2 \left[A_k , A_l^\dagger \right] \left\{ \left[A_k , A_l^\dagger \right] +2  A_l^\dagger  A_k  \right\}=
       \\ \\ \displaystyle
     = -8 i \epsilon_{k l}   A_k^\dagger   A_l +8\sqrt{\frac{\hbar^2}{\kappa\theta}} A_k^\dagger   A_k
     -16 \left(1-\frac{\hbar^2}{\kappa\theta}\right)\,.
    \end{array}
\end{equation}

In particular,  we get
\begin{equation}\label{A3}
    \begin{array}{c} \displaystyle
       \left( A_k^\dagger A_k\right)  \left(A_l^\dagger  A_l \right)=   A_k^\dagger \left[A_k , A_l^\dagger\right]  A_l + A_k^\dagger  A_l^\dagger  A_k  A_l =
       \\ \\ \displaystyle
       =  2 i \epsilon_{k l} A_k^\dagger   A_l  +2\sqrt{\frac{\hbar^2}{\kappa\theta}} \,A_k^\dagger   A_k   + A_k^\dagger  A_l^\dagger  A_k  A_l \,,
    \end{array}
\end{equation}
\begin{equation}\label{A4}
   \begin{array}{c} \displaystyle
      \left(  i \epsilon_{k l} A_k^\dagger A_l  \right)   \left(  A_i^\dagger   A_i \right)  =   i \epsilon_{k l} A_k^\dagger   \left[ A_l ,   A_i^\dagger\right]   A_i
    +i \epsilon_{k l} A_k^\dagger A_i^\dagger A_l    A_i=
    \\ \\ \displaystyle
     = 2  A_k^\dagger A_k + 2 i \sqrt{\frac{\hbar^2}{\kappa\theta}} \epsilon_{k l} A_k^\dagger A_l
     +i \epsilon_{k l} A_k^\dagger A_i^\dagger A_l    A_i \,,
   \end{array}
\end{equation}
\begin{equation}\label{A5}
    \begin{array}{c}
     \left( i \epsilon_{k l} A_k^\dagger   A_l \right)  \left(i \epsilon_{i j}  A_i^\dagger   A_j\right)
     = -  \epsilon_{k l}  \epsilon_{i j} \left\{ A_k^\dagger   \left[ A_l  , A_i^\dagger\right]   A_j +A_k^\dagger  A_i^\dagger   A_l   A_j\right\}
      \\ \\
      = 2i \epsilon_{k l} A_k^\dagger   A_l +2 \sqrt{\frac{\hbar^2}{\kappa\theta}}  A_k^\dagger   A_k
       -  \epsilon_{k l}  \epsilon_{i j}   A_k^\dagger  A_i^\dagger   A_l   A_j \,,
    \end{array}
\end{equation}
and
\begin{equation}\label{A6}
    A_k A_k A_l^\dagger A_l^\dagger=
    A_l^\dagger A_l^\dagger  A_k A_k
     -8 i \epsilon_{k l}   A_k^\dagger   A_l +8\sqrt{\frac{\hbar^2}{\kappa\theta}} A_k^\dagger   A_k
     -16 \left(1-\frac{\hbar^2}{\kappa\theta}\right)\,.
\end{equation}

For $\kappa>\kappa_c$, we can write
\begin{equation}\label{A7}
    \begin{array}{c} \displaystyle
   \frac{ \hat{L}}{\hbar}=\frac{{\sqrt{\frac{\kappa\theta}{\hbar^2}}}}{2\left(1-\frac{\kappa\theta}{\hbar^2}\right)}
    \left\{ -i   \epsilon_{k l} A_k^\dagger  A_l + \sqrt{\frac{\kappa\theta}{\hbar^2}}A_k^\dagger  A_k \right\}\,,
      \\ \\ \displaystyle
     J_3
     = \frac{1}{4\left(1-\frac{\hbar^2}{\kappa\theta} \right)}  \left\{ i   \epsilon_{k l} A_k^\dagger  A_l - \sqrt{\frac{\hbar^2}{\kappa\theta}}A_k^\dagger  A_k \right\} +\frac{1}{2} \,,
       \\ \\ \displaystyle
     J_1 = \frac{1}{8\sqrt{1-\frac{\hbar^2}{\kappa\theta}}} \left\{ A_k^\dagger A_k^\dagger + A_k  A_k \right\}\,,
      \\ \\ \displaystyle
       J_2 = \frac{i}{8\sqrt{1-\frac{\hbar^2}{\kappa\theta}}} \left\{ A_k^\dagger A_k^\dagger - A_k  A_k \right\}\,,
    \end{array}
\end{equation}
from which we get
\begin{equation}\label{A8}
    \begin{array}{c}  \displaystyle
     4 \mathbf{J}^2 = 1  +   \frac{1}{4\left(1-\frac{\hbar^2}{\kappa\theta} \right)^2}  \left\{ i   \epsilon_{k l} A_k^\dagger  A_l - \sqrt{\frac{\hbar^2}{\kappa\theta}}A_k^\dagger  A_k \right\}^2 +
      \\  \\  \displaystyle
           +    \frac{1}{\left(1-\frac{\hbar^2}{\kappa\theta} \right)}  \left\{ i   \epsilon_{k l} A_k^\dagger  A_l - \sqrt{\frac{\hbar^2}{\kappa\theta}}A_k^\dagger  A_k \right\}
       + \frac{1}{8\left(1-\frac{\hbar^2}{\kappa\theta} \right)} \left\{ A_k^\dagger A_k^\dagger  A_l  A_l
           + A_l  A_l A_k^\dagger A_k^\dagger \right\}\,.
    \end{array}
\end{equation}

A straightforward calculation which takes into account Eqs.\ \eqref{A2} - \eqref{A6} shows that
\begin{equation}\label{A9}
    4 \mathbf{J}^2 - \left( \frac{\hat{L}}{\hbar}\right)^2 = -1\,.
\end{equation}

\smallskip

For $\kappa<\kappa_c$, we write
\begin{equation}\label{A10}
    \begin{array}{c} \displaystyle
     \mathcal{J}_0     = \frac{-1}{4\left(\frac{\hbar^2}{\kappa\theta}-1 \right)}
     \left\{ i   \epsilon_{k l} A_k^\dagger  A_l - \sqrt{\frac{\hbar^2}{\kappa\theta}}A_k^\dagger  A_k \right\} +\frac{1}{2} \,,
       \\ \\ \displaystyle
     \mathcal{J}_1 = \frac{1}{8\sqrt{\frac{\hbar^2}{\kappa\theta}-1}} \left\{ A_k^\dagger A_k^\dagger + A_k  A_k \right\}\,,
      \\ \\ \displaystyle
       \mathcal{J}_2 = \frac{i}{8\sqrt{\frac{\hbar^2}{\kappa\theta}-1}} \left\{ A_k^\dagger A_k^\dagger - A_k  A_k \right\}\,,
    \end{array}
\end{equation}
and
\begin{equation}\label{11}
    \begin{array}{c}  \displaystyle
     4 {\mathcal{J}}^2 = 1  +   \frac{1}{4\left(\frac{\hbar^2}{\kappa\theta} -1\right)^2}  \left\{ i   \epsilon_{k l} A_k^\dagger  A_l - \sqrt{\frac{\hbar^2}{\kappa\theta}}A_k^\dagger  A_k \right\}^2 -
      \\  \\  \displaystyle
           -    \frac{1}{\left(\frac{\hbar^2}{\kappa\theta}-1 \right)}  \left\{ i   \epsilon_{k l} A_k^\dagger  A_l - \sqrt{\frac{\hbar^2}{\kappa\theta}}A_k^\dagger  A_k \right\}
       - \frac{1}{8\left(\frac{\hbar^2}{\kappa\theta}-1 \right)} \left\{ A_k^\dagger A_k^\dagger  A_l  A_l
           + A_l  A_l A_k^\dagger A_k^\dagger \right\}\,,
    \end{array}
\end{equation}
from which one gets the same result as in Eq.\ \eqref{A9}.


\section{More algebraic structure}\label{L-Lie}

In Section \ref{NC-2dim-space} we have remarked that, independently of the linear realization of the dynamical variables chosen, the generator of rotations on the noncommutative plane $\hat{L}$ can be incorporated as one of the generators of an $su(2)$ or an $sl(2,\mathbb{R})$ Lie algebras, according to $\kappa$ is less or greater than the critical value $\kappa_c$ \cite{Bellucci,Mik-Hor-2}.

Indeed, let us now define for a vectorial operator $V_\pm:=(\hat{V}_1\pm \imath \hat{V}_2)/\sqrt{2}$ where $\mathbf{\hat{V}}$ stands for $\mathbf{\hat{X}}$, $\mathbf{\hat{P}}$ or $\mathbf{\hat{K}}$ (See Eq.\ (\ref{trnaslations1})).
Then, we get that $\hat{L}$ can be written as
\begin{equation}\label{escalera1}
    \begin{array}{c} \displaystyle
      \hat{L}=\frac{1}{\left(1-\frac{\theta\kappa}{\hbar^2}\right)}
    \left\{ \imath \left( X_+ P_- - X_- P_+\right) + \right.
    \\ \\ \displaystyle
      + \left. \frac{\theta}{2\hbar} \left( P_+ P_- + P_- P_+\right)
    + \frac{\kappa}{2\hbar} \left( X_+ X_- + X_- X_+\right)
    \right\}
    \end{array}
\end{equation}
and satisfy
\begin{equation}\label{escalera2}
    \left[\hat{L} , V_\pm\right] = \pm \hbar V_\pm\,.
\end{equation}

We now introduce the quadratic expressions $P_\pm K_\pm$. We have
\begin{equation}\label{escalera3}
    \begin{array}{c} \displaystyle
    \left[\hat{L} , P_\pm K_\pm\right] = \pm 2\hbar P_\pm K_\pm\,,
    \\ \\ \displaystyle
      \left[ P_+ K_+ , P_- K_- \right]=
      -\frac{\kappa}{\left(1-\frac{\theta\kappa}{\hbar^2}\right)}\, P_+P_- + \kappa K_- K_+ =
      \\ \\ \displaystyle
      =  \frac{\kappa^2}{\hbar \left(1-\frac{\theta\kappa}{\hbar^2}\right)}\, \hat{L}\,.
    \end{array}
\end{equation}

Therefore, for $\kappa<\kappa_c$ we can define the Hermitian operators
\begin{equation}\label{escalera4}
    J_3:=\frac{\hat{L}}{\hbar}\,, \quad
    J_\pm:= \frac{\sqrt{2}\sqrt{1-\frac{\theta\kappa}{\hbar^2}}}{|\kappa|}\, P_\pm K_\pm\,,
\end{equation}
which generate an $su(2)$ Lie algebra,
\begin{equation}\label{escalera5}
    \left[J_3 , J_\pm\right] = \pm J_\pm\,, \quad
    \left[J_+ , J_-\right] = 2 J_3 \,,
\end{equation}
with the quadratic Casimir operator given by
\begin{equation}\label{escalera-casimir1}
    \begin{array}{c} \displaystyle
      {J}^2=J_\pm J_\mp + J_3(J_3\mp 1)=
      \\ \\ \displaystyle
      =\frac{1}{2}\left\{\left(\frac{\mathbf{\hat{P}}^2}{\kappa} + \frac{\hat{L}}{\hbar}\right)^2
      +\frac{\hat{L}^2}{\hbar^2} - 1\right\}\,,
    \end{array}
\end{equation}
as can be straightforwardly verified. Notice that this operator commutes with
$\mathbf{\hat{P}}^2$ but not with $\mathbf{\hat{X}}^2$.

\medskip

On the other hand, for $\kappa>\kappa_c$ we define
\begin{equation}\label{escalera6}
    \mathcal{J}_0:=\frac{\hat{L}}{\hbar}\,, \quad
    \mathcal{J}_\pm:= \frac{\sqrt{2}\sqrt{\frac{\theta\kappa}{\hbar^2}-1}}{\kappa}\, P_\pm K_\pm\,,
\end{equation}
which generate an $sl(2,\mathbb{R})$ (or, equivalently, $su(1,1)$) Lie algebra,
\begin{equation}\label{escalera7}
    \left[\mathcal{J}_0 , \mathcal{J}_\pm\right] = \pm \mathcal{J}_\pm\,, \quad
    \left[\mathcal{J}_+ , \mathcal{J}_-\right] = -2\mathcal{J}_0 \,,
\end{equation}
with the Casimir operator ${\mathcal{J}}^2=\mathcal{J}_0(\mathcal{J}_0\mp 1)-\mathcal{J}_\pm \mathcal{J}_\mp$ taking the same expression as in the second line in Eq.\ (\ref{escalera-casimir1}).

Notice that while the unitary irreducible representations of $su(2)$ are of finite dimension, those of $sl(2,\mathbb{R})$ are infinite-dimensional.

\medskip

Similarly to the introduction of $K_i$, we can define translation operators on the noncommutative momenta plane,
\begin{equation}\label{trnaslations3}
    \hat{M}_i:=\frac{1}{\left(\frac{\theta\kappa}{\hbar^2}-1\right)}\left(
    \hat{X}_i+\frac{\theta}{\hbar}\,\epsilon_{i j} \hat{P}_j \right)\,,
\end{equation}
which satisfy the commutator algebra
\begin{equation}\label{translation4}
\begin{array}{c}\displaystyle
    \left[\hat{M}_i,\hat{X}_j\right]= 0 \,, \quad
    \left[\hat{M}_i,\hat{P}_j\right]=- \imath \hbar \delta_{i j}\,,
   \\ \\ \displaystyle
    \left[\hat{M}_i,\hat{M}_j\right]=\frac{ -\imath\theta\,\epsilon_{i j}}
    {\left(1-\frac{\theta\kappa}{\hbar^2}\right)}\,, \quad
    \left[\hat{L},\hat{M}_i\right]= \imath \hbar \,  \epsilon_{ij} \hat{M}_j \,.
\end{array}
\end{equation}
We also introduce the quadratic expressions  $X_\pm M_\pm$ which satisfy
\begin{equation}\label{escaleraM1}
    \begin{array}{c} \displaystyle
    \left[\hat{L} , X_\pm M_\pm\right] = \pm 2\hbar X_\pm M_\pm\,,
    \\ \\ \displaystyle
      \left[ X_+ M_+ , X_- M_- \right]=
      \frac{\theta^2}{\hbar \left(1-\frac{\theta\kappa}{\hbar^2}\right)}\, \hat{L}\,.
    \end{array}
\end{equation}

Then, for $\kappa<\kappa_c$ we can define the Hermitian operators
\begin{equation}\label{escaleraM2}
    N_3:=\frac{\hat{L}}{\hbar}\,, \quad
    N_\pm:= \frac{\sqrt{2}\sqrt{1-\frac{\theta\kappa}{\hbar^2}}}{\theta}\, X_\pm M_\pm\,,
\end{equation}
which generate an $su(2)$ Lie algebra,
\begin{equation}\label{escaleraM3}
    \left[N_3 , N_\pm\right] = \pm N_\pm\,, \quad
    \left[N_+ , N_-\right] = 2 N_3 \,,
\end{equation}
with the quadratic Casimir operator given by
\begin{equation}\label{escalera-casimir2}
    \begin{array}{c} \displaystyle
      {N}^2=N_\pm N_\mp + N_3(N_3\mp 1)=
      \\ \\ \displaystyle
      =\frac{1}{2}\left\{\left(\frac{\mathbf{\hat{X}}^2}{\theta} + \frac{\hat{L}}{\hbar}\right)^2
      +\frac{\hat{L}^2}{\hbar^2} -1 \right\}\,,
    \end{array}
\end{equation}
which commutes with $\mathbf{\hat{X}}^2$ but not with $\mathbf{\hat{P}}^2$.

On the other hand, for $\kappa>\kappa_c$ we define
\begin{equation}\label{escalera6-1}
    \mathcal{N}_0:=\frac{\hat{L}}{\hbar}\,, \quad
    \mathcal{N}_\pm:= \frac{\sqrt{2}\sqrt{\frac{\theta\kappa}{\hbar^2}-1}}{\kappa}\, P_\pm K_\pm\,,
\end{equation}
which generate an $sl(2,\mathbb{R})$ Lie algebra,
\begin{equation}\label{escalera7-1}
    \left[\mathcal{N}_0 , \mathcal{N}_\pm\right] = \pm \mathcal{N}_\pm\,, \quad
    \left[\mathcal{N}_+ , \mathcal{N}_-\right] = -2\mathcal{N}_0 \,,
\end{equation}
with the Casimir operator ${\mathcal{N}}^2=\mathcal{N}_0(\mathcal{N}_0\mp 1)-\mathcal{N}_\pm \mathcal{N}_\mp$ taking the same expression as in the second line in Eq.\ (\ref{escalera-casimir2}).




\begin{thebibliography}{99}



\bibitem{Peierls}{R.\ Peierls, Z.\ Phys.\ 80, 763 (1933).}

\bibitem{Dunne-Jackiw}{G.\ Dunne and R.\ Jackiw, {Nuclear Physics B (Proc. Suppl.)} 33C (1993) 114-118.}

\bibitem{jcm2}{J. Gamboa,  F. Mendez, M. Loewe and J. C. Rojas, {Mod. Phys. Lett.}{\bf  A 16}, 2075 (2001);  {Int. J. Mod. Phys.}{\bf  A 17}, 2555 (2002).}

\bibitem{Horvathy}{P.A.\ Horv\'{a}thy, Annals of Physics 299, 128–140 (2002).}

\bibitem{Heisenberg-1930}{W.\ Heisenberg,
Z.\ Phys. 65, 4-13.}

\bibitem{snyder1} {H.S.\ Snyder,
Phys.\ Rev.\ 71, 38-41 (1947).}

\bibitem{yang}{C.N.\ Yang, {Phys. Rev.}{\bf  72}, 874 (1947).}

\bibitem{snyder2} {H.S.\ Snyder,
Phys.\ Rev.\ 72, 68-71 (1947).}

\bibitem{Green} {M.\ Green, J.H. Schwarz and E. Witten, {\emph{Superstring theory}}, Cambridge University Press, Cambridge, 1987.}

\bibitem{witten} {N.\ Seiberg y E.\ Witten,
JHEP \textbf{9909} (1999) 032.}

\bibitem{DouNek}{M.\ Douglas and N.\ Nekrasov, {Rev. Mod. Phys.}  {\bf 73}, 977 (2001).}

\bibitem{DFR}{S.\ Doplicher,K.\ Fredenhagen and J.\ Roberts, Commun.\ Math.\ Phys.\ 172 (1995) 187.}

\bibitem{Madore}{J.\ Madore,
Ann.\ Phys.\ 219, 187–198 (1992).}

\bibitem{quant-grav}{D.\ Bigatti and L.\ Susskind, Phys.\ Rev.\ D 62 (2000) 066004.}

\bibitem{quant-grav-1}{N.\ Seiberg, L.\ Susskind and N.\ Toumbas, JHEP 06 (2000) 044.}

\bibitem{quant-grav-2}{ J.W.\ Moffat, Phys.\ Lett.\ B 491 (2000) 345.}

\bibitem{Szabo}{R.J.\ Szabo, {Quantum Field Theory on Noncommutative Spaces}, Physics Reports 378: 207-99 (2003). }

\bibitem{connes}{A.\ Connes, {Non-commutative Geometry}, (Academic Press, London, 1994).}

\bibitem{jackiw} {G.V.\ Dunne, J.\ Jackiw, and C.\ Trugenberger, Phys.\ Rev.\ D 41, 661 (1990).}

\bibitem{mezincescu}{L. Mezincescu, {\emph{Star Product in Quantum Mechanics}},  hep-th/0007046.}

\bibitem{jcm1}{J. Gamboa, M. Loewe and J. C. Rojas,  {Phys.\ Rev.\ }{\bf  D 64}, 067901 (2001).}


\bibitem{Carroll}{S.M.\ Carroll, J.A.\ Harvey, V.A.\ Kosteleck\'{y}, C.D.\ Lane and T.\ Okamoto, Phys.\ Rev.\ Lett.\ 87 (2001) 141601.}

\bibitem{Duval}{C.\ Duval and P.A.\ Horvathy; Phys.\ Lett.\ B 479 (2000) 284; J.\ Phys.\ A: Math.\ Gen.\ 34 (2001) 10097.}

\bibitem{Poly}{V.P.\ Nair and A.P.\ Polychronakos, Phys.\ Lett.\  B505 (2001) 267-274.}

\bibitem{Bellucci}{S.\ Bellucci, A.\ Nersessian and C.\ Sochichiu, Phys.\ Lett.\ B 522 (2001) 345.}

\bibitem{Morariu}{B.\ Morariu and A.P.\ Polychronakos, Nucl.\ Phys.\ B 610 (2001) 531.}

\bibitem{Chai1}{M.\ Chaichian, M.M.\ Sheikh-Jabbari and A.\ Tureanu, Phys.\ Rev.\ Lett.\ 86 (2001) 2716;  Eur.\ Phys.\ J.\ C 36 (2004) 251.}

\bibitem{Scha}{H.R.\ Christiansen and F.A.\ Schaposnik, Phys.\ Rev.\ D 65 (2002) 086005.}

\bibitem{Ho}{P.M.\ Ho and H.C.\ Kho, Phys.\ Rev.\ Lett.\ 88 151602 (2002).}

\bibitem{Smailagic}{A.\ Smailagic and E.\ Spallucci, J.\ Phys.\ A 36 (2003) 467.}

\bibitem{Lukierski}{J.\ Lukierski, P.C.\ Stichel and W.J.\ Zakrzewski, Ann.\ Physics 306 (2003) 78.}

\bibitem{Chakraborty}{B.\ Chakraborty, S.\ Gangopadhyay and A.\ Saha, Phys.\ Rev.\ D 70 107707 (2004).}

\bibitem{Scholtz2}{F.G.\ Scholtz, B.\ Chakraborty, S.\ Gangopadhyay and A.G.\ Hazra, Phys.\ Rev.\ D 71 (2005) 085005.}

\bibitem{Scholtz1}{F.G.\ Scholtz, B.\ Chakraborty, S.\ Gangopadhyay and J.\ Govaerts, J.\ Phys.\ A 38 (2005) 9849.}

\bibitem{Mik-Hor-2}{Peter A.\ Horvathy, Mikhail S.\ Plyushchay. Nucl.\ Phys.\ B714 (2005) 269-291.}

\bibitem{Mikhail1}{ Pedro D.\ Alvarez, Joaquim Gomis, Kiyoshi Kamimura, Mikhail S.\ Plyushchay, Annals of Phys.\ 322 (2007) 1556-1586; Phys.Lett. B659 (2008) 906-912.}

\bibitem{Mik-5}{Pedro D.\ Alvarez, Jose L.\ Cortes, Peter A.\ Horvathy, Mikhail S.\ Plyushchay, JHEP 0903 (2009) 034.}

\bibitem{Saha}{A.\ Saha, Eur.\ Phys.\ J.\ C 51 199 (2007); Phys.\ Rev.\  D 81 125002 (2010); Phys.\ Rev.\ D 89, 025010 (2014).}

\bibitem{Stern}{A.\ Stern, Phys.\ Rev.\ Lett.\ 100 (2008) 061601.}

\bibitem{Scholtz-formulation}{F.G.\ Scholtz, L.\ Gouba, A.\ Hafver and C.M.\ Rohwer, J.\ Phys.\ A: Math.\ Theor.\ 42 (2009) 175303.}



\bibitem{Acatrinei}{C.\ Acatrinei, JHEP {\bf 09} (2001) 007.}

\bibitem{Chai2}{M.\ Chaichian, A.\ Demichec, P.\ Presnajder, M.M.\ Sheikh-Jabbari, and A.\ Tureanu, {Phys.\ Lett.\ }{\bf  B 527}, 149 (2002).}

\bibitem{Falomir1}{H.\ Falomir, J.\ Gamboa, M.\ Loewe, F.\ Mendez, and J.C.\ Rojas, {Phys.\ Rev.\ }{\bf D 66}, 045018 (2002).}

\bibitem{Banerjee}{R.\ Banerjee, Mod.\ Phys.\ Lett.\ A 17, 631 (2002).}

\bibitem{Jonke}{L.\ Jonke and S.\ Meljanac, Eur.\ Phys.\ J.\ {\bf  C 29}, 433 (2003).}

\bibitem{Deriglazov}{A.A.\ Deriglazov, Phys.\ Lett.\ \textbf{B530} (2002), 235243; Phys.\ Lett.\ \textbf{B555} (2003), 8388; J.\ High Energy Phys.\ 2003 (2003), 021.}

\bibitem{Subir-1}{Subir Ghosh, Phys.\ Lett.\ B571 (2003) 97-104.}

\bibitem{Kijanka}{A.\ Kijanka and P.\ Kosinski,  Phys.\ Rev.\  {\bf D 70}, 127702 (2004).}

\bibitem{Kokado}{A.\ Kokado, T.\ Okamura and T.\ Saito, {Phys.\ Rev.\ } {\bf D 69}, 125007 (2004).}

\bibitem{Banerjee1}{R.\ Banerjee, C.\ Lee and H.S.\ Yang, Phys.\ Rev.\ \textbf{D70} (2004), 065015.}

\bibitem{Zhang-PLB-2004}{Jian-zu Zhang,
Physics Letters \textbf{B 584} (2004) 204-209.}

\bibitem{Zhang-PRL-2004}{Jian-zu Zhang,
Phys.\ Rev.\ Lett.\ \textbf{93}, (2004) 043002.}

\bibitem{Hor-Mik-1}{P.A.\ Horvathy and M.S.\ Plyushchay,
Nucl.\ Phys.\ B714 (2005) 269.}

\bibitem{Bertolami2}{O.\ Bertolami, J.G.\ Rosa, C.M.L.\ de Aragao, P.\ Castorina and D.\ Zappala,  Phys.\ Rev.\  {\bf  D 72}, 025010 (2005).}

\bibitem{Bemfica} {F.S.\ Bemfica and H.O.\ Girotti, J.\ Phys.\ A \textbf{38}, L539 (2005).}

\bibitem{Bellucci2}{S.\ Bellucci and A.\ Yeranyan, Phys.\ Rev.\ D72 (2005) 085010.}

\bibitem{Bellucci3}{S.\ Bellucci and A.\ Yeranyan, Phys.\ Lett.\  B609 (2005) 418-423.}

\bibitem{Banerjee2}{R.\ Banerjee and K.\ Kumar,Phys.\ Rev.\ \textbf{D72} (2005), 085012.}

\bibitem{Calmet}{X.\ Calmet and M.\ Selvaggi, Phys.\ Rev.\ {\bf  D 74}, 037901 (2006).}

\bibitem{scholtz} {F.G.\ Scholtz, B.\ Chakraborty, J.\ Govaerts and S.\ Vaidya, 
J.\ Phys.\ A: Math.\ Theor.\ \textbf{40}, 14581 (2007).}

\bibitem{Banerjee4}{R.\ Banerjee, P.\  Mukherjee and S.\ Samanta,  Phys.\ Rev.\ {\bf D 75}, 125020 (2007).}

\bibitem{Bastos}{C.\ Bastos, O.\ Bertolami, N.\ Dias and J.\ Prata,  Phys. Rev.\  {\bf D 78}, 023516 (2008).}

\bibitem{scholtz2} {J.D.\ Thom and F.G.\ Scholtz, 
J.\ Phys.\ A {\bf 42}, 445301 (2009).}

\bibitem{Banerjee3}{R.\ Banerjee, B.\ Chakraborty, Subir Ghosh, P.\ Mukherjee and S.\ Samanta, Found.\ Phys.\ 39 (2009) 1297.}

\bibitem{Gango}{S.\ Gangopadhyay  and F.G.\ Scholtz, Phys.\ Rev.\ Lett.\ 102 (2009), 241602.}

\bibitem{Gomes}{M.\ Gomes and V.G.\ Kupriyanov, {Phys. Rev.}{\bf  D 79}, 125011 (2009).}

\bibitem{Bertolami1}{O.\ Bertolami and C.\ Zarro,  Phys.\  Rev.\  {\bf  D 81}, 025005 (2010)}

\bibitem{Murray}{S.\ Murray and J.\ Govaerts, 
Phys.\ Rev.\ D {\bf 83}, 025009 (2011).}



\bibitem{Mikhail-4}{Mikhail S.\ Plyushchay, Electron.\ J.\ Theor.\ Phys.\ 3N10 (2006) 17-31.}

\bibitem{Mik-6}{
Pedro D.\ Alvarez, Jose L.\ Cortes, Peter A.\ Horvathy and Mikhail S.\ Plyushchay, JHEP 0903 (2009) 034.}

\bibitem{Kupri}{V.G.\ Kupriyanov, Physics Letters B732 (2014) 385–390.}

\bibitem{Panella}{O.\ Panella and P.\ Roy, Phys.\ Rev.\ A 90 (2014), 042111.}


\bibitem{Kupri-Dima}{V.G.\ Kupriyanov, D.V.\ Vassilevich,
Eur.\ Phys.\ J.\ C58 (2008) 627-637.}


\bibitem{Boundaries}{H.\ Falomir, S.A.\ Franchino Vi\~{n}as, P.A.G.\ Pisani, F.\ Vega,
JHEP 1312 (2013) 024.}


\bibitem{Fradkin}{E.\ Fradkin,  \emph{Field Theory of Condensed Matter Systems},
Addison-Wesley Publishing Company  (1991), Redwood City, California, USA.}

\bibitem{Wen}{Xiao-Gang Wen,  \emph{Quantum Field Theory of Many-Body Sistems}, Oxford University Press (2004), Oxford, Great Britain.}

\bibitem{Bargmann}{V.\ Bargmann, {Ann.\ Math.\ }\textbf{48}, 568 (1947).}

\bibitem{Vega}{H.\ Falomir, F.\ Vega, J.\ Gamboa, F.\ M\'{e}ndez and M.\ Loewe, Phys.\ Rev.\ \textbf{D86} (2012) 105035.}

\bibitem{Gango2}{Sunandan Gangopadhyay, Anirban Saha, Aslam Halder, Physics Letters \textbf{A 379} (2015) 2956–2961.}

\bibitem{A-S}{M.\ Abramowitz and I.\ Stegun, {\emph{Handbook of Mathematical Functions}}, Dover Publications, New York, 1972.}

\bibitem{Math6}{\emph{Wolfram Mathematica 6}, Wolfran Research, 2007.}

\bibitem{Bateman1}{{\emph{Higher Trascendental Functions}} Vol.\ 1, Staff of the Bateman Manuscript Proyect, Ed.\ Arthur Erd\'{e}lyi, McGraw HillPublishing Company, Florida, USA, 1953.}
\bibitem{MathWorld}{Weisstein, Eric W. {Jacobi Polynomial}. From MathWorld--A Wolfram Web Resource. http://mathworld.wolfram.com/JacobiPolynomial.html}

\end{thebibliography}
\end{document}